\documentclass[aip,graphicx]{revtex4-1}

\usepackage{graphicx}
\usepackage{dcolumn}
\usepackage{bm}

\usepackage[utf8]{inputenc}
\usepackage[T1]{fontenc}
\usepackage{mathptmx}
\usepackage{etoolbox}
\usepackage{chemformula}
\usepackage{multirow}

\makeatletter
\def\@email#1#2{%
 \endgroup
 \patchcmd{\titleblock@produce}
  {\frontmatter@RRAPformat}
  {\frontmatter@RRAPformat{\produce@RRAP{*#1\href{mailto:#2}{#2}}}\frontmatter@RRAPformat}
  {}{}
}%
\makeatother

\begin{document}

\title
{Relativistic Two-Electron Contributions within Exact Two-Component Theory}

\author{Xubo Wang}
\affiliation{Department of Chemistry, The Johns Hopkins University, Baltimore, MD 21218, USA}

\author{Chaoqun Zhang}
\affiliation{Department of Chemistry, Yale University, New Haven, CT 06520, USA}

\author{Junzi Liu}
\affiliation{Department of Chemistry, The Johns Hopkins University, Baltimore, MD 21218, USA}

\author{Lan Cheng}
\affiliation{Department of Chemistry, The Johns Hopkins University, Baltimore, MD 21218, USA}



\begin{abstract}
  The development of relativistic exact two-component (X2C) theory is briefly reviewed, with an emphasis on cost-effective treatments of relativistic two-electron contributions by means of model potential (MP) techniques and closely related atomic mean-field (AMF) approaches. The correct MP or AMF contribution to the electronic energy is elucidated. 
  The performance of one-center approximations to relativistic two-electron contributions is carefully assessed using benchmark calculations of molecular properties. 
  

\end{abstract}


\maketitle

\section{Introduction}

Relativistic effects \cite{Pyykko88,Dyall07,Autschbach12,Reiher14,Liu17} play important roles in heavy-element chemistry and physics. Four-component (4C) relativistic quantum-chemical methodologies \cite{Grant88,Eliav94,Visscher94,Saue97,Yanai01,Liu03,Storchi10,Komorovsky10,Kelley13,Kadek19,sun_2018_pyscf,Sun21}
in principle can provide accurate treatments of relativistic effects in atoms and molecules. However, the positronic degrees of freedom in four-component theories introduce significant computational overheads and limit the applicability of four-component methods to relatively small molecules.
Two-component (2C) methodologies \cite{VanLenthe96,Reiher06,Liu10,Saue11,Barysz01,Kutzelnigg12,Peng12,Nakajima12} focusing on electronic states thus are favorable options for applications in chemistry and low-energy physics, in which the typical energy scales are far lower than the separation between positronic and electronic states. Two-component approaches apply unitary transformation(s) to block-diagonalize the four-component Hamiltonian into an electronic block and a positronic block, and then focus the calculations on the electronic states. Among two-component approaches, \cite{Hess86,Nakajima12,Wolf02,VanLenthe93,Nakajima99,Barysz97,Barysz02,Filatov05,Liu06a,Liu07,Dyall07,Peng12} the ``exact two-component'' (X2C) \cite{Dyall97,Dyall01,Kutzelnigg05,Ilias07,Liu09} theory has emerged as the most promising method. The X2C theory features a one-step block-diagonalization of the four-component Hamiltonian in its matrix representation. The term ``exact'' in X2C comes from the fact that the resulting electronic Hamiltonian matrix exactly reproduces 
the electronic spectrum of the original four-component Hamiltonian matrix. 
{\ {We also refer the readers to Ref. \onlinecite{Xu09} for an anecdote about the acronym ``X2C''.}} X2C is more rigorous than two-component methods with approximate block-diagonalization schemes, while it is simpler than exact decoupling schemes with multiple transformations. The simple structure of the X2C working equations has enabled development of analytic X2C gradient techniques \cite{Zou11,Filatov14,Zou12,Cheng11b,Cheng11c,Cheng14,Franzke18,Zou20} for efficient calculations of molecular properties. \\

The exact block-diagonalization of the four-component Hamiltonian in the X2C scheme necessitates solving the four-component matrix eigenvalue equation and thus is at least as computationally expensive.
The practical gain in the computational efficiency of X2C calculations comes from the use of an X2C Hamiltonian matrix obtained from a simpler theory in a subsequent more sophisticated calculation to focus the latter on electronic states. The widely used ``X2C-1e'' scheme combines an X2C Hamiltonian obtained by block-diagonalizing the one-electron Dirac Hamiltonian matrix with untransformed two-electron Coulomb interactions in the many-electron treatments. Since scalar-relativistic two-electron interactions typically make small contributions to molecular properties, \cite{vanWullen05,Neese05a,Mastalerz07} the spin-free version of the X2C-1e scheme (the SFX2C-1e scheme) \cite{Dyall01,Liu09,Cheng11b} offers accurate treatments of scalar-relativistic effects and has been established as a standard method of choice for treating scalar-relativistic effects. In contrast, the relativistic two-electron spin-dependent interactions make important contributions to many molecular properties \cite{Blume62,Blume63,Malkina98,Fedorov00,Boettger00,Stopkowicz11b} and cannot be neglected when aiming at accurate treatments of spin-orbit effects. One may include relativistic two-electron interactions by performing the X2C decoupling at the mean-field (MF) level. \cite{Liu06,Sikkema09,Liu10} We focus the present discussion on X2CMF approaches for non-perturbative treatments of spin-orbit coupling. The most rigorous version of the X2CMF scheme, the so-called X2C ``molecular mean-field'' (X2CMMF) scheme, \cite{Liu06,Sikkema09} requires the evaluation of all molecular relativistic two-electron integrals in each iteration of a self-consistent-field calculation; it is as expensive as the corresponding four-component MF calculation and also complicates the computation of analytic gradients for the subsequent electron-correlation calculations. \cite{Kirsch19} It is thus of importance to develop X2C schemes that include relativistic two-electron spin-dependent interactions efficiently while retaining sufficient accuracy for calculations of molecular properties.  \\

Since small-component wave functions are highly localized in the vicinity of nuclei, \cite{deJong02} the small-component contributions to Fock matrix elements are largely transferable from atoms to molecule. Based on this insight, model potential (MP) schemes \cite{vanWullen05, Liu06, Peng07, Knecht22} 
applying atomic approximation to small-component contributions to Fock matrix has been developed aiming at efficient treatments of relativistic two-electron contributions. 
In the four-component framework, this leads to the quasi-four-component (Q4C) approaches. \cite{Liu06,Peng07,Zhao21}
In the two-component framework, the MP approach introduces a correction to the 2C-1e Hamiltonian as the difference between a ``model'' 2CMF Fock matrix and a model 2C-1e Fock matrix. \cite{vanWullen05}
In practice, the model 2CMF Fock matrix is obtained by transforming the model four-component Fock matrix calculated using atomic four-component density matrices and relativistic two-electron integrals into the two-component representation. The model 2C-1e Fock matrix is computed using atomic two-component density matrices and two-electron Coulomb integrals. The difference between them thus represents a two-electron ``picture-change'' (2e-pc) \cite{Samzow92,vanWullen05,Ikabata21} correction. \\

Efficient implementations of the MP approach for density functional theory (DFT) calculations have been reported for a high-order Douglas-Kroll-Hess (DKH) scheme \cite{vanWullen05} and the X2C scheme. \cite{Liu06, Peng07, Knecht22} An MP-based relativistic two-component DFT calculation can include the MP correction to the Fock-like matrix by adding a ``model density'' correction to the molecular two-component density obtained as the difference between atomic four- and two-component densities. \cite{vanWullen05, Liu06, Peng07} Since this density correction can be obtained from atomic calculations, no molecular relativistic two-electron integrals are needed. The MP approach has recently been extended to X2C Hartree-Fock (HF) calculations. \cite{Knecht22} 
The construction of the ``model'' four-component Fock matrix in the resulting ``eamfX2C'' scheme requires one calculation of four-component Fock matrix involving up to three-center relativistic two-electron integrals. Here the use of atomic four-component density matrix reduces the computational cost compared to the calculation of a four-component Fock matrix in a full-fledged four-component HF calculation.
We note that the work on the eamfX2C-HF method \cite{Knecht22} has focused on reproducing the four-component HF orbital energies and total energies. The latter is not compatible with the concept of treating the MP correction as a transferable effective one-electron potential, which, as discussed in the present work, is necessary to accurately recover corrections to molecular properties.  \\

The implementation of the MP-HF method needs to handle molecular relativistic two-electron integrals in uncontracted basis sets.
Therefore, computational approaches with one-center approximations to relativistic two-electron contributions, including the atomic mean-field (AMF) \cite{Hess96a} approaches, are appealing options, since they eliminate molecular relativistic two-electron integrals completely. An AMF approach may be obtained as a one-center approximation to the MMF approach, by using atomic density matrices instead of molecular density matrices and invoking one-center approximation for the relativistic two-electron integrals. Equivalently, an AMF approach can be derived by invoking one-center approximation to the relativistic two-electron integrals in the MP approach. The AMF approaches have been implemented for the construction of two-electron spin-orbit integrals using non-relativistic or scalar-relativistic atomic mean-field density matrices. \cite{Hess96a,Wahlgren97,Ilias01,Wodynski19,Zhang20} Since they use spin-free orbitals in the construction of MF integrals, these schemes are perhaps more compatible with state-interaction treatments of spin-orbit coupling and show relatively large errors in non-perturbative treatments of spin-orbit coupling. \cite{Ilias01,Lin23} To extend the AMF approach to spinor representation with variational treatments of spin-orbit coupling, an X2C AMF approach using atomic four-component spinors to construct AMF integrals (the X2CAMF scheme) \cite{Liu18} has been developed. The X2CAMF scheme was first implemented based on the Dirac-Coulomb Hamiltonian and later extended to include contributions from the Breit term. \cite{Zhang22} The X2CAMF scheme separates the spin-dependent and spin-free contributions in the four-component Coulomb interaction, neglects the scalar 2e-pc contribution, and includes the spin-dependent contributions together with the Breit term in the AMF integrals. 
The extra computational cost in solving the four-component Dirac-Coulomb(-Breit) Hartree-Fock equations for unique atoms has been demonstrated to introduce insignificant overheads. \\

For comparison, an ``amfX2C'' scheme \cite{Knecht22} has recently been derived by applying one-center approximation to relativistic two-electron integrals in eamfX2C scheme. Close inspection reveals that the amfX2C scheme differs from the X2CAMF scheme in the inclusion of scalar 2e-pc contribution in the AMF integrals. 
The inclusion of this correction improves computed core orbital energies. However, it should be noted that scalar-relativistic two-electron integrals are not as localized as the two-electron spin-orbit integrals. Scalar-relativistic integrals decay at a rate of $1/r$ similar to the Coulomb interaction, not as fast as the decay rate of $1/r^2$ of spin-orbit interactions. Potential problems in a one-center approximation to scalar 2e-pc correction have been discussed. \cite{Liu24} Similar concerns about the accuracy of one-center approximations to scalar-relativistic terms in the Gaunt interaction have also been raised. \cite{Liu24} The present work will investigate the numerical performance of these one-center approximations in calculations of molecular energies and properties. \\

In Section \ref{sec:theory} we first briefly summarize the X2C decoupling scheme and cost-effective treatments of relativistic two-electron contributions, including model-potential (MP) techniques and atomic mean-field (AMF) approaches. We then carefully analyze the contributions from the MP or AMF term to electronic energies, with an emphasis on accurate calculations of molecular properties. We further report numerical assessment of one-center approximations to relativistic two-electron contributions in the X2CAMF and amfX2C schemes. We examine the numerical performance of the X2CAMF scheme in treating two-electron spin-orbit interactions and the Gaunt term.
We also study the numerical stability of the inclusion of scalar 2e-pc contributions in the amfX2C scheme. We mention that the X2C-1e schemes with screened nuclear spin-orbit (SNSO) schemes \cite{Filatov13,Franzke18,Wodynski19,Ehrman23} are simple and useful approaches to account for the screening effects of relativistic two-electron spin-orbit interactions. {\ {The computational advantage of the SNSO schemes compared to the X2CAMF scheme is insignificant, since the additional four-component atomic calculations for unique atoms required in the X2CAMF scheme introduce negligible overheads compared to subsequent molecular calculations. \cite{Zhang22} On the other hand, the SNSO schemes can include multi-center contribution of  the two-electron SO screening effects implicitly. Furthermore, the performance of SNSO schemes 
depends on the way that the empirical scaling parameters are obtained. Therefore, a thorough investigation of the performance of SNSO schemes compared to the X2CAMF scheme is beyond the scope of the present study. }}
We thus will not include the SNSO schemes in the present comparison.  
Finally, a summary and an outlook are given in Section \ref{sec:summary}.

\section{Theory} \label{sec:theory}

\subsection{Exact two-component (X2C) decoupling scheme} \label{x2c}

The exact two-component (X2C) decoupling scheme \cite{Dyall97,Liu09} features a one-step transformation that reduces the four-component Dirac Hamiltonian in the matrix representation to a block-diagonal form consisting of an electronic block and a positronic block. The four-component matrix eigenvalue equation can be written as
\begin{eqnarray}
h^{\text{4c}} C^{\text{4c}} = E^{\text{4c}} S^{\text{4c}} C^{\text{4c}}  \label{4ceq}
\end{eqnarray}
in which the four-component wave function coefficients consist of ``large-component'' and ``small component'' contributions
\begin{eqnarray}
    C^{\text{4c}}=
   \begin{pmatrix}
C^{\text{L}} \\
C^{\text{S}} 
\end{pmatrix}, 
\end{eqnarray}
the four-component Hamiltonian and metric matrices are given in generic forms by
\begin{eqnarray}
    h^{\text{4c}}=
   \begin{pmatrix}
h^{\text{LL}} & h^{\text{LS}} \\
h^{\text{SL}} & h^{\text{SS}}
\end{pmatrix}~,~
    S^{\text{4c}}=
   \begin{pmatrix}
S^{\text{LL}} & 0 \\
0 & S^{\text{SS}}
\end{pmatrix}.
\end{eqnarray}
The electronic block of the X2C Hamiltonian matrix
is obtained by means of simple matrix operations
\begin{eqnarray}
    h^{\text{X2C}}_{+} &=& R^\dagger L^{\text{NESC}} R ~,~ \\
        L^{\text{NESC}} &=& h^{\text{LL}} + X^\dagger h^{\text{SL}} + h^{\text{LS}}X + X^\dagger h^{\text{SS}} X.
\end{eqnarray}
The acronym ``NESC'' represents ``normalized elimination of small component''. \cite{Dyall97} The $X$ matrix relates the large- and small-component orbital coefficients of the electronic states
\begin{equation}
    X C^{\text{L}} = C^{\text{S}}. \label{Xeq}
\end{equation}
{\ {The X matrix is in principle determined by a state-universal quadratic condition
\begin{eqnarray}
D^{\text{SL}}+D^{\text{SS}}X-S^{\text{SS}}X(S^{\text{LL}})^{-1}(D^{\text{LL}}+D^{\text{LS}}X)=0.
\end{eqnarray}
In practice, the $X$ matrix is most conveniently obtained by solving the four-component eigenvalue equation Eq. (\ref{4ceq}) directly and then using Eq. (\ref{Xeq}). }}
$R$ is the renormalization matrix \cite{Liu09}
\begin{equation}
    R = [(S^{\text{LL}})^{-1}\tilde{S}]^{-1/2}~,~\tilde{S}=S^{\text{LL}}+X^{\dagger} S^{\text{SS}} X. \label{Req}
\end{equation} 
The resulting two-component matrix eigenvalue equation can be written as
\begin{eqnarray}
h^{\text{X2C}}_{+} C^{\text{2c}} = E S^{\text{LL}} C^{\text{2c}}, \label{2ceq}
\end{eqnarray}
The renormalization matrix $R$ relates the coefficients of the large component wave function and the two-component wave function
\begin{eqnarray}
C^{\text{L}} = R C^{\text{2c}}.
\end{eqnarray}
\\

All two-component theories are based on Foldy-Wouthuysen (FW)-type unitary transformations \cite{Foldy50} to block-diagonalize the four-component Hamiltonian. A unique feature of X2C is to implement FW-type transformation in the matrix representation. Dyall's first study of matrix representation of two-component formulations \cite{Dyall97} inspired by Kutzelnigg's introduction of modified Dirac equation \cite{Kutzelnigg97a} gives the ``NESC'' eigenvalue equation
\begin{eqnarray}
L^{\text{NESC}} C^{\text{L}} = E \tilde{S} C^{\text{L}}. \label{NESCeq}
\end{eqnarray}
The NESC equation can exactly reproduce the four-component eigenvalues for positive energy states but works with a modified metric matrix $\tilde{S}$ and with large-component wave functions instead of two-component wave functions. {\ {The ``NESC'' equation defined as Eq. (\ref{NESCeq}) thus is fundamentally a four-component equation.}} Dyall introduced a renormalization matrix $R=(S^{\text{LL}})^{1/2}\tilde{S}^{-1/2}$ to obtain two-component eigenvalue equations of the form in Eq. (\Ref{2ceq}) with the non-relativistic metric matrix {\ {[(Eq. (21) of Ref. \onlinecite{Dyall01}]. Dyall, Cremer, Filatov, and collaborators continued using the acronym ``NESC'' for Eq. (\ref{2ceq}). \cite{Dyall01,Filatov02} On the other hand, it should be noted that Eq. (\ref{2ceq}) is a genuine two-component equation, different from the NESC equation in Eq. (\ref{NESCeq}).
Importantly, as will be further discussed below, Eq. (\ref{2ceq}) combined with the corrected renormalization matrix in Eq. (\ref{Req}) gives the working equations for the X2C scheme being used nowadays. }} \\

The majority of the work on two-component theories had been focused on multi-step decoupling schemes. Closed analytic form for the decoupling operator is available for the free-electron Dirac equation. The Douglas-Kroll-Hess (DKH) \cite{Hess86,Nakajima00,Wolf02,Reiher04a,Reiher04,Peng09} and Barysz-Sadlej-Snijders (BSS) \cite{Barysz97,Barysz01,Barysz02} methods start with the decoupling of the free-electron Dirac equation and improve the decoupling by applying further unitary transformations. Within the framework of regular approximations, the accuracy of the popular zeroth-order regular approximation (ZORA) \cite{vanLenth93,vanLenthe94} can also be improved by including higher order contributions to obtain, e.g., infinite-order regular approximation (IORA). \cite{Dyall99,Filatov03} Both DKH and BSS approaches can obtain infinite-order two-component formulations through a series of unitary transformations. {\ {Jensen presented an implementation of the two-step BSS decoupling scheme and emphasized the importance of matrix representation for two-component theories in the REHE 2005 conference.}} Kutzelnigg and Liu reported an in-depth revisitation and study of the X2C decoupling scheme. \cite{Kutzelnigg05} Peng and Liu then extended study to X2C DFT calculations with MP techniques, \cite{Liu06,Peng07} while Ilias and Saue reported X2C-1e DFT and HF calculations. \cite{Ilias07} \\

The X2C scheme has been gradually incorporated into various quantum chemistry program packages. Dyall reported a first implementation of the X2C scheme in the NWChem program, \cite{Dyall01,Apra20} followed by Filatov and Cremer's implementation in the COLOGNE program. \cite{Filatov02}
Filatov, Cremer, and Gauss worked on an implementation in the Mainz-Austin-Budapest version of the ACES2 program. This implementation did not find further application, since the computational results were found to vary with respect to unitary rotations among basis functions. It is now understood that, although it can produce the non-relativistic metric matrix, the renormalization matrix $(S^{\text{LL}})^{1/2}\tilde{S}^{-1/2}$ is not consistent with the matrix representation of the corresponding operator formulation. Liu and Peng \cite{Liu09} presented a derivation for the correct form of the renormalization matrix $R=[(S^{\text{LL}})^{-1}\tilde{S}]^{-1/2}$ and corrected the implementation in the BDF program package. \cite{Zhang20c} This renormalization matrix ensures the invariance of X2C energies with respect to rotations within basis functions. We note that Ilias and Saue's implementation of the X2C scheme in the DIRAC program  \cite{Ilias01,Saue20} before the derivation of the renormalization matrix is correct, because this implementation uses orthonormal basis functions and hence $(S^{\text{LL}})^{1/2}\tilde{S}^{-1/2}$ is equivalent to $[(S^{\text{LL}})^{-1}\tilde{S}]^{-1/2}$.
 Since then, the X2C scheme has been implemented in 
many more standard program packages, e.g.,  NWCHEM, \cite {Autschbach12a,Autschbach17,Apra20} CFOUR, \cite{Cheng11b,Cheng11c,Matthews20a} MOLCAS, \cite{Peng12,Peng13,Feng21,Aquilante20}  MOLPRO, \cite{Peng12,Peng13,Werner20} TURBOMOLE, \cite{Peng12,Peng13,Franzke18,Wodynski19,Franzke23,Balasubramani20} Psi4, \cite{Verma16,Smith20} Gaussian, \cite{Goings16,Egidi17} Chronus Quantum, \cite{Koulias19,SharmaP22,ZhangT24,Kovtun24,WilliamsYoung20} PySCF, \cite{Guo16, Mussard18, Yeh22, Majumder24,Sun20} {\ {ReSpect, \cite{Repisky20} Hyperion, \cite{Birnoschi22}}} Q-Chem, \cite{Cunha22,Epifanovsky21} and ORCA, \cite{Neese22} enabling extensive chemical and spectroscopic applications. 
Representative extensions in theory and algorithms include analytic X2C energy derivative techniques \cite{Zou11,Filatov14,Zou12,Cheng11b,Cheng11c,Cheng14,Franzke18,Zou20} to enable efficient calculations of molecular properties, algorithms based on local unitary transformations to extend X2C calculations to large molecules, \cite{Peng07,Peng12,Seino12,Peng13,Franzke18,ZhangT20} and implementation for periodic boundary conditions to enable calculations of solids. \cite{Zhao16,Yeh22}  
The tremendous joint efforts of quantum chemistry community have established the X2C scheme as a standard method for routine molecular applications.

\subsection{The X2C-1e scheme} \label{x2c1e}

In the representation of kinetically-balanced basis sets \cite{Stanton84} using $\{ f_\mu\}$ and $\{\frac{\vec{\sigma}\cdot \vec{p}}{2c}f_\mu\}$ as the large- and small-component basis functions, the four-component metric matrices are given by
\begin{eqnarray}
    S^{\text{LL}}=S~,~ S^{\text{SS}}=\frac{T}{2c^2},
\end{eqnarray}
in which $S$ is the non-relativistic overlap matrix $S_{\mu\nu}=\langle f_\mu|f_\nu\rangle$ and $T$ is the kinetic energy matrix. 
In the X2C-1e scheme, the four-component Hamiltonian matrix is chosen as the matrix representation of the one-electron Dirac Hamiltonian 
\begin{eqnarray}
    h^{\text{LL, 1e}}=V~,~ h^{\text{LS, 1e}}=T~,~ 
    h^{\text{SL, 1e}}=T~,~ h^{\text{SS, 1e}}=\frac{W}{4c^2}-T,
\end{eqnarray}
in which $V$ is the nuclear attraction matrix and $W$ is the small-component nuclear attraction matrix with $W_{\mu\nu}=\langle f_\mu |(\vec{\sigma}\cdot{\vec{p}})\hat{V}(\vec{\sigma}\cdot{\vec{p}})| f_\nu\rangle$.
Block-diagonalization of this four-component Hamiltonian matrix using the X2C decoupling scheme gives an electronic block $h^{\text{X2C-1e}}_{+}$.   
The X2C-1e scheme combines $h^{\text{X2C-1e}}_{+}$ with the non-relativistic Coulomb interaction in the subsequent two-component many-electron calculations. \\

The X2C-1e scheme is perhaps the simplest practical X2C scheme for molecular calculations. It takes into account the one-electron relativistic effects accurately, while neglecting the contributions from the relativistic two-electron interactions. The errors of the X2C-1e scheme are often termed as the ``two-electron picture-change'' (2e-pc) errors, because the X2C-1e scheme uses the operator form of the Coulomb interaction in the four-component picture directly in two-component calculations, without performing a transformation from the four-component representation to the two-component representation. The usefulness of 2C-1e schemes have been extensively investigated in calculations using approximate decoupling schemes including ZORA and DKH approaches, e.g., see Refs. \onlinecite{VanLenthe96a,Hess00,Neese05a,Mastalerz07}. The scalar 2e-pc corrections to molecular properties have been demonstrated to be small. \cite{vanWullen05,Neese05a,Mastalerz07} The spin-free version of the X2C-1e scheme (the SFX2C-1e scheme) \cite{Dyall01,Liu09,Cheng11b} thus has been established as a standard approach for treating scalar-relativistic effects. In contrast, the two-electron spin-orbit interactions make important contributions to molecular properties. \cite{Blume62,Blume63,Malkina98,Fedorov00,Stopkowicz11b} Therefore, the spin-orbit mean-field approaches \cite{Hess96a,Neese05,Epifanovsky15,Cao17,Perera17,Mussard18,Cheng18a,Berning00,Netz21,Majumder24} have been developed to enable cost-effective treatments of two-electron spin-orbit contributions. The next subsection summarizes the development within exact two-component theory.

\subsection{The X2C mean-field (X2CMF) approaches}  \label{x2cmmf}

An X2C mean-field Hamiltonian $h^{\text{X2CMF}}$  is obtained by transforming the four-component mean-field Hamiltonian $h^{\text{4c, MF}}$ comprising the one-electron Dirac Hamiltonian $h^{\text{4c,1e}}$ and a mean-field approximation $h^{\text{4c, MF2e}}$
to the four-component two-electron interactions  \cite{Liu06,Sikkema09,Liu10}
\begin{eqnarray}
    h^{\text{4c, MF}}=h^{\text{4c, 1e}}+h^{\text{4c, MF2e}}.
\end{eqnarray}
$h^{\text{4c, MF2e}}$ in a Hartree-Fock calculation can be written as
\begin{eqnarray}
    h^{\text{LL, MF2e}} &=& J^{\text{LL, C}}-K^{\text{LL, C}}-K^{\text{LL, G}}, \label{MF2eLL}\\
 h^{\text{LS, MF2e}} &=& -K^{\text{LS, C}}+J^{\text{LS, G}}-K^{\text{LS, G}} , \label{MF2eLS}\\
    h^{\text{SL, MF2e}} &=& -K^{\text{SL, C}}+J^{\text{SL, G}}-K^{\text{SL, G}},  \label{MF2eSL}\\
     h^{\text{SS, MF2e}} &=& J^{\text{SS, C}}-K^{\text{SS, C}}-K^{\text{SS, G}}. \label{MF2eSS}
    \end{eqnarray}  
    ``$J$'' and ``$K$'' denote Coulomb and exchange contributions to the Fock matrix, while ``C'' and ``G" represent contributions from the instantaneous Coulomb interaction and the Gaunt term.
    In Dirac-Coulomb density-functional theory calculations, the ``LL'' and ``SS'' blocks of $h^{\text{4c, MF2e}}$ contain the Coulomb terms and the contributions from the exchange-correlation (XC) potential ($V_{\text{XC}}$) 
    \begin{eqnarray}
    h^{\text{LL, MF2e}} &=& J^{\text{LL, C}}+V_{\text{XC}}^{\text{LL}}, \label{MF2eLL_DFT}\\
    h^{\text{LS, MF2e}} &=& h^{\text{SL, MF2e}} =0, \label{MF2eLS_DFT}\\
    h^{\text{SS, MF2e}} &=& J^{\text{SS, C}}+V_{\text{XC}}^{\text{SS}}. \label{MF2eSS_DFT}
    \end{eqnarray}

The rigorous version of X2CMF approach \cite{Liu06,Peng07,Sikkema09} solves the four-component HF or DFT equations and then transforms the four-component Fock(-like) matrix into two-component representation using the X2C decoupling scheme. This version of the X2CMF approach is as accurate as and, not surprisingly, as expensive as the corresponding four-component calculation at the MF level. This X2CMF scheme has been called the ``X2C molecular mean-field'' (X2CMMF) approach. We follow this nomenclature in the present discussion. The X2CMMF scheme for DFT calculations serve as an intermediate approach leading to efficient ``model potential'' schemes. \cite{Liu06,Peng07} For the electron correlation calculations, the X2CMMF scheme gains computational efficiency by neglecting the picture change for the fluctuation potential; the integral transformation involves only the non-relativistic two-electron integrals. \cite{Sikkema09}
We mention that two-component electron-correlation calculations with picture change of fluctuation potential taken into account have been reported. \cite{Seino12} \\

Several approximate versions of X2CMF approaches have been studied. \cite{Ilias01,Cao17,Mussard18,Cheng18a,Zhang20,Majumder24} 
Similar to many earlier calculations of spin-orbit coupling based on
non-relativistic and scalar-relativistic wave functions, \cite{Hess96a,Neese05,Epifanovsky15,Cao17,Perera17,Cheng18a}
these approaches augment the SFX2C-1e Hamiltonian with SO-MF contributions constructed using scalar-relativistic orbitals. 
Note that the construction of MF integrals using scalar-relativistic orbitals neglects spin-orbit corrections to the orbitals. The spin-orbit Hamiltonian matrix elements thus have intrinsic errors in higher-order spin-orbit contributions.
Therefore, for calculations of heavy elements, these schemes are perhaps more compatible with state-interaction calculations. 
\cite{Ilias01,Liu18b,Lin23}
    
\subsection{The X2C model potential (X2CMP) scheme}  \label{x2cmp}

Since the X2CMMF approach is essentially identical to the corresponding four-component mean-field method in terms of efficiency, it is of significant interest to develop computational schemes with reduced computational cost that retain the accuracy of the X2CMMF approach. The underlying idea of a ``model potential'' (MP) approach here is to identify computationally expensive contributions from relativistic two-electron integrals to Fock matrices that are transferable from atoms to molecule, and approximate them using atomic information to reduce computational costs. 
An MP approach in general calculates two MP Fock matrices, an X2CMF one and an X2C-1e one, and takes the difference \cite{vanWullen05}
\begin{eqnarray}
    h^{\text{MP}}=h^{\text{X2CMF(MP)}}-h^{\text{X2C-1e(MP)}}
\end{eqnarray}
as a correction to the X2C-1e Hamiltonian. 
The subtraction of $h^{\text{X2C-1e(MP)}}$ avoids double counting of the Coulomb interaction in the two-component picture.
The resulting Hamiltonian $h^{\text{X2CMP}}$, the X2C-1e Hamiltonian augmented with an MP correction,
\begin{eqnarray}
    h^{\text{X2CMP}}=h^{\text{X2C-1e}}+h^{\text{MP}} \label{HX2CMP}
\end{eqnarray}
is used together with the non-relativistic Coulomb interaction in subsequent two-component molecular calculations. 
Since small component wave functions are highly localized in the vicinity of nuclei, a natural choice for an MP approach is to approximate the contributions from small-component wave functions to the four-component Fock matrix [Eqns. (\ref{MF2eLL})-(\ref{MF2eSS})] using atomic information. Efficient algorithms have been developed and implemented for X2CMP-DFT and HF calculations. \cite{vanWullen05,Liu06,Peng07,Knecht22}\\

The underlying idea is to approximate the molecular small-component density matrices with a direct sum of the atomic ones.
$h^{\text{LL, MF2e}}$ and $h^{\text{SS, MF2e}}$ involve contractions between the small-small block of the four-component density matrix and relativistic two-electron integrals. $h^{\text{LL, MF2e}}$ and $h^{\text{SL, MF2e}}$ involve contractions between the small-large block of the density matrix and relativistic two-electron integrals. Approximating these blocks of density matrix involving small component using the atomic values can effectively reduce the number of relativistic two-electron integrals needed for the calculations
because of the sparsity of the atomic density matrices.  
Note that $h^{\text{SS, MF2e}}$ also involves a contraction between the large-large block of the four-component density matrix and the (LL|SS)- and (LS|LS)-type relativistic two-electron integrals. In order to obtain an approach with meaningful computational benefit, it is necessary to invoke an atomic approximation for the large-large block of the density matrix in these contributions. We mention that the implementation of these approximations within the four-component framework gives rise to a quasi-four-component (Q4C) approach for HF calculations. \\

Within a two-component MP approach, to avoid double counting of the Coulomb interaction in the two-component representation, the density matrix used to construct $h^{\text{X2C-1e(MP)}}$ needs to match that used to calculate the contributions from the spin-free Dirac-Coulomb integrals to $h^{\text{X2CMF(MP)}}$. 
Therefore, it is necessary to calculate the contributions from the non-relativistic two-electron integrals to $h^{\text{X2C-1e(MP)}}$ and $h^{\text{X2CMF(MP)}}$ using the corresponding atomic density matrices. Atomic two-component density matrix and atomic large-large density matrix are used to construct $h^{\text{X2C-1e(MP)}}$ and the contribution from the non-relativistic two-electron integrals to $h^{\text{LL, MF2e}}$, respectively. 
The MP scheme designed in this way has the favorable asymptotic property
that it exactly recovers the X2CMMF result for atomic calculations. In the asymptotic limit that one uses molecular density matrices in the construction of the MP correction, the X2CMP calculation also recovers the X2CMMF result for the molecule.  \\

The MP approach invokes atomic approximation for all blocks of density matrices in the calculations of $h^{\text{X2CMF(MP)}}$ and $h^{\text{X2C-1e(MP)}}$. 
The robustness of the X2CMP scheme is based on the atomic nature of the small-component density matrices; atomic small-component density matrices are excellent approximations to the molecular ones. We should note that the atomic approximation to the large-large block of the four-component density matrix is used in the evaluation of $h^{\text{SS, MF2e}}$. Such contributions involving spin-free Dirac-Coulomb integrals largely cancel $h^{\text{X2C-1e(MP)}}$ calculated using atomic two-component density matrix. They thus do not introduce significantg errors. However, the contributions from the spin-dependent relativistic two-electron integrals and those in the Gaunt term calculated using the atomic large-large density matrix may be less accurate. \\

In the implementation of MP correction in density-functional theory (DFT) calculations, efforts have been devoted to avoiding calculations of multi-center relativistic two-electron integrals. Since Coulomb and exchange-correlation (XC) contributions to Fock-like matrix in DFT calculations entirely depend on electron density, a ``model density'' approach \cite{vanWullen05,Peng07} has been developed to incorporate the MP correction efficiently. This model density correction is obtained as the difference between the four- and two-component atomic densities. In the calculation of the XC contribution, this density correction can be added to the two-component density with no increase of the computational cost. This also applies to the evaluation of the Coulomb matrix elements using quadrature. \cite{Peng07} In the calculations of Coulomb contribution using analytically evaluated two-electron integrals, a scheme has been developed to fit the density correction using a linear combination of $s$-type Gaussian functions. \cite{vanWullen05}
The contribution to Coulomb matrix elements can then be evaluated using standard algorithms 
for analytic evaluation of nuclear attraction integrals.  
This introduces negligible overheads in computational costs. \\

For Hartree-Fock calculations, the use of atomic density matrices in the construction the X2CMF(MP) Fock matrix gives rise to the ``eamfX2C'' prescription \cite{Knecht22} for $h^{\text{X2CMP}}$, which requires up to three-center relativistic two-electron integrals. The number of relativistic two-electron integrals is reduced by the spherical symmetry of atomic density matrix elements. 
On the other hand, the relativistic two-electron integrals required for the construction of MP correction need to be evaluated for uncontracted basis functions. It thus is not possible to give a simple factor for the computational overhead introduced by calculations of these relativistic two-electron integrals compared to 
the two-component SCF calculations, except that one may safely conclude that the construction of the MP correction is significantly less expensive than one iteration of the corresponding four-component SCF calculation using uncontracted basis functions. It is still of substantial interest to develop computational schemes to eliminate multi-center relativistic two-electron integrals entirely while retaining computational accuracy. This motivates the development of atomic mean-field approaches.

\subsection{The X2C atomic mean-field (AMF) schemes} \label{x2camf}

The atomic mean-field (AMF) approaches are straightforwardly related to the molecular mean-field approaches by replacing molecular density matrices with atomic ones and keeping only one-center integrals. \cite{Hess96a} An alternative way to obtain the AMF approaches is to apply one-center integral approximation to the MP approach. The X2CAMF scheme was derived following the former procedure. The X2CAMF scheme \cite{Liu18} performs a spin separation for the four-component Coulomb integrals and includes only the spin-dependent Coulomb (SDC) integrals in the calculations of AMF integrals. The spin-free Coulomb (SFC) matrix elements in the four-component theory are reduced to the Coulomb interaction in the two-component theory. \cite{Liu18} This corresponds to the neglect of scalar two-electron picture-change correction. The subsequent development of the X2CAMF scheme based on the Dirac-Coulomb-Breit Hamiltonian  \cite{Zhang22} includes the contributions from the Breit term into the AMF integrals. In short, the X2CAMF scheme augments the X2C-1e Hamiltonian with a MP-like correction consisting of AMF integrals of the SDC interaction and the Breit term
\begin{eqnarray}
    h^{\text{MP, X2CAMF}}=h^{\text{X2CMF(MP),SDC}}+h^{\text{X2CMF(MP),B}}.
\end{eqnarray}
For comparison, an amfX2C scheme \cite{Knecht22} has been developed by applying one-center approximation to the two-electron integrals in the eamfX2C scheme. The X2CAMF and amfX2C schemes are closely related to each other, despite the seemingly distinct appearances and derivations. A separation of the MP correction in the amfX2C scheme into SFC, SDC, and the Gaunt term
\begin{eqnarray}
    h^{\text{MP, amfX2C}} &\approx& [h^{\text{X2CMF(MP),SFC}}-h^{\text{X2C-1e(MP)}}]+h^{\text{X2CMF(MP),SDC}}+h^{\text{X2CMF(MP),G}} \\
    &=& h^{\text{MP, X2CAMF}}+[h^{\text{X2CMF(MP),SFC}}-h^{\text{X2C-1e(MP)}}] 
\end{eqnarray}
 shows that the amfX2C scheme further includes an additional correction $[h^{\text{X2CMF(MP),SFC}}-h^{\text{X2C-1e(MP)}}]$.
This correction represents the difference between the SFC interaction in the four-component representation and the Coulomb interaction in the two-component representation and thus corresponds to the ``scalar 2e-pc'' correction. \\

In the limit of an atomic calculation, the amfX2C scheme recovers the four-component mean-field result exactly. The X2CAMF scheme still has the underlying approximation of neglecting the scalar 2e-pc correction. However, the scalar-relativistic two-electron integrals are not as localized as the spin-orbit two-electron integrals.  The accuracy of the amfX2C scheme in molecular calculations thus has recently been questioned. \cite{Liu24} Similar concerns have been raised for including the spin-free contributions of the Breit term in the X2CAMF scheme. \cite{Liu24}
It is  of interest to assess the performance of the X2CAMF and amfX2C schemes in calculations of molecular energies and properties. Since the errors of one-center integral approximations may be sensitive to the size of basis sets, it is necessary to use extended basis sets for the benchmark calculations. 

\subsection{The MP/AMF contributions to electronic energies} \label{amfene}

We now focus our discussion on a subtle point concerning the MP/AMF contributions to electronic energies. The MP correction to the Hamiltonian is an effective one-electron approximation to a two-electron contribution. Note that the contribution from a two-electron interaction to the Hartree-Fock energy can be written as
\begin{eqnarray}
    E^{\text{MF2e}}_{\text{HF}}=\frac{1}{2}\sum_{\mu\nu}h^{\text{MF2e}}_{\mu\nu}\rho^{\text{2c}}_{\mu\nu}, \label{MF2eene}
\end{eqnarray}
in which $\rho$ represents the one-electron HF density matrix. The factor of $\frac{1}{2}$ avoids a double counting for the two-electron interaction energy. 
A similar expression has been used to calculate the Hartree-Fock energy for 
the (e)amfX2C scheme
\begin{eqnarray}
    E^{\text{eamfX2C}}_{\text{HF}}=\sum_{\mu\nu}[h^{\text{X2C-1e}}+\frac{1}{2}h^{\text{MP}}+\frac{1}{2}(J^{\text{2c}}-K^{\text{2c}})]_{\mu\nu}\rho^{\text{2c}}_{\mu\nu}. \label{eamfX2Cene}
\end{eqnarray}
The resulting (e)amfX2C electronic energies agree closely with the corresponding four-component HF energies. \\

However, if $h^{\text{MP}}$ is to be treated as a transferable effective one-electron matrix, \cite{Huzinaga71} the corresponding contribution to the Hartree-Fock energy should be given by
\begin{eqnarray}
    E^{\text{X2CMP}}_{\text{HF}}&=&\sum_{\mu\nu}[h^{\text{X2CMP}}+\frac{1}{2}(J^{\text{2c}}-K^{\text{2c}})]_{\mu\nu}\rho^{\text{2c}}_{\mu\nu}\nonumber\\
    &=&\sum_{\mu\nu}[h^{\text{X2C-1e}}+h^{\text{MP}}+\frac{1}{2}(J^{\text{2c}}-K^{\text{2c}})]_{\mu\nu}\rho^{\text{2c}}_{\mu\nu}. \label{MP2eene}
\end{eqnarray}
The difference in a factor of two for the contributions from $h^{\text{MP}}$ in Eq. (\ref{eamfX2Cene}) and (\ref{MP2eene}) originates from treating this contribution as a genuine two-electron contribution in Eq. (\ref{eamfX2Cene}) and as an effective one-electron interaction in Eq. (\ref{MP2eene}).  Although Eq. (\ref{eamfX2Cene}) gives absolute energies close to the four-component energies, Eq. (\ref{MP2eene}) is consistent with the physical concept of MP as a transferable effective one-electron potential. Furthermore, the energy expression in Eq. (\ref{MP2eene}) is stationary for the X2CMP-HF wave functions. In contrast, the energy expression in Eq. (\ref{eamfX2Cene}) is not stationary for
the X2CMP-HF wave function. Therefore, if one uses the expectation value formulation to calculate the X2CMF-HF first-order properties, one tacitly assumes the energy expression in Eq. (\ref{MP2eene}). This favors Eq. (\ref{MP2eene}) as the X2CMP-HF energy expression.\\

Importantly, Eq. (\ref{MP2eene}) accounts for the variation of electronic energies correctly. 
The energy expressions in Eq. (\ref{MF2eene}) and (\ref{eamfX2Cene}) depend quadratically with respect to density matrices. It requires full variation of the energy with respect to molecular density matrices to properly account for energy differences. However, 
the energy expression in Eq. (\ref{eamfX2Cene}) only sees the variation through $\rho^{\text{2c}}$ but not through model density matrices, 
since model density matrices are pre-chosen and fixed. 
Consequently, Eq. (\ref{eamfX2Cene}) only covers about half of the contributions from relativistic two-electron interactions to energy differences. 
When treating $h^{\text{MP}}$ as a transferable one-electron matrix, 
 Eq. (\ref{MP2eene}) takes care of the variation of the electronic energies properly.
Let us take the contribution from the (LL|SS)-type relativistic two-electron integrals denoted as $G^{\text{LLSS}}$ to the HF Coulomb energy as an example. In the four-component formulation and equivalently in the X2CMMF method, this contribution can be written as
\begin{eqnarray}
    \frac{1}{2}\sum_{\mu\nu\sigma\rho}G_{\mu\nu\sigma\rho}^{\text{LLSS}}\rho^{\text{LL,mol}}_{\mu\nu}\rho^{\text{SS,mol}}_{\sigma\rho}
    +\frac{1}{2}\sum_{\mu\nu\sigma\rho}G_{\mu\nu\sigma\rho}^{\text{SSLL}}\rho^{\text{SS,mol}}_{\mu\nu}\rho^{\text{LL,mol}}_{\sigma\rho}
    =\sum_{\mu\nu\sigma\rho}G_{\mu\nu\sigma\rho}^{\text{LLSS}}\rho^{\text{LL,mol}}_{\mu\nu}\rho^{\text{SS,mol}}_{\sigma\rho},
\end{eqnarray}
in which ``mol'' represents the molecular values. Assume that a perturbation leads to variation of the molecular density matrices
\begin{eqnarray}
    \rho^{\text{LL,mol}}\to \rho^{\text{LL,mol}}+\Delta\rho^{\text{LL,mol}}~,~
        \rho^{\text{SS,mol}}\to \rho^{\text{SS,mol}}+\Delta\rho^{\text{SS,mol}}. \label{rhoper}
\end{eqnarray}
The corresponding linear variation of the energy is then given by
\begin{eqnarray}
\sum_{\mu\nu\sigma\rho}G_{\mu\nu\sigma\rho}^{\text{LLSS}}\Delta\rho^{\text{LL,mol}}_{\mu\nu}\rho^{\text{SS,mol}}_{\sigma\rho}
+\sum_{\mu\nu\sigma\rho}G_{\mu\nu\sigma\rho}^{\text{LLSS}}\rho^{\text{LL,mol}}_{\mu\nu}\Delta\rho^{\text{SS,mol}}_{\sigma\rho}, \label{4cenevar}
\end{eqnarray}
For comparison, in an MP calculation using the energy expression of Eq. (\ref{eamfX2Cene}), 
the corresponding contribution to the electronic energy can be written as
\begin{eqnarray}
    \frac{1}{2}\sum_{\mu\nu\sigma\rho}G_{\mu\nu\sigma\rho}^{\text{LLSS}}\rho^{\text{LL,mod}}_{\mu\nu}\rho^{\text{SS,mol}}_{\sigma\rho}
    +\frac{1}{2}\sum_{\mu\nu\sigma\rho}G_{\mu\nu\sigma\rho}^{\text{SSLL}}\rho^{\text{SS,mod}}_{\mu\nu}\rho^{\text{LL,mol}}_{\sigma\rho},
\end{eqnarray}
in which the superscript ``mod'' of $\rho$ denotes the ``model'' density matrices, chosen as the atomic values in the present X2CMP scheme.
Assuming the same perturbation as in Eq. (\ref{rhoper}), 
the linear variation of the energy contribution can be written as
\begin{eqnarray}
\frac{1}{2}\sum_{\mu\nu\sigma\rho}G_{\mu\nu\sigma\rho}^{\text{LLSS}}\Delta\rho^{\text{LL,mol}}_{\mu\nu}\rho^{\text{SS,mod}}_{\sigma\rho}
+\frac{1}{2}\sum_{\mu\nu\sigma\rho}G_{\mu\nu\sigma\rho}^{\text{LLSS}}\rho^{\text{LL,mod}}_{\mu\nu}\Delta\rho^{\text{SS,mol}}_{\sigma\rho}, \label{halfcon}
\end{eqnarray}
When $\rho^{\text{LL,mod}}$ and $\rho^{\text{S,mod}}$ are good approximations to
$\rho^{\text{LL,mol}}$ and $\rho^{\text{LL,mol}}$, Eq. (\ref{halfcon})
recovers about half of the contribution in Eq. (\ref{4cenevar}). In other words, 
the energy expression in Eq. (\ref{eamfX2Cene}) only accounts for roughly half of the 
relativistic two-electron corrections to energy differences.
In contrast, the energy expression of Eq. (\ref{MP2eene}) gives a linear variation of the energy contribution 
\begin{eqnarray}
\sum_{\mu\nu\sigma\rho}G_{\mu\nu\sigma\rho}^{\text{LLSS}}\Delta\rho^{\text{LL,mol}}_{\mu\nu}\rho^{\text{SS,mod}}_{\sigma\rho}
+\sum_{\mu\nu\sigma\rho}G_{\mu\nu\sigma\rho}^{\text{LLSS}}\rho^{\text{LL,mod}}_{\mu\nu}\Delta\rho^{\text{SS,mol}}_{\sigma\rho}, 
\end{eqnarray}
which is an accurate approximation to the four-component one in Eq. (\ref{4cenevar}).
Therefore, Eq. (\ref{MP2eene}) produces the correct variation of the electronic energies and hence can give accurate contributions to energy differences. \\

To facilitate the discussion of benchmark results, 
hereafter we refer to the methods using the X2CMP Hamiltonian in Eq. (\ref{HX2CMP}) and the energy expression in Eq. (\ref{MP2eene})
as the X2CMP schemes. The X2CMP schemes based on the Dirac-Coulomb Hamiltonian and the Dirac-Coulomb-Gaunt Hamiltonian are called X2CMP(DC) and X2CMP(DCG), respectively.
The X2CMP scheme presented here shares the same prescription as the eamfX2C scheme for the calculation of the effective one-electron correction to the one-electron Hamiltonian matrix. It differs from the eamfX2C scheme in the choice of the energy expression. We mention that the X2CAMF calculations reported so far have used Eq. (\ref{MP2eene}) as the energy expression. \\

Finally, a simple atomic correction to the X2CMP electronic energy can 
defined as 
\begin{eqnarray}
    \Delta E^{\text{X2CMP,atom}}_{\text{HF}}&=&
    -\sum_A \Delta E^{\text{X2CMP,A}}_{\text{HF}}, \label{deltaMP2eene} \\
\Delta E^{\text{X2CMP,A}}_{\text{HF}}&=&\frac{1}{2}\sum_{\mu\nu}h^{\text{MP,A}}_{\mu\nu}\rho^{\text{2c,A}}_{\mu\nu},
\end{eqnarray}
in which $\sum_A$ denotes a sum over all atoms and $\Delta E^{\text{X2CMP,A}}_{\text{HF}}$ is obtained using the model potential correction and two-component density matrix of atom $\text{A}$.  
Adding this correction to Eq. (\ref{MP2eene}) gives a corrected X2CMP-HF energy
\begin{eqnarray}
    E^{\text{X2CMP, corrected}}_{\text{HF}}=E^{\text{X2CMP}}_{\text{HF}}+\Delta E^{\text{X2CMP,atom}}_{\text{HF}}. \label{MP02eene}
\end{eqnarray}
This correction improves the agreement of the absolute value of the X2CMP-HF energy with the parent four-component HF energy. On the other hand, since it takes a constant value for a molecule, this correction does not affect molecular properties. 


\section{Computational results and discussions} \label{sec:results}

{\ {The computational study in Section \ref{MPeneSec} aims to elucidate the correct model potential/atomic mean field contributions to electronic energies. Section \ref{Num1c} is devoted to assessing the performance of the one-center approximations, especially the effects due to diffuse basis functions.}} The present demonstration uses simple and important molecular parameters of chemical interest, including structural parameters and thermochemical parameters. The calculations have employed the recent implementations of the X2CMP schemes in the PySCF \cite{Sun20} and CFOUR \cite{Matthews20a,CFOURfull} program packages using the X2CAMF program \cite{Zhang22} to provide atomic density matrices. The one-center approximations have used the X2CAMF program to provide AMF integrals. {\ {As shown in Ref. \onlinecite{Zhang22}, an efficient implementation with the use of spherical symmetry in the X2CAMF module enables four-component calculations of unique atoms to be carried out with insignificant overheads compared to subsequent molecular calculations.}} The X2CMP calculations using the PySCF program have used the libcint package \cite{Sun24} for the molecular relativistic two-electron integrals. 
The calculations of uranium thermochemical parameters have used basis sets constructed in Ref. \onlinecite{Zhang22b} by augmenting the primitive functions of the uncontracted ANO-RCC sets \cite{Faegri01,Roos05a} and correlation-consistent sets. \cite{Peterson15}
The calculations of spin-orbit splittings and structural parameters have used the combination of Dyall's VTZ (or aVTZ) basis sets for Ag, Au, I, At \cite{Dyall02_dyalltz_4p5p6p,Dyall06_dyalltz_revise_4p5p6p,Dyall04_dz_tz_qz_5d, Dyall07_4d_dtqz,Dyall09_revised_5d,Dyall23_dyatz_5d} and the uncontracted cc-pVTZ (or aug-cc-pVTZ) basis sets for H and F \cite{Dunning89,Kendall92}, which is denoted as ``VTZ'' (or ``aVTZ'') basis.

\begin{table}
  \caption{Ionization energies (IEs) of U, UO, and UO$_2$ and dissociation energies (D$_\text{e}$'s) of UO and UO$_2$ in kJ/mol computed at the Kramers unrestricted Hartree-Fock level using the four-component Dirac-Coulomb (4c-DC) approach, the X2C-1e scheme, as well as the X2CAMF, eamfX2C, X2CMP schemes. Triple-zeta basis sets as described in Ref. \onlinecite{Zhang22b} have been used. The eamfX2C calculations used the energy expression in Eq. (\ref{eamfX2Cene}), while X2CAMF and X2CMP calculations used Eq. (\ref{MP2eene}). The X2C-1e, X2CAMF, eamfX2C, and X2CMP results are shown as the differences from the 4c-DC values.  }
  \begin{tabular}{cccccccc}
  \hline
  \hline
 & ~ ~ X2C-1e ~ ~ & ~ ~ X2CAMF ~ ~ & ~ ~ eamfX2C ~ ~ & ~ ~ X2CMP ~ ~ & ~~ 4c-DC ~~  \\
 \hline
IE(U)                 &  5.3   &  1.0  & 2.9 & 0.0 &  576.3     \\
IE(UO)                &  -2.3  &  -0.7  & -1.2 & 0.0 &  517.4     \\
IE(UO$_2$)            &  -1.2  &  -1.2  &  -0.6 & 0.0 &  523.1     \\
D$_\text{e}$(UO)      &  -12.2  &  -0.2  & -6.3  & 0.0 &  404.2     \\
D$_\text{e}$(UO$_2$)  &  -30.7  &  -2.6  &  -14.1 & 0.0 &  356.6     \\
 \hline
 \hline
\end{tabular}
\label{tab-utherm}
\end{table}

\subsection{The MP/AMF contributions to electronic energies} \label{MPeneSec}

We first focus our discussions on the thermochemical parameters of uranium-containing atomic and molecular species, using a benchmark set assembled in Ref. \onlinecite{Zhang22b} including the ionization energies of U, UO, and UO$_2$, and the bond dissociation energies of UO and UO$_2$. The two-electron spin-orbit contributions (2e-SO) in the Coulomb interaction can be represented as the difference between the four-component Dirac-Coulomb (DC) and X2C-1e results. They often make important contributions to molecular properties. For example, the 2e-SO contribution to the bond dissociation energy of UO$_2$ amounts to more than 30 kJ/mol. The X2CAMF scheme using the energy expression in Eq. (\ref{MP2eene}) has been shown to recover the majority of the 2e-SO contributions, with remaining errors smaller than 1 kcal/mol. In Table 1, the ``MP'' and ``eamfX2C'' columns summarize results obtained from calculations using the ``eamfX2C'' prescription for $h^{\text{MP}}$ and using the energy expressions in Eq. (\ref{MP2eene}) and Eq. (\ref{eamfX2Cene}), respectively. The MP results with the energy expression in Eq. (\ref{MP2eene}) accurately reproduce the 4c-DC results with remaining errors less than 0.1 kJ/mol. In contrast, the ``eamfX2C'' results using Eq. (\ref{eamfX2Cene}) recovers around half of the 2e-SO contributions for all the parameters presented here. This observation confirms the analysis in Section II.F.\\

\begin{table}
\caption{Spin-orbit splitting (cm$^{-1}$) in $^2\Pi$ radicals computed at the Kramers unrestricted Hartree-Fock level using Dyall's VTZ basis sets for Se, Te, Br, I \cite{Dyall02_dyalltz_4p5p6p,Dyall06_dyalltz_revise_4p5p6p} and uncontracted cc-pVTZ basis sets for H, O, S, F and Cl. \cite{Dunning89,Kendall92,Woon93} The eamfX2C calculations used the energy expression in Eq. (\ref{eamfX2Cene}), while X2CAMF and X2CMP calculations used Eq. (\ref{MP2eene}). The X2C-1e, X2CAMF, eamfX2C, and X2CMP results are shown as the differences from the 4c-DC values. }
\begin{tabular}{ccccccc}
\hline
  \hline
  Molecule & ~ ~ X2C-1e ~ ~ & ~ ~ X2CAMF ~ ~ & ~ ~ eamfX2C ~ ~ & ~ ~ X2CMP ~ ~ & ~~ 4c-DC ~~ \\
  \hline
  TeH & 200.6 & -2.2 & 100.4 & 0.0 & 3826.2 \\
  SeH & 148.7 & -0.2 & 74.4  & 0.3 & 1766.2 \\
  IO  & 88.5  &  5.6 & 44.7  & 0.6 & 716.7 \\
  BrO & 86.1  &  3.6 & 43.4  & 0.5 & 479.9 \\
  SH  & 70.9  &  0.6 & 35.6  & 0.2 & 389.7 \\
  ClO & 74.9  &  1.3 & 37.7  & 0.4 & 251.3 \\
  FO  & 72.7  &  0.0 & 36.2  & -0.3& 194.0 \\
  OH  & 57.9  &  2.3 & 29.4  & 0.8 & 144.1 \\
   \hline
 \hline
\end{tabular}
\label{tab-SO}
\end{table}

The observation for thermochemistry discussed here holds for the 2e-SO contributions to spin-orbit splittings, equilibrium bond lengths, and harmonic vibrational frequencies. As shown in Table \ref{tab-SO}, the MP calculations accurately recover the 2e-SO contributions to spin-orbit splittings in representative diatomic radicals with remaining errors less than 1 cm$^{-1}$. 
In contrast, the ``eamfX2C'' results recover around half of the 2e-SO contributions, essentially halving the errors of the X2C-1e scheme. 
Similarly, as shown in Table \ref{tab-freq}, the X2CMP calculations provide accurate equilibrium bond lengths and harmonic vibrational frequencies, with remaining errors smaller than 0.0001 {\AA} for bond lengths and 1 cm$^{-1}$ for vibrational frequencies. For comparison, the ``eamfX2C'' results recover around half of the 2e-SO contributions in most cases. For example, the error of the eamfX2C bond length for At$_2$ is 0.004 {\AA}, around half of the error of 0.007 {\AA} for the X2C-1e scheme. Therefore, all molecular properties studied here support the use of Eq. (\ref{MP2eene}) as the MP contribution to electronic energies. 



\begin{table}
\caption{Bond lengths (\AA) and harmonic vibrational frequencies (cm$^{-1}$) computed at the Kramers unrestricted Hartree-Fock level using Dyall's VTZ basis sets for Ag, Au, I, At \cite{Dyall02_dyalltz_4p5p6p,Dyall06_dyalltz_revise_4p5p6p,Dyall04_dz_tz_qz_5d, Dyall07_4d_dtqz,Dyall09_revised_5d,Dyall23_dyatz_5d} and uncontracted cc-pVTZ basis sets for H and F. \cite{Dunning89,Kendall92} The eamfX2C calculations used the energy expression in Eq. (\ref{eamfX2Cene}), while X2CMP calculations used Eq. (\ref{MP2eene}). The X2C-1e, eamfX2C, and X2CMP results are shown as the differences from the 4c-DC values. }
\begin{tabular}{cccccccccc}
\hline\hline
& \multicolumn{4}{c}{bond lengths} & ~~ & \multicolumn{4}{c}{harmonic frequencies} \\
\hline
Molecule & ~X2C-1e~ & ~eamfX2C~ & ~X2CMP~ & ~4c-DC~ & ~~ & ~X2C-1e~ & ~eamfX2C~ & ~X2CMP~ & ~4c-DC~ \\
\hline
Ag$_2$ & 0.00066 & 0.00032 & -0.00005 & 2.70318 & ~~ &  -0.1 & -0.1 &  0.0 & 148.7 \\
Au$_2$ & 0.00071 & 0.00035 & -0.00005 & 2.58768 & ~~ & -0.1 & -0.1 &  0.0 & 159.7 \\
I$_2$  & 0.00160 & 0.00076 & -0.00004 & 2.68020 & ~~ & -0.8 & -0.4&  0.0 & 228.3 \\
At$_2$ & 0.00741 & 0.00368 & -0.00005 & 2.96714 & ~~ & -1.4& -0.7&  0.0 & 130.5 \\
HI     & 0.00038 & 0.00018 & -0.00002 & 1.60486 & ~~ & -2.1 & -1.0 & 0.0& 2424.6 \\
HAt    & 0.00187 & 0.00091 & -0.00002 & 1.71064 & ~~ & -8.4 & -4.0 &  0.0 & 2120.7 \\
AgF    & 0.00022 & 0.00011 & -0.00002 & 2.01975 & ~~ & -0.1 & -0.1 &  0.0 & 501.0 \\
AuF    &-0.00089 &-0.00038 & -0.00003 & 1.95716 & ~~ &  0.5 &  0.2&  0.0 & 537.3 \\
AgH    & 0.00049 & 0.00023 & -0.00003 & 1.69924 & ~~ & -0.7 & -0.4& 0.0 & 1603.7 \\
AuH    & 0.00091 & 0.00045 & -0.00004 & 1.56711 & ~~ & -1.0 &  -0.7 & -0.1 & 2105.4 \\
\hline\hline
\end{tabular}
\label{tab-freq}
\end{table}


\subsection{Numerical performance of the one-center approximations} \label{Num1c}

\subsubsection{{\ {One-center approximation to scalar two-electron picture-change correction in the amfX2C scheme: numerical instability and a possible solution}}}

The scalar-relativistic two-electron integrals are less localized than the two-electron spin-orbit integrals. Therefore, the one-center approximation for the scalar 2e-pc contribution is expected to be less accurate than for two-electron spin-orbit contributions. In particular, a diffuse function may receive as significant contributions to scalar 2e-pc correction from other atoms as from its own center. Neglect of multi-center contributions may lead to numerical instability. 
This numerical instability of the amfX2C scheme is demonstrated in Table \ref{tab-ch4} using a simple example calculation of methane (CH$_4$) with standard basis sets. The amfX2C calculations could produce accurate energies when using cc-pVXZ (X=T, Q, 5) and aug-cc-pVXZ (X=T, Q) basis sets, but could not converge when using the aug-cc-pV5Z basis set. Note that this example only involves an organic molecule with no significant role of relativistic effects. Therefore, the one-center approximation to scalar 2e-pc correction in the amfX2C scheme as described in section II.D and II.E. is not recommended. \\

\begin{table}
  \caption{Hartree-Fock energies as well as the HOMO and LUMO spinor energies (Hartree) for the methane molecule (CH$_4$) with the C-H distances of 1.091 \AA. }
  \begin{tabular}{ccccccccc}
  \hline
  \hline
 & \multicolumn{3}{c}{cc-pV5Z-unc} & ~ ~ ~ ~ & \multicolumn{3}{c}{aug-cc-pV5Z-unc} \\
 & E$_{\text{HF}}$ & E$_{\text{HOMO}}$ & E$_{\text{LUMO}}$ & & E$_{\text{HF}}$ & E$_{\text{HOMO}}$ & E$_{\text{LUMO}}$ \\
 \hline
4c-DC                  &  -40.232878   & -0.544330   & 0.094706 & & -40.232833 &  -0.544329   & 0.023430  \\
X2C-1e                 &  -40.231496   & -0.544292   & 0.094709 & & -40.231502 &  -0.544291   & 0.023430 \\
X2CAMF                 &  -40.231506   & -0.544319   & 0.094708 & & -40.231526 &  -0.544319  & 0.023430 \\
X2CAMF [Eq. (\ref{eamfX2Cene})]                 &  -40.231496   & -0.544319   & 0.094708 & & -40.231502 &  -0.544319  & 0.023430 \\
amfX2C                 &  -40.232813   & -0.544314   & 0.093572 & &  \multicolumn{3}{c}{Not converged} \\
amfX2C$^a$                 &  ~ -40.232814  ~ & ~ -0.544314 ~ & ~ 0.094503 ~ & & ~ -40.232811 ~ & ~ -0.544317 ~  & ~ 0.023429 ~ \\
 \hline
 \hline
 \multicolumn{7}{l}{$^a$Excluding amfX2C integrals for the diffuse basis functions.}
\end{tabular}
\label{tab-ch4}
\end{table}

The numerical instability of the amfX2C scheme can in principle be alleviated by excluding the contributions to AMF integrals of diffuse basis functions. The last row in Table \ref{tab-ch4} presents a calculation with the exclusion of the amfX2C correction to diffuse basis functions in the aug-cc-pV5Z sets. This scheme successfully converges the calculation and produces accurate HF energy as well as the HOMO and LUMO spinor energies. On the other hand, the HF energies computed using this scheme do not show the right trend in that the HF/aug-cc-pV5Z energy is higher than the HF/cc-pV5Z energy. Nevertheless, this is a potential solution to the problem and is definitely worth further investigation. However, we should still stress the non-local nature of scalar-relativistic two-electron integrals and that special care should be taken when using one-center approximations to these contributions. The following discussion will not further consider the amfX2C scheme.

\begin{table}[h!]
\centering
\caption{The spin-orbit contributions to the bond lengths (\AA) and harmonic vibrational frequencies (cm$^{-1}$).
The ``four-component'' values are obtained as the difference between 4c-DC and SFDC results.
The X2CAMF values are obtained as the difference between X2CAMF and SFX2C-1e results. The X2CAMF values are given as the difference from the four-component values. The combination of Dyall's VTZ (or aVTZ) basis sets for Ag, Au, I, At \cite{Dyall02_dyalltz_4p5p6p,Dyall06_dyalltz_revise_4p5p6p,Dyall04_dz_tz_qz_5d, Dyall07_4d_dtqz,Dyall09_revised_5d,Dyall23_dyatz_5d} and the uncontracted cc-pVTZ (or aug-cc-pVTZ) basis sets for H and F \cite{Dunning89,Kendall92} are denoted as VTZ (or aVTZ) basis sets in the table.  }
\label{tab:comparison1}
\begin{tabular}{lccccccccc} 
\hline\hline
 & \multicolumn{4}{c}{four-component} & ~ ~ ~ ~ & \multicolumn{4}{c}{X2CAMF} \\ 
  \hline
 & \multicolumn{2}{c}{VTZ} & \multicolumn{2}{c}{aVTZ} &&  \multicolumn{2}{c}{VTZ} & \multicolumn{2}{c}{aVTZ} \\
 \hline
 & $R_e$ & $\omega_e$ &  $R_e$ & $\omega_e$ &&  $R_e$ & $\omega_e$ &  $R_e$ & $\omega_e$ \\
 \hline
Ag$_2$ & ~ -0.00076 ~ & ~ 0.2 ~ & ~ -0.00076 ~ & ~ 0.2 ~ && ~ -0.00005 ~ & ~ 0.0 ~ & ~ -0.00010 ~ & ~ 0.0 ~ \\
Au$_2$ & -0.00725 & 1.9 & -0.00723 & 1.9 && -0.00009 & 0.0 & -0.00030 & 0.0 \\
I$_2$ & 0.01352 & -8.2 & 0.01354 & -8.2 && -0.00002 & 0.0 & -0.00008 & 0.0 \\
At$_2$ & 0.12733 & -38.4 & 0.12705 & -38.0 && -0.00009 & 0.0 & -0.00020 & 0.0 \\
HI & 0.00262 & -19.2 & 0.00261 & -19.2 && -0.00001 & 0.0 & -0.00003 & 0.0 \\
HAt & 0.02733 & -160.4 & 0.02731 & -160.7 && -0.00006 & 0.4 & -0.00011 & 0.1 \\
AgF & -0.00038 & 0.2 & -0.00039 & 0.2 && 0.00000 & 0.0 & -0.00003 & 0.0 \\
AuF & -0.00853 & 5.1 & -0.00872 & 5.2 && -0.00006 & -0.1 & 0.00001 & -0.1 \\
AgH & -0.00053 & 1.3 & -0.00055 & 1.3 && -0.00002 & -0.1 & -0.00005 & -0.1 \\
AuH & -0.00294 & 19.4 & -0.00293 & 19.5 && -0.00009 & 0.0 & -0.00023 & -0.3 \\ 
\hline\hline
\end{tabular}
\end{table}

\subsubsection{One-center approximation to two-electron spin-orbit contributions and the Gaunt term in the X2CAMF scheme}

We now focus the discussion on the one-center approximation to two-electron spin-orbit (2e-SO) contributions in the X2CAMF scheme based on the Dirac-Coulomb Hamiltonian. As shown in Tables \ref{tab:comparison1}, the X2CAMF scheme can recover SO contributions accurately when using the VTZ and aVTZ basis sets. The SO contributions obtained using the X2CAMF scheme agree well with the 4c-DC results for equilibrium bond lengths and harmonic vibrational frequencies. The errors for bond lengths and harmonic frequencies are smaller than 0.0001 {\AA} and 0.5 cm$^{-1}$, respectively, when using VTZ basis sets. We note that, although the errors increase when including diffuse functions in the calculations, the absolute values for the errors are still small. For example, the largest error for bond lengths when using aVTZ basis sets amounts to 0.0003 {\AA} in the case of Au$_2$. As shown Tables \ref{tab-utherm} and \ref{tab-SO}, the X2CAMF scheme recovers 2e-SO contributions to thermochemical parameters and SO splittings accurately. The X2CAMF results are only slightly less accurate than the X2CMP ones. Therefore, it seems safe to conclude that the X2CAMF one-center approximation provides accurate treatment for the 2e-SO contributions to these molecular properties.  \\

\begin{table}[h!]
\centering
\caption{The contributions from the Gaunt term to the bond lengths (\AA) and harmonic vibrational frequencies (cm$^{-1}$).
The ``four-component'' values are obtained as the difference between 4c-DCG and 4c-DC results.
The X2CAMF values are obtained as the difference between X2CAMF(DCG) and X2CAMF(DC) results. The X2CAMF values are given as the difference from the four-component values. The combination of Dyall's VTZ (or aVTZ) basis sets for Ag, Au, I, At \cite{Dyall02_dyalltz_4p5p6p,Dyall06_dyalltz_revise_4p5p6p,Dyall04_dz_tz_qz_5d, Dyall07_4d_dtqz,Dyall09_revised_5d,Dyall23_dyatz_5d} and the uncontracted cc-pVTZ (or aug-cc-pVTZ) basis sets for H and F \cite{Dunning89,Kendall92} are denoted as VTZ (or aVTZ) basis sets in the table.}
\label{tab:comparison2}
\begin{tabular}{lccccccccc} 
\hline\hline
 & \multicolumn{4}{c}{four-component} & ~ ~ ~ ~ & \multicolumn{4}{c}{X2CAMF} \\ 
  \hline
 & \multicolumn{2}{c}{VTZ} & \multicolumn{2}{c}{aVTZ} &&  \multicolumn{2}{c}{VTZ} & \multicolumn{2}{c}{aVTZ} \\
 \hline
 & $R_e$ & $\omega_e$ &  $R_e$ & $\omega_e$ &&  $R_e$ & $\omega_e$ &  $R_e$ & $\omega_e$ \\
 \hline
 Ag$_2$& ~ 0.00191 ~ & ~ -0.3 ~  & ~ 0.00191 ~ & ~ -0.3 ~  && ~ -0.00034 ~  & ~ 0.0 ~ & ~ -0.00331 ~  & ~ 0.1 ~        \\ 
 Au$_2$&0.00319 & -0.6   & 0.00319 & -0.6   && -0.00053   & 0.0         & -0.00454   & 0.4         \\ 
 I$_2$ &0.00128 & 0.0    & 0.00129 & 0.0    && -0.00003   & 0.0         & -0.00078   & 0.0         \\ 
 At$_2$&0.00112 & 0.2    & 0.00113 & 0.2    && -0.00006   & 0.0         & -0.00131   & 0.0         \\
 HI &0.00078 & -1.1   & 0.00078 & -1.1   && -0.00002   & 0.0         & -0.00023   & 0.2         \\
 HAt&0.00113 & -0.3   & 0.00113 & -0.3   && -0.00001   & 0.1         & -0.00034   & 0.3         \\
 AgF&0.00119 & -0.5   & 0.00123 & -0.5   && -0.00001   & 0.1         & -0.00005   & 0.0         \\
 AuF&0.00279 & -1.6   & 0.00287 & -1.6   && -0.00003   & 0.0         & -0.00025   & 0.0         \\
 AgH&0.00138 & -2.6   & 0.00140 & -2.6   && -0.00005   & -0.1        & -0.00050   & 0.0         \\
 AuH&0.00180 & -7.0   & 0.00180 & -7.0   && -0.00019   & -0.7        & -0.00150   & -0.9        \\
\hline\hline
\end{tabular}
\end{table}

The spin-other-orbit interaction in the Gaunt term is as localized as the spin-same-orbit interaction in the Coulomb interaction. On the other hand, the scalar Gaunt contributions are not as localized as the spin-same-orbit contributions. The performance of the X2CAMF(DCG) scheme to recover the contributions from the Gaunt term thus is expected to be somewhat less robust than the X2CAMF(DC) scheme to recover 2e-SO contributions in the Coulomb interaction. As shown in Table \ref{tab:comparison2}, X2CAMF(DCG) calculations using VTZ basis sets can recover the Gaunt-term contributions to equilibrium bond lengths and harmonic frequencies accurately. The largest error for equilibrium bond lengths takes a value of 0.0005 {\AA} in the case of Au$_2$, which amounts to around 20\% of the total correction. This magnitude of an error is insignificant for chemical and spectroscopic applications. For comparison, the X2CAMF(DCG) calculations with aVTZ basis sets produce larger errors for the Gaunt-term contributions. The error for the equilibrium bond length of Au$_2$ increases to around 0.005 {\AA} and becomes significant for calculations aiming at high accuracy. We mention that the contributions from the Gaunt term to these properties are in general small and the overall errors of the X2CAMF(DCG) calculations are still small. Nevertheless, only calculations without diffuse functions are recommended when aiming at accurate treatments of the Gaunt-term contributions to structural parameters. \\

Finally, we mention that the increased errors in the one-center approximation with respect to the inclusion of diffuse functions can in principle be alleviated by excluding the AMF contributions to diffuse functions. This is similar to the approach to eliminate numerical instability of the amfX2C scheme discussed earlier. This idea is worth exploring in order to develop the X2CAMF scheme as a generally applicable approach of black-box nature. The research along this direction will be reported separately.

\section{Summary and Outlook} \label{sec:summary}

Cost-effective treatments of relativistic two-electron contributions within exact two-component theory are reviewed, with a focus on model potential (MP) and closely related atomic mean-field (AMF) approaches. 
It has been shown formally and numerically that the MP or AMF contributions to the electronic energies should be evaluated with the MP or AMF term in the Fock matrix treated as a transferable effective potential. 
The present study also analyzes the performance of one-center approximations to relativistic two-electron contributions. One-center approximations are demonstrated to be more reliable for contributions highly localized in the vicinity of nuclei, e.g., the two-electron spin-orbit interactions, and less accurate for the scalar Gaunt term that is less localized. The one-center approximation to the non-local scalar two-electron picture change (scalar 2e-pc) correction is shown to be numerically unstable. 
A simple scheme of excluding AMF contributions to diffuse basis functions is proposed to alleviate the numerical problems of the one-center approximations to non-local contributions. 

\section*{Acknowledgments}

The work at the Johns Hopkins University 
was supported by the Department
of Energy, Office and Science, Office of Basic Energy Sciences under Award Number
DE-SC0020317.
The computations were carried out at
Advanced Research Computing at Hopkins (ARCH) core
facility (rockfish.jhu.edu), which is supported by the NSF 
under Grant OAC-1920103.

\section*{Supplementary Material}


\section*{Declaration of interest statement}

The authors report that there are no competing interests to declare.

\newpage  

\clearpage

\bibliography{reference0,reference1,ref-laser-cooling,referenceKB,referencesDy,referencesMQMNSMintro,reference-nrh}

\begin{thebibliography}{149}%
\makeatletter
\providecommand \@ifxundefined [1]{%
 \@ifx{#1\undefined}
}%
\providecommand \@ifnum [1]{%
 \ifnum #1\expandafter \@firstoftwo
 \else \expandafter \@secondoftwo
 \fi
}%
\providecommand \@ifx [1]{%
 \ifx #1\expandafter \@firstoftwo
 \else \expandafter \@secondoftwo
 \fi
}%
\providecommand \natexlab [1]{#1}%
\providecommand \enquote  [1]{``#1''}%
\providecommand \bibnamefont  [1]{#1}%
\providecommand \bibfnamefont [1]{#1}%
\providecommand \citenamefont [1]{#1}%
\providecommand \href@noop [0]{\@secondoftwo}%
\providecommand \href [0]{\begingroup \@sanitize@url \@href}%
\providecommand \@href[1]{\@@startlink{#1}\@@href}%
\providecommand \@@href[1]{\endgroup#1\@@endlink}%
\providecommand \@sanitize@url [0]{\catcode `\\12\catcode `\$12\catcode
  `\&12\catcode `\#12\catcode `\^12\catcode `\_12\catcode `\%12\relax}%
\providecommand \@@startlink[1]{}%
\providecommand \@@endlink[0]{}%
\providecommand \url  [0]{\begingroup\@sanitize@url \@url }%
\providecommand \@url [1]{\endgroup\@href {#1}{\urlprefix }}%
\providecommand \urlprefix  [0]{URL }%
\providecommand \Eprint [0]{\href }%
\providecommand \doibase [0]{http://dx.doi.org/}%
\providecommand \selectlanguage [0]{\@gobble}%
\providecommand \bibinfo  [0]{\@secondoftwo}%
\providecommand \bibfield  [0]{\@secondoftwo}%
\providecommand \translation [1]{[#1]}%
\providecommand \BibitemOpen [0]{}%
\providecommand \bibitemStop [0]{}%
\providecommand \bibitemNoStop [0]{.\EOS\space}%
\providecommand \EOS [0]{\spacefactor3000\relax}%
\providecommand \BibitemShut  [1]{\csname bibitem#1\endcsname}%
\let\auto@bib@innerbib\@empty
\bibitem [{\citenamefont {Pyykk{\"o}}(1988)}]{Pyykko88}%
  \BibitemOpen
  \bibfield  {author} {\bibinfo {author} {\bibfnamefont {P.}~\bibnamefont
  {Pyykk{\"o}}},\ }\bibfield  {title} {\enquote {\bibinfo {title}
  {{Relativistic effects in structural chemistry}},}\ }\href@noop {} {\bibfield
   {journal} {\bibinfo  {journal} {Chem. Rev.}\ }\textbf {\bibinfo {volume}
  {88}},\ \bibinfo {pages} {563--594} (\bibinfo {year} {1988})}\BibitemShut
  {NoStop}%
\bibitem [{\citenamefont {Dyall}\ and\ \citenamefont
  {F{\ae}gri~Jr}(2007)}]{Dyall07}%
  \BibitemOpen
  \bibfield  {author} {\bibinfo {author} {\bibfnamefont {K.~G.}\ \bibnamefont
  {Dyall}}\ and\ \bibinfo {author} {\bibfnamefont {K.}~\bibnamefont
  {F{\ae}gri~Jr}},\ }\href@noop {} {\emph {\bibinfo {title} {Introduction to
  relativistic quantum chemistry}}}\ (\bibinfo  {publisher} {Oxford University
  Press},\ \bibinfo {year} {2007})\BibitemShut {NoStop}%
\bibitem [{\citenamefont {Autschbach}(2012)}]{Autschbach12}%
  \BibitemOpen
  \bibfield  {author} {\bibinfo {author} {\bibfnamefont {J.}~\bibnamefont
  {Autschbach}},\ }\bibfield  {title} {\enquote {\bibinfo {title} {Perspective:
  relativistic effects},}\ }\href@noop {} {\bibfield  {journal} {\bibinfo
  {journal} {J. Chem. Phys.}\ }\textbf {\bibinfo {volume} {136}},\ \bibinfo
  {pages} {150902} (\bibinfo {year} {2012})}\BibitemShut {NoStop}%
\bibitem [{\citenamefont {Reiher}\ and\ \citenamefont {Wolf}(2014)}]{Reiher14}%
  \BibitemOpen
  \bibfield  {author} {\bibinfo {author} {\bibfnamefont {M.}~\bibnamefont
  {Reiher}}\ and\ \bibinfo {author} {\bibfnamefont {A.}~\bibnamefont {Wolf}},\
  }\href@noop {} {\emph {\bibinfo {title} {Relativistic quantum chemistry: the
  fundamental theory of molecular science}}}\ (\bibinfo  {publisher} {John
  Wiley \& Sons},\ \bibinfo {year} {2014})\BibitemShut {NoStop}%
\bibitem [{\citenamefont {Liu}(2017)}]{Liu17}%
  \BibitemOpen
  \bibinfo {editor} {\bibfnamefont {W.}~\bibnamefont {Liu}},\ ed.,\ \href@noop
  {} {}Handbook of Relativistic Quantum Chemsitry\ (\bibinfo  {publisher}
  {Springer},\ \bibinfo {address} {Berlin},\ \bibinfo {year}
  {2017})\BibitemShut {NoStop}%
\bibitem [{\citenamefont {Grant}\ and\ \citenamefont {Quiney}(1988)}]{Grant88}%
  \BibitemOpen
  \bibfield  {author} {\bibinfo {author} {\bibfnamefont {I.~P.}\ \bibnamefont
  {Grant}}\ and\ \bibinfo {author} {\bibfnamefont {H.~M.}\ \bibnamefont
  {Quiney}},\ }\bibfield  {title} {\enquote {\bibinfo {title} {Foundations of
  the relativistic theory of atomic and molecular structure},}\ }in\ \href@noop
  {} {\emph {\bibinfo {booktitle} {Advances in atomic and molecular
  physics}}},\ Vol.~\bibinfo {volume} {23}\ (\bibinfo  {publisher} {Elsevier},\
  \bibinfo {year} {1988})\ pp.\ \bibinfo {pages} {37--86}\BibitemShut {NoStop}%
\bibitem [{\citenamefont {Eliav}, \citenamefont {Kaldor},\ and\ \citenamefont
  {Ishikawa}(1994)}]{Eliav94}%
  \BibitemOpen
  \bibfield  {author} {\bibinfo {author} {\bibfnamefont {E.}~\bibnamefont
  {Eliav}}, \bibinfo {author} {\bibfnamefont {U.}~\bibnamefont {Kaldor}}, \
  and\ \bibinfo {author} {\bibfnamefont {Y.}~\bibnamefont {Ishikawa}},\
  }\bibfield  {title} {\enquote {\bibinfo {title} {{Relativistic coupled
  cluster theory based on the no-pair dirac--coulomb--breit hamiltonian:
  Relativistic pair correlation energies of the Xe atom}},}\ }\href {\doibase
  https://doi.org/10.1002/qua.560520821} {\bibfield  {journal} {\bibinfo
  {journal} {Int. J. Quantum Chem.}\ }\textbf {\bibinfo {volume} {52}},\
  \bibinfo {pages} {205--214} (\bibinfo {year} {1994})}\BibitemShut {NoStop}%
\bibitem [{\citenamefont {Visscher}\ \emph {et~al.}(1994)\citenamefont
  {Visscher}, \citenamefont {Visser}, \citenamefont {Aerts}, \citenamefont
  {Merenga},\ and\ \citenamefont {Nieuwpoort}}]{Visscher94}%
  \BibitemOpen
  \bibfield  {author} {\bibinfo {author} {\bibfnamefont {L.}~\bibnamefont
  {Visscher}}, \bibinfo {author} {\bibfnamefont {O.}~\bibnamefont {Visser}},
  \bibinfo {author} {\bibfnamefont {P.~J.~C.}\ \bibnamefont {Aerts}}, \bibinfo
  {author} {\bibfnamefont {H.}~\bibnamefont {Merenga}}, \ and\ \bibinfo
  {author} {\bibfnamefont {W.~C.}\ \bibnamefont {Nieuwpoort}},\ }\bibfield
  {title} {\enquote {\bibinfo {title} {{Relativistic quantum chemistry: the
  MOLFDIR program package}},}\ }\href {\doibase
  http://dx.doi.org/10.1016/0010-4655(94)90115-5} {\bibfield  {journal}
  {\bibinfo  {journal} {Comput. Phys. Commun.}\ }\textbf {\bibinfo {volume}
  {81}},\ \bibinfo {pages} {120--144} (\bibinfo {year} {1994})}\BibitemShut
  {NoStop}%
\bibitem [{\citenamefont {Saue}\ \emph {et~al.}(1997)\citenamefont {Saue},
  \citenamefont {Faegri}, \citenamefont {Helgaker},\ and\ \citenamefont
  {Gropen}}]{Saue97}%
  \BibitemOpen
  \bibfield  {author} {\bibinfo {author} {\bibfnamefont {T.}~\bibnamefont
  {Saue}}, \bibinfo {author} {\bibfnamefont {K.}~\bibnamefont {Faegri}},
  \bibinfo {author} {\bibfnamefont {T.~U.}\ \bibnamefont {Helgaker}}, \ and\
  \bibinfo {author} {\bibfnamefont {O.}~\bibnamefont {Gropen}},\ }\bibfield
  {title} {\enquote {\bibinfo {title} {{Principles of Direct 4-component
  Relativistic SCF: Application to Caesium Auride}},}\ }\href@noop {}
  {\bibfield  {journal} {\bibinfo  {journal} {Mol. Phys.}\ }\textbf {\bibinfo
  {volume} {91}},\ \bibinfo {pages} {937} (\bibinfo {year} {1997})}\BibitemShut
  {NoStop}%
\bibitem [{\citenamefont {Yanai}\ \emph {et~al.}(2001)\citenamefont {Yanai},
  \citenamefont {Nakajima}, \citenamefont {Ishikawa},\ and\ \citenamefont
  {Hirao}}]{Yanai01}%
  \BibitemOpen
  \bibfield  {author} {\bibinfo {author} {\bibfnamefont {T.}~\bibnamefont
  {Yanai}}, \bibinfo {author} {\bibfnamefont {T.}~\bibnamefont {Nakajima}},
  \bibinfo {author} {\bibfnamefont {Y.}~\bibnamefont {Ishikawa}}, \ and\
  \bibinfo {author} {\bibfnamefont {K.}~\bibnamefont {Hirao}},\ }\bibfield
  {title} {\enquote {\bibinfo {title} {{A new computational scheme for the
  Dirac-Hartree-Fock method employing an efficient integral algorithm}},}\
  }\href@noop {} {\bibfield  {journal} {\bibinfo  {journal} {J. Chem. Phys.}\
  }\textbf {\bibinfo {volume} {114}},\ \bibinfo {pages} {6526} (\bibinfo {year}
  {2001})}\BibitemShut {NoStop}%
\bibitem [{\citenamefont {Liu}, \citenamefont {Wang},\ and\ \citenamefont
  {Li}(2003)}]{Liu03}%
  \BibitemOpen
  \bibfield  {author} {\bibinfo {author} {\bibfnamefont {W.}~\bibnamefont
  {Liu}}, \bibinfo {author} {\bibfnamefont {F.}~\bibnamefont {Wang}}, \ and\
  \bibinfo {author} {\bibfnamefont {L.}~\bibnamefont {Li}},\ }\bibfield
  {title} {\enquote {\bibinfo {title} {{The Beijing Density Functional (BDF)
  Program Package: Methodologies and Applications}},}\ }\href {\doibase
  10.1142/S0219633603000471} {\bibfield  {journal} {\bibinfo  {journal} {J.
  Theor. Comput. Chem.}\ }\textbf {\bibinfo {volume} {02}},\ \bibinfo {pages}
  {257--272} (\bibinfo {year} {2003})}\BibitemShut {NoStop}%
\bibitem [{\citenamefont {Storchi}\ \emph {et~al.}(2010)\citenamefont
  {Storchi}, \citenamefont {Belpassi}, \citenamefont {Tarantelli},
  \citenamefont {Sgamellotti},\ and\ \citenamefont {Quiney}}]{Storchi10}%
  \BibitemOpen
  \bibfield  {author} {\bibinfo {author} {\bibfnamefont {L.}~\bibnamefont
  {Storchi}}, \bibinfo {author} {\bibfnamefont {L.}~\bibnamefont {Belpassi}},
  \bibinfo {author} {\bibfnamefont {F.}~\bibnamefont {Tarantelli}}, \bibinfo
  {author} {\bibfnamefont {A.}~\bibnamefont {Sgamellotti}}, \ and\ \bibinfo
  {author} {\bibfnamefont {H.~M.}\ \bibnamefont {Quiney}},\ }\bibfield  {title}
  {\enquote {\bibinfo {title} {{An Efficient Parallel All-Electron
  Four-Component Dirac-Kohn-Sham Program Using a Distributed Matrix
  Approach}},}\ }\href {\doibase 10.1021/ct900539m} {\bibfield  {journal}
  {\bibinfo  {journal} {J. Chem. Theory Comput.}\ }\textbf {\bibinfo {volume}
  {6}},\ \bibinfo {pages} {384--394} (\bibinfo {year} {2010})}\BibitemShut
  {NoStop}%
\bibitem [{\citenamefont {Komorovsky}\ \emph {et~al.}(2010)\citenamefont
  {Komorovsky}, \citenamefont {Repisky}, \citenamefont {Malkina},\ and\
  \citenamefont {Malkin}}]{Komorovsky10}%
  \BibitemOpen
  \bibfield  {author} {\bibinfo {author} {\bibfnamefont {S.}~\bibnamefont
  {Komorovsky}}, \bibinfo {author} {\bibfnamefont {M.}~\bibnamefont {Repisky}},
  \bibinfo {author} {\bibfnamefont {O.~L.}\ \bibnamefont {Malkina}}, \ and\
  \bibinfo {author} {\bibfnamefont {V.~G.}\ \bibnamefont {Malkin}},\ }\bibfield
   {title} {\enquote {\bibinfo {title} {{Fully Relativistic Calculations of NMR
  Shielding Tensors Using Restricted Magnetically Balanced Basis and Gauge
  Including Atomic Orbitals}},}\ }\href@noop {} {\bibfield  {journal} {\bibinfo
   {journal} {J. Chem. Phys.}\ }\textbf {\bibinfo {volume} {132}},\ \bibinfo
  {pages} {154101} (\bibinfo {year} {2010})}\BibitemShut {NoStop}%
\bibitem [{\citenamefont {Kelley}\ and\ \citenamefont
  {Shiozaki}(2013)}]{Kelley13}%
  \BibitemOpen
  \bibfield  {author} {\bibinfo {author} {\bibfnamefont {M.~S.}\ \bibnamefont
  {Kelley}}\ and\ \bibinfo {author} {\bibfnamefont {T.}~\bibnamefont
  {Shiozaki}},\ }\bibfield  {title} {\enquote {\bibinfo {title} {{Large-scale
  Dirac--Fock--Breit method using density fitting and 2-spinor basis
  functions}},}\ }\href {\doibase 10.1063/1.4807612} {\bibfield  {journal}
  {\bibinfo  {journal} {J. Chem. Phys.}\ }\textbf {\bibinfo {volume} {138}},\
  \bibinfo {pages} {204113} (\bibinfo {year} {2013})}\BibitemShut {NoStop}%
\bibitem [{\citenamefont {Kadek}, \citenamefont {Repisky},\ and\ \citenamefont
  {Ruud}(2019)}]{Kadek19}%
  \BibitemOpen
  \bibfield  {author} {\bibinfo {author} {\bibfnamefont {M.}~\bibnamefont
  {Kadek}}, \bibinfo {author} {\bibfnamefont {M.}~\bibnamefont {Repisky}}, \
  and\ \bibinfo {author} {\bibfnamefont {K.}~\bibnamefont {Ruud}},\ }\bibfield
  {title} {\enquote {\bibinfo {title} {{All-electron fully relativistic
  Kohn-Sham theory for solids based on the Dirac-Coulomb Hamiltonian and
  Gaussian-type functions}},}\ }\href {\doibase 10.1103/PhysRevB.99.205103}
  {\bibfield  {journal} {\bibinfo  {journal} {Phys. Rev. B}\ }\textbf {\bibinfo
  {volume} {99}},\ \bibinfo {pages} {205103} (\bibinfo {year}
  {2019})}\BibitemShut {NoStop}%
\bibitem [{\citenamefont {Sun}\ \emph {et~al.}(2018)\citenamefont {Sun},
  \citenamefont {Berkelbach}, \citenamefont {Blunt}, \citenamefont {Booth},
  \citenamefont {Guo}, \citenamefont {Li}, \citenamefont {Liu}, \citenamefont
  {McClain}, \citenamefont {Sayfutyarova}, \citenamefont {Sharma},
  \citenamefont {Wouters},\ and\ \citenamefont {Chan}}]{sun_2018_pyscf}%
  \BibitemOpen
  \bibfield  {author} {\bibinfo {author} {\bibfnamefont {Q.}~\bibnamefont
  {Sun}}, \bibinfo {author} {\bibfnamefont {T.~C.}\ \bibnamefont {Berkelbach}},
  \bibinfo {author} {\bibfnamefont {N.~S.}\ \bibnamefont {Blunt}}, \bibinfo
  {author} {\bibfnamefont {G.~H.}\ \bibnamefont {Booth}}, \bibinfo {author}
  {\bibfnamefont {S.}~\bibnamefont {Guo}}, \bibinfo {author} {\bibfnamefont
  {Z.}~\bibnamefont {Li}}, \bibinfo {author} {\bibfnamefont {J.}~\bibnamefont
  {Liu}}, \bibinfo {author} {\bibfnamefont {J.~D.}\ \bibnamefont {McClain}},
  \bibinfo {author} {\bibfnamefont {E.~R.}\ \bibnamefont {Sayfutyarova}},
  \bibinfo {author} {\bibfnamefont {S.}~\bibnamefont {Sharma}}, \bibinfo
  {author} {\bibfnamefont {S.}~\bibnamefont {Wouters}}, \ and\ \bibinfo
  {author} {\bibfnamefont {G.~K.-L.}\ \bibnamefont {Chan}},\ }\bibfield
  {title} {\enquote {\bibinfo {title} {{{PySCF}}: The {{Python-based}}
  simulations of chemistry framework},}\ }\href {\doibase 10.1002/wcms.1340}
  {\bibfield  {journal} {\bibinfo  {journal} {WIREs Computational Molecular
  Science}\ }\textbf {\bibinfo {volume} {8}},\ \bibinfo {pages} {e1340}
  (\bibinfo {year} {2018})}\BibitemShut {NoStop}%
\bibitem [{\citenamefont {Sun}\ \emph {et~al.}(2021)\citenamefont {Sun},
  \citenamefont {Stetina}, \citenamefont {Zhang}, \citenamefont {Hu},
  \citenamefont {Valeev}, \citenamefont {Sun},\ and\ \citenamefont
  {Li}}]{Sun21}%
  \BibitemOpen
  \bibfield  {author} {\bibinfo {author} {\bibfnamefont {S.}~\bibnamefont
  {Sun}}, \bibinfo {author} {\bibfnamefont {T.~F.}\ \bibnamefont {Stetina}},
  \bibinfo {author} {\bibfnamefont {T.}~\bibnamefont {Zhang}}, \bibinfo
  {author} {\bibfnamefont {H.}~\bibnamefont {Hu}}, \bibinfo {author}
  {\bibfnamefont {E.~F.}\ \bibnamefont {Valeev}}, \bibinfo {author}
  {\bibfnamefont {Q.}~\bibnamefont {Sun}}, \ and\ \bibinfo {author}
  {\bibfnamefont {X.}~\bibnamefont {Li}},\ }\bibfield  {title} {\enquote
  {\bibinfo {title} {{Efficient Four-Component Dirac--Coulomb--Gaunt
  Hartree--Fock in the Pauli Spinor Representation}},}\ }\href {\doibase
  10.1021/acs.jctc.1c00137} {\bibfield  {journal} {\bibinfo  {journal} {J.
  Chem. Theory Comput.}\ }\textbf {\bibinfo {volume} {17}},\ \bibinfo {pages}
  {3388--3402} (\bibinfo {year} {2021})}\BibitemShut {NoStop}%
\bibitem [{\citenamefont {Van~Lenthe}\ \emph {et~al.}(1996)\citenamefont
  {Van~Lenthe}, \citenamefont {Van~Leeuwen}, \citenamefont {Baerends},\ and\
  \citenamefont {Snijders}}]{VanLenthe96}%
  \BibitemOpen
  \bibfield  {author} {\bibinfo {author} {\bibfnamefont {E.}~\bibnamefont
  {Van~Lenthe}}, \bibinfo {author} {\bibfnamefont {R.}~\bibnamefont
  {Van~Leeuwen}}, \bibinfo {author} {\bibfnamefont {E.}~\bibnamefont
  {Baerends}}, \ and\ \bibinfo {author} {\bibfnamefont {J.}~\bibnamefont
  {Snijders}},\ }\bibfield  {title} {\enquote {\bibinfo {title} {Relativistic
  regular two-component hamiltonians},}\ }\href@noop {} {\bibfield  {journal}
  {\bibinfo  {journal} {Int. J. Quantum Chem.}\ }\textbf {\bibinfo {volume}
  {57}},\ \bibinfo {pages} {281--293} (\bibinfo {year} {1996})}\BibitemShut
  {NoStop}%
\bibitem [{\citenamefont {Reiher}(2006)}]{Reiher06}%
  \BibitemOpen
  \bibfield  {author} {\bibinfo {author} {\bibfnamefont {M.}~\bibnamefont
  {Reiher}},\ }\bibfield  {title} {\enquote {\bibinfo {title}
  {{Douglas–Kroll–Hess Theory: a relativistic electrons-only theory for
  chemistry}},}\ }\href {\doibase 10.1007/s00214-005-0003-2} {\bibfield
  {journal} {\bibinfo  {journal} {Theor. Chem. Acc.}\ }\textbf {\bibinfo
  {volume} {116}},\ \bibinfo {pages} {241--252} (\bibinfo {year}
  {2006})}\BibitemShut {NoStop}%
\bibitem [{\citenamefont {Liu}(2010)}]{Liu10}%
  \BibitemOpen
  \bibfield  {author} {\bibinfo {author} {\bibfnamefont {W.}~\bibnamefont
  {Liu}},\ }\bibfield  {title} {\enquote {\bibinfo {title} {{Ideas of
  relativistic quantum chemistry}},}\ }\href {\doibase
  10.1080/00268971003781571} {\bibfield  {journal} {\bibinfo  {journal} {Mol.
  Phys.}\ }\textbf {\bibinfo {volume} {108}},\ \bibinfo {pages} {1679--1706}
  (\bibinfo {year} {2010})}\BibitemShut {NoStop}%
\bibitem [{\citenamefont {Saue}(2011)}]{Saue11}%
  \BibitemOpen
  \bibfield  {author} {\bibinfo {author} {\bibfnamefont {T.}~\bibnamefont
  {Saue}},\ }\bibfield  {title} {\enquote {\bibinfo {title} {{Relativistic
  Hamiltonians for Chemistry: A Primer}},}\ }\href {\doibase
  10.1002/cphc.201100682} {\bibfield  {journal} {\bibinfo  {journal}
  {ChemPhysChem}\ }\textbf {\bibinfo {volume} {12}},\ \bibinfo {pages}
  {3077--3094} (\bibinfo {year} {2011})}\BibitemShut {NoStop}%
\bibitem [{\citenamefont {Barysz}\ and\ \citenamefont
  {Sadlej}(2001)}]{Barysz01}%
  \BibitemOpen
  \bibfield  {author} {\bibinfo {author} {\bibfnamefont {M.}~\bibnamefont
  {Barysz}}\ and\ \bibinfo {author} {\bibfnamefont {A.~J.}\ \bibnamefont
  {Sadlej}},\ }\bibfield  {title} {\enquote {\bibinfo {title} {Two-component
  methods of relativistic quantum chemistry: from the douglas--kroll
  approximation to the exact two-component formalism},}\ }\href@noop {}
  {\bibfield  {journal} {\bibinfo  {journal} {J. Mol. Struct.}\ }\textbf
  {\bibinfo {volume} {573}},\ \bibinfo {pages} {181--200} (\bibinfo {year}
  {2001})}\BibitemShut {NoStop}%
\bibitem [{\citenamefont {Kutzelnigg}(2012)}]{Kutzelnigg12}%
  \BibitemOpen
  \bibfield  {author} {\bibinfo {author} {\bibfnamefont {W.}~\bibnamefont
  {Kutzelnigg}},\ }\bibfield  {title} {\enquote {\bibinfo {title} {{Solved and
  unsolved problems in relativistic quantum chemistry}},}\ }\href {\doibase
  10.1016/j.chemphys.2011.06.001} {\bibfield  {journal} {\bibinfo  {journal}
  {Chem. Phys.}\ }\textbf {\bibinfo {volume} {395}},\ \bibinfo {pages} {16--34}
  (\bibinfo {year} {2012})}\BibitemShut {NoStop}%
\bibitem [{\citenamefont {Peng}\ and\ \citenamefont {Reiher}(2012)}]{Peng12}%
  \BibitemOpen
  \bibfield  {author} {\bibinfo {author} {\bibfnamefont {D.}~\bibnamefont
  {Peng}}\ and\ \bibinfo {author} {\bibfnamefont {M.}~\bibnamefont {Reiher}},\
  }\bibfield  {title} {\enquote {\bibinfo {title} {{Exact decoupling of the
  relativistic Fock operator}},}\ }\href@noop {} {\bibfield  {journal}
  {\bibinfo  {journal} {Theor. Chem. Acc.}\ }\textbf {\bibinfo {volume}
  {131}},\ \bibinfo {pages} {1--20} (\bibinfo {year} {2012})}\BibitemShut
  {NoStop}%
\bibitem [{\citenamefont {Nakajima}\ and\ \citenamefont
  {Hirao}(2012)}]{Nakajima12}%
  \BibitemOpen
  \bibfield  {author} {\bibinfo {author} {\bibfnamefont {T.}~\bibnamefont
  {Nakajima}}\ and\ \bibinfo {author} {\bibfnamefont {K.}~\bibnamefont
  {Hirao}},\ }\bibfield  {title} {\enquote {\bibinfo {title} {{The
  Douglas-Kroll-Hess Approach}},}\ }\href {\doibase 10.1021/cr200040s}
  {\bibfield  {journal} {\bibinfo  {journal} {Chem. Rev.}\ }\textbf {\bibinfo
  {volume} {112}},\ \bibinfo {pages} {385--402} (\bibinfo {year}
  {2012})}\BibitemShut {NoStop}%
\bibitem [{\citenamefont {He{\ss}}(1986)}]{Hess86}%
  \BibitemOpen
  \bibfield  {author} {\bibinfo {author} {\bibfnamefont {B.~A.}\ \bibnamefont
  {He{\ss}}},\ }\bibfield  {title} {\enquote {\bibinfo {title} {Relativistic
  electronic-structure calculations employing a two-component no-pair formalism
  with external-field projection operators},}\ }\href@noop {} {\bibfield
  {journal} {\bibinfo  {journal} {Phys. Rev. A}\ }\textbf {\bibinfo {volume}
  {33}},\ \bibinfo {pages} {3742} (\bibinfo {year} {1986})}\BibitemShut
  {NoStop}%
\bibitem [{\citenamefont {Wolf}, \citenamefont {Reiher},\ and\ \citenamefont
  {Hess}(2002)}]{Wolf02}%
  \BibitemOpen
  \bibfield  {author} {\bibinfo {author} {\bibfnamefont {A.}~\bibnamefont
  {Wolf}}, \bibinfo {author} {\bibfnamefont {M.}~\bibnamefont {Reiher}}, \ and\
  \bibinfo {author} {\bibfnamefont {B.~A.}\ \bibnamefont {Hess}},\ }\bibfield
  {title} {\enquote {\bibinfo {title} {{The generalized Douglas–Kroll
  transformation}},}\ }\href {\doibase 10.1063/1.1515314} {\bibfield  {journal}
  {\bibinfo  {journal} {J. Chem. Phys.}\ }\textbf {\bibinfo {volume} {117}},\
  \bibinfo {pages} {9215--9226} (\bibinfo {year} {2002})}\BibitemShut {NoStop}%
\bibitem [{\citenamefont {van Lenthe}, \citenamefont {Baerends},\ and\
  \citenamefont {Snijders}(1993{\natexlab{a}})}]{VanLenthe93}%
  \BibitemOpen
  \bibfield  {author} {\bibinfo {author} {\bibfnamefont {E.}~\bibnamefont {van
  Lenthe}}, \bibinfo {author} {\bibfnamefont {E.~J.}\ \bibnamefont {Baerends}},
  \ and\ \bibinfo {author} {\bibfnamefont {J.~G.}\ \bibnamefont {Snijders}},\
  }\bibfield  {title} {\enquote {\bibinfo {title} {{Relativistic regular
  two‐component Hamiltonians}},}\ }\href {\doibase 10.1063/1.466059}
  {\bibfield  {journal} {\bibinfo  {journal} {J. Chem. Phys.}\ }\textbf
  {\bibinfo {volume} {99}},\ \bibinfo {pages} {4597--4610} (\bibinfo {year}
  {1993}{\natexlab{a}})}\BibitemShut {NoStop}%
\bibitem [{\citenamefont {Nakajima}\ and\ \citenamefont
  {Hirao}(1999)}]{Nakajima99}%
  \BibitemOpen
  \bibfield  {author} {\bibinfo {author} {\bibfnamefont {T.}~\bibnamefont
  {Nakajima}}\ and\ \bibinfo {author} {\bibfnamefont {K.}~\bibnamefont
  {Hirao}},\ }\bibfield  {title} {\enquote {\bibinfo {title} {A new
  relativistic theory: a relativistic scheme by eliminating small components
  (resc)},}\ }\href@noop {} {\bibfield  {journal} {\bibinfo  {journal} {Chem.
  Phys. Lett.}\ }\textbf {\bibinfo {volume} {302}},\ \bibinfo {pages}
  {383--391} (\bibinfo {year} {1999})}\BibitemShut {NoStop}%
\bibitem [{\citenamefont {Barysz}\ and\ \citenamefont
  {Sadlej}(1997)}]{Barysz97}%
  \BibitemOpen
  \bibfield  {author} {\bibinfo {author} {\bibfnamefont {M.}~\bibnamefont
  {Barysz}}\ and\ \bibinfo {author} {\bibfnamefont {A.~J.}\ \bibnamefont
  {Sadlej}},\ }\bibfield  {title} {\enquote {\bibinfo {title} {{Expectation
  values of operators in approximate two-component relativistic theories}},}\
  }\href {\doibase 10.1007/s002140050260} {\bibfield  {journal} {\bibinfo
  {journal} {Theor. Chem. Acc.}\ }\textbf {\bibinfo {volume} {97}},\ \bibinfo
  {pages} {260--270} (\bibinfo {year} {1997})}\BibitemShut {NoStop}%
\bibitem [{\citenamefont {Barysz}\ and\ \citenamefont
  {Sadlej}(2002)}]{Barysz02}%
  \BibitemOpen
  \bibfield  {author} {\bibinfo {author} {\bibfnamefont {M.}~\bibnamefont
  {Barysz}}\ and\ \bibinfo {author} {\bibfnamefont {A.~J.}\ \bibnamefont
  {Sadlej}},\ }\bibfield  {title} {\enquote {\bibinfo {title} {{Infinite-order
  two-component theory for relativistic quantum chemistry}},}\ }\href {\doibase
  10.1063/1.1436462} {\bibfield  {journal} {\bibinfo  {journal} {J. Chem.
  Phys.}\ }\textbf {\bibinfo {volume} {116}},\ \bibinfo {pages} {2696--2704}
  (\bibinfo {year} {2002})}\BibitemShut {NoStop}%
\bibitem [{\citenamefont {Filatov}\ and\ \citenamefont
  {Cremer}(2005)}]{Filatov05}%
  \BibitemOpen
  \bibfield  {author} {\bibinfo {author} {\bibfnamefont {M.}~\bibnamefont
  {Filatov}}\ and\ \bibinfo {author} {\bibfnamefont {D.}~\bibnamefont
  {Cremer}},\ }\bibfield  {title} {\enquote {\bibinfo {title} {{Connection
  between the regular approximation and the normalized elimination of the small
  component in relativistic quantum theory}},}\ }\href {\doibase
  10.1063/1.1844298} {\bibfield  {journal} {\bibinfo  {journal} {J. Chem.
  Phys.}\ }\textbf {\bibinfo {volume} {122}},\ \bibinfo {pages} {64104}
  (\bibinfo {year} {2005})}\BibitemShut {NoStop}%
\bibitem [{\citenamefont {Kutzelnigg}\ and\ \citenamefont
  {Liu}(2006)}]{Liu06a}%
  \BibitemOpen
  \bibfield  {author} {\bibinfo {author} {\bibfnamefont {W.}~\bibnamefont
  {Kutzelnigg}}\ and\ \bibinfo {author} {\bibfnamefont {W.}~\bibnamefont
  {Liu}},\ }\bibfield  {title} {\enquote {\bibinfo {title} {{Quasirelativistic
  theory I. Theory in terms of a quasi-relativistic operator}},}\ }\href
  {\doibase 10.1080/00268970600662481} {\bibfield  {journal} {\bibinfo
  {journal} {Mol. Phys.}\ }\textbf {\bibinfo {volume} {104}},\ \bibinfo {pages}
  {2225--2240} (\bibinfo {year} {2006})}\BibitemShut {NoStop}%
\bibitem [{\citenamefont {Liu}\ and\ \citenamefont {Kutzelnigg}(2007)}]{Liu07}%
  \BibitemOpen
  \bibfield  {author} {\bibinfo {author} {\bibfnamefont {W.}~\bibnamefont
  {Liu}}\ and\ \bibinfo {author} {\bibfnamefont {W.}~\bibnamefont
  {Kutzelnigg}},\ }\bibfield  {title} {\enquote {\bibinfo {title}
  {{Quasirelativistic theory. II. Theory at matrix level}},}\ }\href {\doibase
  10.1063/1.2710258} {\bibfield  {journal} {\bibinfo  {journal} {J. Chem.
  Phys.}\ }\textbf {\bibinfo {volume} {126}},\ \bibinfo {pages} {114107}
  (\bibinfo {year} {2007})}\BibitemShut {NoStop}%
\bibitem [{\citenamefont {Dyall}(1997)}]{Dyall97}%
  \BibitemOpen
  \bibfield  {author} {\bibinfo {author} {\bibfnamefont {K.~G.}\ \bibnamefont
  {Dyall}},\ }\bibfield  {title} {\enquote {\bibinfo {title} {{Interfacing
  relativistic and nonrelativistic methods. I. Normalized elimination of the
  small component in the modified Dirac equation}},}\ }\href@noop {} {\bibfield
   {journal} {\bibinfo  {journal} {J. Chem. Phys.}\ }\textbf {\bibinfo {volume}
  {106}},\ \bibinfo {pages} {9618--9626} (\bibinfo {year} {1997})}\BibitemShut
  {NoStop}%
\bibitem [{\citenamefont {Dyall}(2001)}]{Dyall01}%
  \BibitemOpen
  \bibfield  {author} {\bibinfo {author} {\bibfnamefont {K.~G.}\ \bibnamefont
  {Dyall}},\ }\bibfield  {title} {\enquote {\bibinfo {title} {{Interfacing
  relativistic and nonrelativistic methods. IV. One- and two-electron scalar
  approximations}},}\ }\href@noop {} {\bibfield  {journal} {\bibinfo  {journal}
  {J. Chem. Phys.}\ }\textbf {\bibinfo {volume} {115}},\ \bibinfo {pages}
  {9136--9143} (\bibinfo {year} {2001})}\BibitemShut {NoStop}%
\bibitem [{\citenamefont {Kutzelnigg}\ and\ \citenamefont
  {Liu}(2005)}]{Kutzelnigg05}%
  \BibitemOpen
  \bibfield  {author} {\bibinfo {author} {\bibfnamefont {W.}~\bibnamefont
  {Kutzelnigg}}\ and\ \bibinfo {author} {\bibfnamefont {W.}~\bibnamefont
  {Liu}},\ }\bibfield  {title} {\enquote {\bibinfo {title} {{Quasirelativistic
  theory equivalent to fully relativistic theory}},}\ }\href@noop {} {\bibfield
   {journal} {\bibinfo  {journal} {J. Chem. Phys.}\ }\textbf {\bibinfo {volume}
  {123}},\ \bibinfo {pages} {241102} (\bibinfo {year} {2005})}\BibitemShut
  {NoStop}%
\bibitem [{\citenamefont {Ilia{\v{s}}}\ and\ \citenamefont
  {Saue}(2007)}]{Ilias07}%
  \BibitemOpen
  \bibfield  {author} {\bibinfo {author} {\bibfnamefont {M.}~\bibnamefont
  {Ilia{\v{s}}}}\ and\ \bibinfo {author} {\bibfnamefont {T.}~\bibnamefont
  {Saue}},\ }\bibfield  {title} {\enquote {\bibinfo {title} {{An infinite-order
  two-component relativistic Hamiltonian by a simple one-step
  transformation}},}\ }\href@noop {} {\bibfield  {journal} {\bibinfo  {journal}
  {J. Chem. Phys.}\ }\textbf {\bibinfo {volume} {126}},\ \bibinfo {pages}
  {064102} (\bibinfo {year} {2007})}\BibitemShut {NoStop}%
\bibitem [{\citenamefont {Liu}\ and\ \citenamefont {Peng}(2009)}]{Liu09}%
  \BibitemOpen
  \bibfield  {author} {\bibinfo {author} {\bibfnamefont {W.}~\bibnamefont
  {Liu}}\ and\ \bibinfo {author} {\bibfnamefont {D.}~\bibnamefont {Peng}},\
  }\bibfield  {title} {\enquote {\bibinfo {title} {Exact two-component
  hamiltonians revisited},}\ }\href@noop {} {\bibfield  {journal} {\bibinfo
  {journal} {J. Chem. Phys.}\ }\textbf {\bibinfo {volume} {131}},\ \bibinfo
  {pages} {031104} (\bibinfo {year} {2009})}\BibitemShut {NoStop}%
\bibitem [{\citenamefont {Xu}\ \emph {et~al.}(2009)\citenamefont {Xu},
  \citenamefont {Ma}, \citenamefont {Peng}, \citenamefont {Zou}, \citenamefont
  {Liu},\ and\ \citenamefont {Staemmler}}]{Xu09}%
  \BibitemOpen
  \bibfield  {author} {\bibinfo {author} {\bibfnamefont {W.}~\bibnamefont
  {Xu}}, \bibinfo {author} {\bibfnamefont {J.}~\bibnamefont {Ma}}, \bibinfo
  {author} {\bibfnamefont {D.}~\bibnamefont {Peng}}, \bibinfo {author}
  {\bibfnamefont {W.}~\bibnamefont {Zou}}, \bibinfo {author} {\bibfnamefont
  {W.}~\bibnamefont {Liu}}, \ and\ \bibinfo {author} {\bibfnamefont
  {V.}~\bibnamefont {Staemmler}},\ }\bibfield  {title} {\enquote {\bibinfo
  {title} {{Excited states of ReO4-: A comprehensive time-dependent
  relativistic density functional theory study}},}\ }\href {\doibase
  https://doi.org/10.1016/j.chemphys.2008.10.011} {\bibfield  {journal}
  {\bibinfo  {journal} {Chem. Phys.}\ }\textbf {\bibinfo {volume} {356}},\
  \bibinfo {pages} {219--228} (\bibinfo {year} {2009})}\BibitemShut {NoStop}%
\bibitem [{\citenamefont {Zou}, \citenamefont {Filatov},\ and\ \citenamefont
  {Cremer}(2011)}]{Zou11}%
  \BibitemOpen
  \bibfield  {author} {\bibinfo {author} {\bibfnamefont {W.}~\bibnamefont
  {Zou}}, \bibinfo {author} {\bibfnamefont {M.}~\bibnamefont {Filatov}}, \ and\
  \bibinfo {author} {\bibfnamefont {D.}~\bibnamefont {Cremer}},\ }\bibfield
  {title} {\enquote {\bibinfo {title} {Development and application of the
  analytical energy gradient for the normalized elimination of the small
  component method},}\ }\href@noop {} {\bibfield  {journal} {\bibinfo
  {journal} {J. Chem. Phys.}\ }\textbf {\bibinfo {volume} {134}},\ \bibinfo
  {pages} {244117} (\bibinfo {year} {2011})}\BibitemShut {NoStop}%
\bibitem [{\citenamefont {Filatov}, \citenamefont {Zou},\ and\ \citenamefont
  {Cremer}(2014)}]{Filatov14}%
  \BibitemOpen
  \bibfield  {author} {\bibinfo {author} {\bibfnamefont {M.}~\bibnamefont
  {Filatov}}, \bibinfo {author} {\bibfnamefont {W.}~\bibnamefont {Zou}}, \ and\
  \bibinfo {author} {\bibfnamefont {D.}~\bibnamefont {Cremer}},\ }\bibfield
  {title} {\enquote {\bibinfo {title} {{Calculation of response properties with
  the normalized elimination of the small component method}},}\ }\href
  {\doibase 10.1002/qua.24578} {\bibfield  {journal} {\bibinfo  {journal} {Int.
  J. Quantum Chem.}\ }\textbf {\bibinfo {volume} {114}},\ \bibinfo {pages}
  {993--1005} (\bibinfo {year} {2014})}\BibitemShut {NoStop}%
\bibitem [{\citenamefont {Zou}, \citenamefont {Filatov},\ and\ \citenamefont
  {Cremer}(2012)}]{Zou12}%
  \BibitemOpen
  \bibfield  {author} {\bibinfo {author} {\bibfnamefont {W.}~\bibnamefont
  {Zou}}, \bibinfo {author} {\bibfnamefont {M.}~\bibnamefont {Filatov}}, \ and\
  \bibinfo {author} {\bibfnamefont {D.}~\bibnamefont {Cremer}},\ }\bibfield
  {title} {\enquote {\bibinfo {title} {{Development, Implementation, and
  Application of an Analytic Second Derivative Formalism for the Normalized
  Elimination of the Small Component Method}},}\ }\href {\doibase
  10.1021/ct300127e} {\bibfield  {journal} {\bibinfo  {journal} {J. Chem.
  Theory Comput.}\ }\textbf {\bibinfo {volume} {8}},\ \bibinfo {pages}
  {2617--2629} (\bibinfo {year} {2012})}\BibitemShut {NoStop}%
\bibitem [{\citenamefont {Cheng}\ and\ \citenamefont
  {Gauss}(2011{\natexlab{a}})}]{Cheng11b}%
  \BibitemOpen
  \bibfield  {author} {\bibinfo {author} {\bibfnamefont {L.}~\bibnamefont
  {Cheng}}\ and\ \bibinfo {author} {\bibfnamefont {J.}~\bibnamefont {Gauss}},\
  }\bibfield  {title} {\enquote {\bibinfo {title} {Analytic energy gradients
  for the spin-free exact two-component theory using an exact block
  diagonalization for the one-electron {D}irac {H}amiltonian},}\ }\href@noop {}
  {\bibfield  {journal} {\bibinfo  {journal} {J. Chem. Phys.}\ }\textbf
  {\bibinfo {volume} {135}},\ \bibinfo {pages} {084114} (\bibinfo {year}
  {2011}{\natexlab{a}})}\BibitemShut {NoStop}%
\bibitem [{\citenamefont {Cheng}\ and\ \citenamefont
  {Gauss}(2011{\natexlab{b}})}]{Cheng11c}%
  \BibitemOpen
  \bibfield  {author} {\bibinfo {author} {\bibfnamefont {L.}~\bibnamefont
  {Cheng}}\ and\ \bibinfo {author} {\bibfnamefont {J.}~\bibnamefont {Gauss}},\
  }\bibfield  {title} {\enquote {\bibinfo {title} {Analytic second derivatives
  for the spin-free exact two-component theory},}\ }\href@noop {} {\bibfield
  {journal} {\bibinfo  {journal} {J. Chem. Phys.}\ }\textbf {\bibinfo {volume}
  {135}},\ \bibinfo {pages} {244104} (\bibinfo {year}
  {2011}{\natexlab{b}})}\BibitemShut {NoStop}%
\bibitem [{\citenamefont {Cheng}, \citenamefont {Stopkowicz},\ and\
  \citenamefont {Gauss}(2014)}]{Cheng14}%
  \BibitemOpen
  \bibfield  {author} {\bibinfo {author} {\bibfnamefont {L.}~\bibnamefont
  {Cheng}}, \bibinfo {author} {\bibfnamefont {S.}~\bibnamefont {Stopkowicz}}, \
  and\ \bibinfo {author} {\bibfnamefont {J.}~\bibnamefont {Gauss}},\ }\bibfield
   {title} {\enquote {\bibinfo {title} {Analytic energy derivatives in
  relativistic quantum chemistry},}\ }\href@noop {} {\bibfield  {journal}
  {\bibinfo  {journal} {Int. J. Quantum Chem.}\ }\textbf {\bibinfo {volume}
  {114}},\ \bibinfo {pages} {1108--1127} (\bibinfo {year} {2014})}\BibitemShut
  {NoStop}%
\bibitem [{\citenamefont {Franzke}, \citenamefont {Middendorf},\ and\
  \citenamefont {Weigend}(2018)}]{Franzke18}%
  \BibitemOpen
  \bibfield  {author} {\bibinfo {author} {\bibfnamefont {Y.~J.}\ \bibnamefont
  {Franzke}}, \bibinfo {author} {\bibfnamefont {N.}~\bibnamefont {Middendorf}},
  \ and\ \bibinfo {author} {\bibfnamefont {F.}~\bibnamefont {Weigend}},\
  }\bibfield  {title} {\enquote {\bibinfo {title} {Efficient implementation of
  one-and two-component analytical energy gradients in exact two-component
  theory},}\ }\href@noop {} {\bibfield  {journal} {\bibinfo  {journal} {J.
  Chem. Phys.}\ }\textbf {\bibinfo {volume} {148}},\ \bibinfo {pages} {104110}
  (\bibinfo {year} {2018})}\BibitemShut {NoStop}%
\bibitem [{\citenamefont {Zou}\ \emph {et~al.}(2020)\citenamefont {Zou},
  \citenamefont {Guo}, \citenamefont {Suo},\ and\ \citenamefont {Liu}}]{Zou20}%
  \BibitemOpen
  \bibfield  {author} {\bibinfo {author} {\bibfnamefont {W.}~\bibnamefont
  {Zou}}, \bibinfo {author} {\bibfnamefont {G.}~\bibnamefont {Guo}}, \bibinfo
  {author} {\bibfnamefont {B.}~\bibnamefont {Suo}}, \ and\ \bibinfo {author}
  {\bibfnamefont {W.}~\bibnamefont {Liu}},\ }\bibfield  {title} {\enquote
  {\bibinfo {title} {{Analytic Energy Gradients and Hessians of Exact
  Two-Component Relativistic Methods: Efficient Implementation and Extensive
  Applications}},}\ }\href {\doibase 10.1021/acs.jctc.9b01120} {\bibfield
  {journal} {\bibinfo  {journal} {J. Chem. Theory Comput.}\ }\textbf {\bibinfo
  {volume} {16}},\ \bibinfo {pages} {1541--1554} (\bibinfo {year}
  {2020})}\BibitemShut {NoStop}%
\bibitem [{\citenamefont {van W{\"u}llen}\ and\ \citenamefont
  {Michauk}(2005)}]{vanWullen05}%
  \BibitemOpen
  \bibfield  {author} {\bibinfo {author} {\bibfnamefont {C.}~\bibnamefont {van
  W{\"u}llen}}\ and\ \bibinfo {author} {\bibfnamefont {C.}~\bibnamefont
  {Michauk}},\ }\bibfield  {title} {\enquote {\bibinfo {title} {Accurate and
  efficient treatment of two-electron contributions in quasirelativistic
  high-order douglas-kroll density-functional calculations},}\ }\href@noop {}
  {\bibfield  {journal} {\bibinfo  {journal} {J. Chem. Phys.}\ }\textbf
  {\bibinfo {volume} {123}},\ \bibinfo {pages} {204113} (\bibinfo {year}
  {2005})}\BibitemShut {NoStop}%
\bibitem [{\citenamefont {Neese}\ \emph {et~al.}(2005)\citenamefont {Neese},
  \citenamefont {Wolf}, \citenamefont {Fleig}, \citenamefont {Reiher},\ and\
  \citenamefont {Hess}}]{Neese05a}%
  \BibitemOpen
  \bibfield  {author} {\bibinfo {author} {\bibfnamefont {F.}~\bibnamefont
  {Neese}}, \bibinfo {author} {\bibfnamefont {A.}~\bibnamefont {Wolf}},
  \bibinfo {author} {\bibfnamefont {T.}~\bibnamefont {Fleig}}, \bibinfo
  {author} {\bibfnamefont {M.}~\bibnamefont {Reiher}}, \ and\ \bibinfo {author}
  {\bibfnamefont {B.~A.}\ \bibnamefont {Hess}},\ }\bibfield  {title} {\enquote
  {\bibinfo {title} {{Calculation of electric-field gradients based on
  higher-order generalized Douglas-Kroll transformations}},}\ }\href {\doibase
  10.1063/1.1904589} {\bibfield  {journal} {\bibinfo  {journal} {J. Chem.
  Phys.}\ }\textbf {\bibinfo {volume} {122}},\ \bibinfo {pages} {204107}
  (\bibinfo {year} {2005})}\BibitemShut {NoStop}%
\bibitem [{\citenamefont {Mastalerz}\ \emph {et~al.}(2007)\citenamefont
  {Mastalerz}, \citenamefont {Barone}, \citenamefont {Lindh},\ and\
  \citenamefont {Reiher}}]{Mastalerz07}%
  \BibitemOpen
  \bibfield  {author} {\bibinfo {author} {\bibfnamefont {R.}~\bibnamefont
  {Mastalerz}}, \bibinfo {author} {\bibfnamefont {G.}~\bibnamefont {Barone}},
  \bibinfo {author} {\bibfnamefont {R.}~\bibnamefont {Lindh}}, \ and\ \bibinfo
  {author} {\bibfnamefont {M.}~\bibnamefont {Reiher}},\ }\bibfield  {title}
  {\enquote {\bibinfo {title} {{Analytic high-order Douglas-Kroll-Hess electric
  field gradients}},}\ }\href {\doibase 10.1063/1.2761880} {\bibfield
  {journal} {\bibinfo  {journal} {J. Chem. Phys.}\ }\textbf {\bibinfo {volume}
  {127}},\ \bibinfo {pages} {74105} (\bibinfo {year} {2007})}\BibitemShut
  {NoStop}%
\bibitem [{\citenamefont {Blume}, \citenamefont {Watson},\ and\ \citenamefont
  {Peierls}(1962)}]{Blume62}%
  \BibitemOpen
  \bibfield  {author} {\bibinfo {author} {\bibfnamefont {M.}~\bibnamefont
  {Blume}}, \bibinfo {author} {\bibfnamefont {R.~E.}\ \bibnamefont {Watson}}, \
  and\ \bibinfo {author} {\bibfnamefont {R.~E.}\ \bibnamefont {Peierls}},\
  }\bibfield  {title} {\enquote {\bibinfo {title} {{Theory of spin-orbit
  coupling in atoms I. Derivation of the spin-orbit coupling constant}},}\
  }\href {\doibase 10.1098/rspa.1962.0207} {\bibfield  {journal} {\bibinfo
  {journal} {Proc. R. Soc. London. Ser. A. Math. Phys. Sci.}\ }\textbf
  {\bibinfo {volume} {270}},\ \bibinfo {pages} {127--143} (\bibinfo {year}
  {1962})}\BibitemShut {NoStop}%
\bibitem [{\citenamefont {Blume}\ and\ \citenamefont {Watson}(1963)}]{Blume63}%
  \BibitemOpen
  \bibfield  {author} {\bibinfo {author} {\bibfnamefont {M.}~\bibnamefont
  {Blume}}\ and\ \bibinfo {author} {\bibfnamefont {R.}~\bibnamefont {Watson}},\
  }\bibfield  {title} {\enquote {\bibinfo {title} {Theory of spin-orbit
  coupling in atoms, ii. comparison of theory with experiment},}\ }\href@noop
  {} {\bibfield  {journal} {\bibinfo  {journal} {Proc. R. Soc. A: Math. Phys.
  Eng. Sci.}\ }\textbf {\bibinfo {volume} {271}},\ \bibinfo {pages} {565--578}
  (\bibinfo {year} {1963})}\BibitemShut {NoStop}%
\bibitem [{\citenamefont {Malkina}\ \emph {et~al.}(1998)\citenamefont
  {Malkina}, \citenamefont {Schimmelpfennig}, \citenamefont {Kaupp},
  \citenamefont {Hess}, \citenamefont {Chandra}, \citenamefont {Wahlgren},\
  and\ \citenamefont {Malkin}}]{Malkina98}%
  \BibitemOpen
  \bibfield  {author} {\bibinfo {author} {\bibfnamefont {O.~L.}\ \bibnamefont
  {Malkina}}, \bibinfo {author} {\bibfnamefont {B.}~\bibnamefont
  {Schimmelpfennig}}, \bibinfo {author} {\bibfnamefont {M.}~\bibnamefont
  {Kaupp}}, \bibinfo {author} {\bibfnamefont {B.~A.}\ \bibnamefont {Hess}},
  \bibinfo {author} {\bibfnamefont {P.}~\bibnamefont {Chandra}}, \bibinfo
  {author} {\bibfnamefont {U.}~\bibnamefont {Wahlgren}}, \ and\ \bibinfo
  {author} {\bibfnamefont {V.~G.}\ \bibnamefont {Malkin}},\ }\bibfield  {title}
  {\enquote {\bibinfo {title} {{Spin–orbit corrections to NMR shielding
  constants from density functional theory. How important are the two-electron
  terms?}}}\ }\href {\doibase https://doi.org/10.1016/S0009-2614(98)00998-1}
  {\bibfield  {journal} {\bibinfo  {journal} {Chem. Phys. Lett.}\ }\textbf
  {\bibinfo {volume} {296}},\ \bibinfo {pages} {93--104} (\bibinfo {year}
  {1998})}\BibitemShut {NoStop}%
\bibitem [{\citenamefont {Fedorov}\ and\ \citenamefont
  {Gordon}(2000)}]{Fedorov00}%
  \BibitemOpen
  \bibfield  {author} {\bibinfo {author} {\bibfnamefont {D.~G.}\ \bibnamefont
  {Fedorov}}\ and\ \bibinfo {author} {\bibfnamefont {M.~S.}\ \bibnamefont
  {Gordon}},\ }\bibfield  {title} {\enquote {\bibinfo {title} {{A study of the
  relative importance of one and two-electron contributions to spin--orbit
  coupling}},}\ }\href {\doibase 10.1063/1.481136} {\bibfield  {journal}
  {\bibinfo  {journal} {J. Chem. Phys.}\ }\textbf {\bibinfo {volume} {112}},\
  \bibinfo {pages} {5611--5623} (\bibinfo {year} {2000})}\BibitemShut {NoStop}%
\bibitem [{\citenamefont {Boettger}(2000)}]{Boettger00}%
  \BibitemOpen
  \bibfield  {author} {\bibinfo {author} {\bibfnamefont {J.}~\bibnamefont
  {Boettger}},\ }\bibfield  {title} {\enquote {\bibinfo {title} {Approximate
  two-electron spin-orbit coupling term for density-functional-theory dft
  calculations using the douglas-kroll-hess transformation},}\ }\href@noop {}
  {\bibfield  {journal} {\bibinfo  {journal} {Phys. Rev. B}\ }\textbf {\bibinfo
  {volume} {62}},\ \bibinfo {pages} {7809} (\bibinfo {year}
  {2000})}\BibitemShut {NoStop}%
\bibitem [{\citenamefont {Stopkowicz}\ and\ \citenamefont
  {Gauss}(2011)}]{Stopkowicz11b}%
  \BibitemOpen
  \bibfield  {author} {\bibinfo {author} {\bibfnamefont {S.}~\bibnamefont
  {Stopkowicz}}\ and\ \bibinfo {author} {\bibfnamefont {J.}~\bibnamefont
  {Gauss}},\ }\bibfield  {title} {\enquote {\bibinfo {title} {Fourth-order
  relativistic corrections to electrical properties using {D}irect
  {P}erturbation {T}heory},}\ }\href@noop {} {\bibfield  {journal} {\bibinfo
  {journal} {J. Chem. Phys.}\ }\textbf {\bibinfo {volume} {134}},\ \bibinfo
  {pages} {204106} (\bibinfo {year} {2011})}\BibitemShut {NoStop}%
\bibitem [{\citenamefont {Liu}\ and\ \citenamefont {Peng}(2006)}]{Liu06}%
  \BibitemOpen
  \bibfield  {author} {\bibinfo {author} {\bibfnamefont {W.}~\bibnamefont
  {Liu}}\ and\ \bibinfo {author} {\bibfnamefont {D.}~\bibnamefont {Peng}},\
  }\bibfield  {title} {\enquote {\bibinfo {title} {Infinite-order
  quasirelativistic density functional method based on the exact matrix
  quasirelativistic theory},}\ }\href@noop {} {\bibfield  {journal} {\bibinfo
  {journal} {J. Chem. Phys.}\ }\textbf {\bibinfo {volume} {125}},\ \bibinfo
  {pages} {044102} (\bibinfo {year} {2006})}\BibitemShut {NoStop}%
\bibitem [{\citenamefont {Sikkema}\ \emph {et~al.}(2009)\citenamefont
  {Sikkema}, \citenamefont {Visscher}, \citenamefont {Saue},\ and\
  \citenamefont {Ilia\v{s}}}]{Sikkema09}%
  \BibitemOpen
  \bibfield  {author} {\bibinfo {author} {\bibfnamefont {J.}~\bibnamefont
  {Sikkema}}, \bibinfo {author} {\bibfnamefont {L.}~\bibnamefont {Visscher}},
  \bibinfo {author} {\bibfnamefont {T.}~\bibnamefont {Saue}}, \ and\ \bibinfo
  {author} {\bibfnamefont {M.}~\bibnamefont {Ilia\v{s}}},\ }\bibfield  {title}
  {\enquote {\bibinfo {title} {The molecular mean-field approach for correlated
  relativistic calculations},}\ }\href@noop {} {\bibfield  {journal} {\bibinfo
  {journal} {J. Chem. Phys.}\ }\textbf {\bibinfo {volume} {131}},\ \bibinfo
  {pages} {124116} (\bibinfo {year} {2009})}\BibitemShut {NoStop}%
\bibitem [{\citenamefont {Kirsch}, \citenamefont {Engel},\ and\ \citenamefont
  {Gauss}(2019)}]{Kirsch19}%
  \BibitemOpen
  \bibfield  {author} {\bibinfo {author} {\bibfnamefont {T.}~\bibnamefont
  {Kirsch}}, \bibinfo {author} {\bibfnamefont {F.}~\bibnamefont {Engel}}, \
  and\ \bibinfo {author} {\bibfnamefont {J.}~\bibnamefont {Gauss}},\ }\bibfield
   {title} {\enquote {\bibinfo {title} {{Analytic evaluation of first-order
  properties within the mean-field variant of spin-free exact two-component
  theory}},}\ }\href@noop {} {\bibfield  {journal} {\bibinfo  {journal} {J.
  Chem. Phys.}\ }\textbf {\bibinfo {volume} {150}},\ \bibinfo {pages} {204115}
  (\bibinfo {year} {2019})}\BibitemShut {NoStop}%
\bibitem [{\citenamefont {de~Jong}\ and\ \citenamefont
  {Visscher}(2002)}]{deJong02}%
  \BibitemOpen
  \bibfield  {author} {\bibinfo {author} {\bibfnamefont {G.~T.}\ \bibnamefont
  {de~Jong}}\ and\ \bibinfo {author} {\bibfnamefont {L.}~\bibnamefont
  {Visscher}},\ }\bibfield  {title} {\enquote {\bibinfo {title} {Using the
  locality of the small-component density in molecular dirac-hartree-fock
  calculations},}\ }\href@noop {} {\bibfield  {journal} {\bibinfo  {journal}
  {Theor. Chem. Acc.}\ }\textbf {\bibinfo {volume} {107}},\ \bibinfo {pages}
  {304} (\bibinfo {year} {2002})}\BibitemShut {NoStop}%
\bibitem [{\citenamefont {Peng}\ \emph {et~al.}(2007)\citenamefont {Peng},
  \citenamefont {Liu}, \citenamefont {Xiao},\ and\ \citenamefont
  {Cheng}}]{Peng07}%
  \BibitemOpen
  \bibfield  {author} {\bibinfo {author} {\bibfnamefont {D.}~\bibnamefont
  {Peng}}, \bibinfo {author} {\bibfnamefont {W.}~\bibnamefont {Liu}}, \bibinfo
  {author} {\bibfnamefont {Y.}~\bibnamefont {Xiao}}, \ and\ \bibinfo {author}
  {\bibfnamefont {L.}~\bibnamefont {Cheng}},\ }\bibfield  {title} {\enquote
  {\bibinfo {title} {Making four-and two-component relativistic density
  functional methods fully equivalent based on the idea of ``from atoms to
  molecule''},}\ }\href@noop {} {\bibfield  {journal} {\bibinfo  {journal} {J.
  Chem. Phys.}\ }\textbf {\bibinfo {volume} {127}},\ \bibinfo {pages} {104106}
  (\bibinfo {year} {2007})}\BibitemShut {NoStop}%
\bibitem [{\citenamefont {Knecht}\ \emph {et~al.}(2022)\citenamefont {Knecht},
  \citenamefont {Repisky}, \citenamefont {Jensen},\ and\ \citenamefont
  {Saue}}]{Knecht22}%
  \BibitemOpen
  \bibfield  {author} {\bibinfo {author} {\bibfnamefont {S.}~\bibnamefont
  {Knecht}}, \bibinfo {author} {\bibfnamefont {M.}~\bibnamefont {Repisky}},
  \bibinfo {author} {\bibfnamefont {H.~J.~{\relax Aa}.}\ \bibnamefont
  {Jensen}}, \ and\ \bibinfo {author} {\bibfnamefont {T.}~\bibnamefont
  {Saue}},\ }\bibfield  {title} {\enquote {\bibinfo {title} {{Exact
  two-component Hamiltonians for relativistic quantum chemistry: Two-electron
  picture-change corrections made simple}},}\ }\href {\doibase
  10.1063/5.0095112} {\bibfield  {journal} {\bibinfo  {journal} {J. Chem.
  Phys.}\ }\textbf {\bibinfo {volume} {157}},\ \bibinfo {pages} {114106}
  (\bibinfo {year} {2022})}\BibitemShut {NoStop}%
\bibitem [{\citenamefont {Zhao}\ \emph {et~al.}(2021)\citenamefont {Zhao},
  \citenamefont {Yu}, \citenamefont {Zhang}, \citenamefont {Xiao},
  \citenamefont {Zhang},\ and\ \citenamefont {Blum}}]{Zhao21}%
  \BibitemOpen
  \bibfield  {author} {\bibinfo {author} {\bibfnamefont {R.}~\bibnamefont
  {Zhao}}, \bibinfo {author} {\bibfnamefont {V.~W.-z.}\ \bibnamefont {Yu}},
  \bibinfo {author} {\bibfnamefont {K.}~\bibnamefont {Zhang}}, \bibinfo
  {author} {\bibfnamefont {Y.}~\bibnamefont {Xiao}}, \bibinfo {author}
  {\bibfnamefont {Y.}~\bibnamefont {Zhang}}, \ and\ \bibinfo {author}
  {\bibfnamefont {V.}~\bibnamefont {Blum}},\ }\bibfield  {title} {\enquote
  {\bibinfo {title} {{Quasi-four-component method with numeric atom-centered
  orbitals for relativistic density functional simulations of molecules and
  solids}},}\ }\href {\doibase 10.1103/PhysRevB.103.245144} {\bibfield
  {journal} {\bibinfo  {journal} {Phys. Rev. B}\ }\textbf {\bibinfo {volume}
  {103}},\ \bibinfo {pages} {245144} (\bibinfo {year} {2021})}\BibitemShut
  {NoStop}%
\bibitem [{\citenamefont {Samzow}, \citenamefont {Hess},\ and\ \citenamefont
  {Jansen}(1992)}]{Samzow92}%
  \BibitemOpen
  \bibfield  {author} {\bibinfo {author} {\bibfnamefont {R.}~\bibnamefont
  {Samzow}}, \bibinfo {author} {\bibfnamefont {B.~A.}\ \bibnamefont {Hess}}, \
  and\ \bibinfo {author} {\bibfnamefont {G.}~\bibnamefont {Jansen}},\
  }\bibfield  {title} {\enquote {\bibinfo {title} {{The two‐electron terms of
  the no‐pair Hamiltonian}},}\ }\href {\doibase 10.1063/1.462210} {\bibfield
  {journal} {\bibinfo  {journal} {J. Chem. Phys.}\ }\textbf {\bibinfo {volume}
  {96}},\ \bibinfo {pages} {1227--1231} (\bibinfo {year} {1992})}\BibitemShut
  {NoStop}%
\bibitem [{\citenamefont {Ikabata}\ and\ \citenamefont
  {Nakai}(2021)}]{Ikabata21}%
  \BibitemOpen
  \bibfield  {author} {\bibinfo {author} {\bibfnamefont {Y.}~\bibnamefont
  {Ikabata}}\ and\ \bibinfo {author} {\bibfnamefont {H.}~\bibnamefont
  {Nakai}},\ }\bibfield  {title} {\enquote {\bibinfo {title} {{Picture-change
  correction in relativistic density functional theory}},}\ }\href {\doibase
  10.1039/D1CP01773J} {\bibfield  {journal} {\bibinfo  {journal} {Phys. Chem.
  Chem. Phys.}\ }\textbf {\bibinfo {volume} {23}},\ \bibinfo {pages}
  {15458--15474} (\bibinfo {year} {2021})}\BibitemShut {NoStop}%
\bibitem [{\citenamefont {He{\ss}}\ \emph {et~al.}(1996)\citenamefont
  {He{\ss}}, \citenamefont {Marian}, \citenamefont {Wahlgren},\ and\
  \citenamefont {Gropen}}]{Hess96a}%
  \BibitemOpen
  \bibfield  {author} {\bibinfo {author} {\bibfnamefont {B.~A.}\ \bibnamefont
  {He{\ss}}}, \bibinfo {author} {\bibfnamefont {C.~M.}\ \bibnamefont {Marian}},
  \bibinfo {author} {\bibfnamefont {U.}~\bibnamefont {Wahlgren}}, \ and\
  \bibinfo {author} {\bibfnamefont {O.}~\bibnamefont {Gropen}},\ }\bibfield
  {title} {\enquote {\bibinfo {title} {{A mean-field spin-orbit method
  applicable to correlated wavefunctions}},}\ }\href@noop {} {\bibfield
  {journal} {\bibinfo  {journal} {Chem. Phys. Lett.}\ }\textbf {\bibinfo
  {volume} {251}},\ \bibinfo {pages} {365--371} (\bibinfo {year}
  {1996})}\BibitemShut {NoStop}%
\bibitem [{\citenamefont {Wahlgren}\ \emph {et~al.}(1997)\citenamefont
  {Wahlgren}, \citenamefont {Sj{\o}voll}, \citenamefont {Fagerli},
  \citenamefont {Gropen},\ and\ \citenamefont {Schimmelpfennig}}]{Wahlgren97}%
  \BibitemOpen
  \bibfield  {author} {\bibinfo {author} {\bibfnamefont {U.}~\bibnamefont
  {Wahlgren}}, \bibinfo {author} {\bibfnamefont {M.}~\bibnamefont
  {Sj{\o}voll}}, \bibinfo {author} {\bibfnamefont {H.}~\bibnamefont {Fagerli}},
  \bibinfo {author} {\bibfnamefont {O.}~\bibnamefont {Gropen}}, \ and\ \bibinfo
  {author} {\bibfnamefont {B.}~\bibnamefont {Schimmelpfennig}},\ }\bibfield
  {title} {\enquote {\bibinfo {title} {{Ab initio calculations of the
  $^2P_{\frac{1}{2}}$-$^2P_{\frac{3}{2}}$ splitting in the thallium atom}},}\
  }\href {\doibase 10.1007/s002140050268} {\bibfield  {journal} {\bibinfo
  {journal} {Theor. Chem. Acc.}\ }\textbf {\bibinfo {volume} {97}},\ \bibinfo
  {pages} {324--330} (\bibinfo {year} {1997})}\BibitemShut {NoStop}%
\bibitem [{\citenamefont {Ilia{\v{s}}}\ \emph {et~al.}(2001)\citenamefont
  {Ilia{\v{s}}}, \citenamefont {Kell{\"{o}}}, \citenamefont {Visscher},\ and\
  \citenamefont {Schimmelpfennig}}]{Ilias01}%
  \BibitemOpen
  \bibfield  {author} {\bibinfo {author} {\bibfnamefont {M.}~\bibnamefont
  {Ilia{\v{s}}}}, \bibinfo {author} {\bibfnamefont {V.}~\bibnamefont
  {Kell{\"{o}}}}, \bibinfo {author} {\bibfnamefont {L.}~\bibnamefont
  {Visscher}}, \ and\ \bibinfo {author} {\bibfnamefont {B.}~\bibnamefont
  {Schimmelpfennig}},\ }\bibfield  {title} {\enquote {\bibinfo {title}
  {{Inclusion of mean-field spin--orbit effects based on all-electron
  two-component spinors: Pilot calculations on atomic and molecular
  properties}},}\ }\href {\doibase 10.1063/1.1413510} {\bibfield  {journal}
  {\bibinfo  {journal} {J. Chem. Phys.}\ }\textbf {\bibinfo {volume} {115}},\
  \bibinfo {pages} {9667--9674} (\bibinfo {year} {2001})}\BibitemShut {NoStop}%
\bibitem [{\citenamefont {Wody{\'{n}}ski}\ and\ \citenamefont
  {Kaupp}(2019)}]{Wodynski19}%
  \BibitemOpen
  \bibfield  {author} {\bibinfo {author} {\bibfnamefont {A.}~\bibnamefont
  {Wody{\'{n}}ski}}\ and\ \bibinfo {author} {\bibfnamefont {M.}~\bibnamefont
  {Kaupp}},\ }\bibfield  {title} {\enquote {\bibinfo {title} {{Density
  Functional Calculations of Electron Paramagnetic Resonance g- and
  Hyperfine-Coupling Tensors Using the Exact Two-Component (X2C) Transformation
  and Efficient Approximations to the Two-Electron Spin–Orbit Terms}},}\
  }\href {\doibase 10.1021/acs.jpca.9b03979} {\bibfield  {journal} {\bibinfo
  {journal} {J. Phys. Chem. A}\ }\textbf {\bibinfo {volume} {123}},\ \bibinfo
  {pages} {5660--5672} (\bibinfo {year} {2019})}\BibitemShut {NoStop}%
\bibitem [{\citenamefont {Zhang}\ and\ \citenamefont {Cheng}(2020)}]{Zhang20}%
  \BibitemOpen
  \bibfield  {author} {\bibinfo {author} {\bibfnamefont {C.}~\bibnamefont
  {Zhang}}\ and\ \bibinfo {author} {\bibfnamefont {L.}~\bibnamefont {Cheng}},\
  }\bibfield  {title} {\enquote {\bibinfo {title} {Performance of an atomic
  mean-field spin--orbit approach within exact two-component theory for
  perturbative treatment of spin--orbit coupling},}\ }\href@noop {} {\bibfield
  {journal} {\bibinfo  {journal} {Mol. Phys.}\ }\textbf {\bibinfo {volume}
  {118}},\ \bibinfo {pages} {e1768313} (\bibinfo {year} {2020})}\BibitemShut
  {NoStop}%
\bibitem [{\citenamefont {Lin}, \citenamefont {Zhang},\ and\ \citenamefont
  {Cheng}(2023)}]{Lin23}%
  \BibitemOpen
  \bibfield  {author} {\bibinfo {author} {\bibfnamefont {Z.}~\bibnamefont
  {Lin}}, \bibinfo {author} {\bibfnamefont {C.}~\bibnamefont {Zhang}}, \ and\
  \bibinfo {author} {\bibfnamefont {L.}~\bibnamefont {Cheng}},\ }\bibfield
  {title} {\enquote {\bibinfo {title} {{Comparison of state-interaction and
  spinor-representation calculations of spin-orbit coupling within exact
  two-component coupled-cluster theories}},}\ }\href {\doibase
  10.1080/00268976.2023.2256423} {\bibfield  {journal} {\bibinfo  {journal}
  {Mol. Phys.}\ ,\ \bibinfo {pages} {e2256423}} (\bibinfo {year}
  {2023})}\BibitemShut {NoStop}%
\bibitem [{\citenamefont {Liu}\ and\ \citenamefont {Cheng}(2018)}]{Liu18}%
  \BibitemOpen
  \bibfield  {author} {\bibinfo {author} {\bibfnamefont {J.}~\bibnamefont
  {Liu}}\ and\ \bibinfo {author} {\bibfnamefont {L.}~\bibnamefont {Cheng}},\
  }\bibfield  {title} {\enquote {\bibinfo {title} {{An atomic mean-field
  spin-orbit approach within exact two-component theory for a non-perturbative
  treatment of spin-orbit coupling}},}\ }\href@noop {} {\bibfield  {journal}
  {\bibinfo  {journal} {J. Chem. Phys.}\ }\textbf {\bibinfo {volume} {148}},\
  \bibinfo {pages} {144108} (\bibinfo {year} {2018})}\BibitemShut {NoStop}%
\bibitem [{\citenamefont {Zhang}\ and\ \citenamefont
  {Cheng}(2022{\natexlab{a}})}]{Zhang22}%
  \BibitemOpen
  \bibfield  {author} {\bibinfo {author} {\bibfnamefont {C.}~\bibnamefont
  {Zhang}}\ and\ \bibinfo {author} {\bibfnamefont {L.}~\bibnamefont {Cheng}},\
  }\bibfield  {title} {\enquote {\bibinfo {title} {{Atomic Mean-Field Approach
  within Exact Two-Component Theory Based on the Dirac--Coulomb--Breit
  Hamiltonian}},}\ }\href {\doibase 10.1021/acs.jpca.2c02181} {\bibfield
  {journal} {\bibinfo  {journal} {J. Phys. Chem. A}\ }\textbf {\bibinfo
  {volume} {126}},\ \bibinfo {pages} {4537--4553} (\bibinfo {year}
  {2022}{\natexlab{a}})}\BibitemShut {NoStop}%
\bibitem [{\citenamefont {Liu}(2024)}]{Liu24}%
  \BibitemOpen
  \bibfield  {author} {\bibinfo {author} {\bibfnamefont {W.}~\bibnamefont
  {Liu}},\ }\bibfield  {title} {\enquote {\bibinfo {title} {{Unified
  construction of relativistic Hamiltonians}},}\ }\href {\doibase
  10.1063/5.0188794} {\bibfield  {journal} {\bibinfo  {journal} {J. Chem.
  Phys.}\ }\textbf {\bibinfo {volume} {160}},\ \bibinfo {pages} {84111}
  (\bibinfo {year} {2024})}\BibitemShut {NoStop}%
\bibitem [{\citenamefont {Filatov}, \citenamefont {Zou},\ and\ \citenamefont
  {Cremer}(2013)}]{Filatov13}%
  \BibitemOpen
  \bibfield  {author} {\bibinfo {author} {\bibfnamefont {M.}~\bibnamefont
  {Filatov}}, \bibinfo {author} {\bibfnamefont {W.}~\bibnamefont {Zou}}, \ and\
  \bibinfo {author} {\bibfnamefont {D.}~\bibnamefont {Cremer}},\ }\bibfield
  {title} {\enquote {\bibinfo {title} {Spin-orbit coupling calculations with
  the two-component normalized elimination of the small component method},}\
  }\href@noop {} {\bibfield  {journal} {\bibinfo  {journal} {J. Chem. Phys.}\
  }\textbf {\bibinfo {volume} {139}},\ \bibinfo {pages} {014106} (\bibinfo
  {year} {2013})}\BibitemShut {NoStop}%
\bibitem [{\citenamefont {Ehrman}\ \emph {et~al.}(2023)\citenamefont {Ehrman},
  \citenamefont {Martinez-Baez}, \citenamefont {Jenkins},\ and\ \citenamefont
  {Li}}]{Ehrman23}%
  \BibitemOpen
  \bibfield  {author} {\bibinfo {author} {\bibfnamefont {J.}~\bibnamefont
  {Ehrman}}, \bibinfo {author} {\bibfnamefont {E.}~\bibnamefont
  {Martinez-Baez}}, \bibinfo {author} {\bibfnamefont {A.~J.}\ \bibnamefont
  {Jenkins}}, \ and\ \bibinfo {author} {\bibfnamefont {X.}~\bibnamefont {Li}},\
  }\bibfield  {title} {\enquote {\bibinfo {title} {{Improving One-Electron
  Exact-Two-Component Relativistic Methods with the
  Dirac–Coulomb–Breit-Parameterized Effective Spin–Orbit Coupling}},}\
  }\href {\doibase 10.1021/acs.jctc.3c00479} {\bibfield  {journal} {\bibinfo
  {journal} {J. Chem. Theory Comput.}\ }\textbf {\bibinfo {volume} {19}},\
  \bibinfo {pages} {5785--5790} (\bibinfo {year} {2023})}\BibitemShut {NoStop}%
\bibitem [{\citenamefont {Foldy}\ and\ \citenamefont
  {Wouthuysen}(1950)}]{Foldy50}%
  \BibitemOpen
  \bibfield  {author} {\bibinfo {author} {\bibfnamefont {L.~L.}\ \bibnamefont
  {Foldy}}\ and\ \bibinfo {author} {\bibfnamefont {S.~A.}\ \bibnamefont
  {Wouthuysen}},\ }\bibfield  {title} {\enquote {\bibinfo {title} {{On the
  Dirac Theory of Spin 1/2 Particles and Its Non-Relativistic Limit}},}\ }\href
  {\doibase 10.1103/PhysRev.78.29} {\bibfield  {journal} {\bibinfo  {journal}
  {Phys. Rev.}\ }\textbf {\bibinfo {volume} {78}},\ \bibinfo {pages} {29--36}
  (\bibinfo {year} {1950})}\BibitemShut {NoStop}%
\bibitem [{\citenamefont {Kutzelnigg}(1997)}]{Kutzelnigg97a}%
  \BibitemOpen
  \bibfield  {author} {\bibinfo {author} {\bibfnamefont {W.}~\bibnamefont
  {Kutzelnigg}},\ }\bibfield  {title} {\enquote {\bibinfo {title}
  {{Relativistic one-electron Hamiltonians `for electrons only' and the
  variational treatment of the Dirac equation}},}\ }\href {\doibase
  https://doi.org/10.1016/S0301-0104(97)00240-1} {\bibfield  {journal}
  {\bibinfo  {journal} {Chem. Phys.}\ }\textbf {\bibinfo {volume} {225}},\
  \bibinfo {pages} {203--222} (\bibinfo {year} {1997})}\BibitemShut {NoStop}%
\bibitem [{\citenamefont {Filatov}\ and\ \citenamefont
  {Cremer}(2002)}]{Filatov02}%
  \BibitemOpen
  \bibfield  {author} {\bibinfo {author} {\bibfnamefont {M.}~\bibnamefont
  {Filatov}}\ and\ \bibinfo {author} {\bibfnamefont {D.}~\bibnamefont
  {Cremer}},\ }\bibfield  {title} {\enquote {\bibinfo {title} {{A new
  quasi-relativistic approach for density functional theory based on the
  normalized elimination of the small component}},}\ }\href {\doibase
  https://doi.org/10.1016/S0009-2614(01)01357-4} {\bibfield  {journal}
  {\bibinfo  {journal} {Chem. Phys. Lett.}\ }\textbf {\bibinfo {volume}
  {351}},\ \bibinfo {pages} {259--266} (\bibinfo {year} {2002})}\BibitemShut
  {NoStop}%
\bibitem [{\citenamefont {Nakajima}\ and\ \citenamefont
  {Hirao}(2000)}]{Nakajima00}%
  \BibitemOpen
  \bibfield  {author} {\bibinfo {author} {\bibfnamefont {T.}~\bibnamefont
  {Nakajima}}\ and\ \bibinfo {author} {\bibfnamefont {K.}~\bibnamefont
  {Hirao}},\ }\bibfield  {title} {\enquote {\bibinfo {title} {{The higher-order
  Douglas–Kroll transformation}},}\ }\href {\doibase 10.1063/1.1316037}
  {\bibfield  {journal} {\bibinfo  {journal} {J. Chem. Phys.}\ }\textbf
  {\bibinfo {volume} {113}},\ \bibinfo {pages} {7786--7789} (\bibinfo {year}
  {2000})}\BibitemShut {NoStop}%
\bibitem [{\citenamefont {Reiher}\ and\ \citenamefont
  {Wolf}(2004{\natexlab{a}})}]{Reiher04a}%
  \BibitemOpen
  \bibfield  {author} {\bibinfo {author} {\bibfnamefont {M.}~\bibnamefont
  {Reiher}}\ and\ \bibinfo {author} {\bibfnamefont {A.}~\bibnamefont {Wolf}},\
  }\bibfield  {title} {\enquote {\bibinfo {title} {{Exact decoupling of the
  Dirac Hamiltonian. I. General theory}},}\ }\href {\doibase 10.1063/1.1768160}
  {\bibfield  {journal} {\bibinfo  {journal} {J. Chem. Phys.}\ }\textbf
  {\bibinfo {volume} {121}},\ \bibinfo {pages} {2037--2047} (\bibinfo {year}
  {2004}{\natexlab{a}})}\BibitemShut {NoStop}%
\bibitem [{\citenamefont {Reiher}\ and\ \citenamefont
  {Wolf}(2004{\natexlab{b}})}]{Reiher04}%
  \BibitemOpen
  \bibfield  {author} {\bibinfo {author} {\bibfnamefont {M.}~\bibnamefont
  {Reiher}}\ and\ \bibinfo {author} {\bibfnamefont {A.}~\bibnamefont {Wolf}},\
  }\bibfield  {title} {\enquote {\bibinfo {title} {{Exact decoupling of the
  Dirac Hamiltonian. II. The generalized Douglas–Kroll–Hess transformation
  up to arbitrary order}},}\ }\href {\doibase 10.1063/1.1818681} {\bibfield
  {journal} {\bibinfo  {journal} {J. Chem. Phys.}\ }\textbf {\bibinfo {volume}
  {121}},\ \bibinfo {pages} {10945--10956} (\bibinfo {year}
  {2004}{\natexlab{b}})}\BibitemShut {NoStop}%
\bibitem [{\citenamefont {Peng}\ and\ \citenamefont {Hirao}(2009)}]{Peng09}%
  \BibitemOpen
  \bibfield  {author} {\bibinfo {author} {\bibfnamefont {D.}~\bibnamefont
  {Peng}}\ and\ \bibinfo {author} {\bibfnamefont {K.}~\bibnamefont {Hirao}},\
  }\bibfield  {title} {\enquote {\bibinfo {title} {{An arbitrary order
  Douglas–Kroll method with polynomial cost}},}\ }\href {\doibase
  10.1063/1.3068310} {\bibfield  {journal} {\bibinfo  {journal} {J. Chem.
  Phys.}\ }\textbf {\bibinfo {volume} {130}},\ \bibinfo {pages} {44102}
  (\bibinfo {year} {2009})}\BibitemShut {NoStop}%
\bibitem [{\citenamefont {van Lenthe}, \citenamefont {Baerends},\ and\
  \citenamefont {Snijders}(1993{\natexlab{b}})}]{vanLenth93}%
  \BibitemOpen
  \bibfield  {author} {\bibinfo {author} {\bibfnamefont {E.}~\bibnamefont {van
  Lenthe}}, \bibinfo {author} {\bibfnamefont {E.~J.}\ \bibnamefont {Baerends}},
  \ and\ \bibinfo {author} {\bibfnamefont {J.~G.}\ \bibnamefont {Snijders}},\
  }\bibfield  {title} {\enquote {\bibinfo {title} {{Relativistic regular
  two-component Hamiltonians}},}\ }\href {\doibase
  http://dx.doi.org/10.1063/1.466059} {\bibfield  {journal} {\bibinfo
  {journal} {The Journal of Chemical Physics}\ }\textbf {\bibinfo {volume}
  {99}},\ \bibinfo {pages} {4597--4610} (\bibinfo {year}
  {1993}{\natexlab{b}})}\BibitemShut {NoStop}%
\bibitem [{\citenamefont {van Lenthe}, \citenamefont {Baerends},\ and\
  \citenamefont {Snijders}(1994)}]{vanLenthe94}%
  \BibitemOpen
  \bibfield  {author} {\bibinfo {author} {\bibfnamefont {E.}~\bibnamefont {van
  Lenthe}}, \bibinfo {author} {\bibfnamefont {E.~J.}\ \bibnamefont {Baerends}},
  \ and\ \bibinfo {author} {\bibfnamefont {J.~G.}\ \bibnamefont {Snijders}},\
  }\bibfield  {title} {\enquote {\bibinfo {title} {{Relativistic total energy
  using regular approximations}},}\ }\href {\doibase
  http://dx.doi.org/10.1063/1.467943} {\bibfield  {journal} {\bibinfo
  {journal} {The Journal of Chemical Physics}\ }\textbf {\bibinfo {volume}
  {101}},\ \bibinfo {pages} {9783--9792} (\bibinfo {year} {1994})}\BibitemShut
  {NoStop}%
\bibitem [{\citenamefont {Dyall}\ and\ \citenamefont {van
  Lenthe}(1999)}]{Dyall99}%
  \BibitemOpen
  \bibfield  {author} {\bibinfo {author} {\bibfnamefont {K.~G.}\ \bibnamefont
  {Dyall}}\ and\ \bibinfo {author} {\bibfnamefont {E.}~\bibnamefont {van
  Lenthe}},\ }\bibfield  {title} {\enquote {\bibinfo {title} {{Relativistic
  regular approximations revisited: An infinite-order relativistic
  approximation}},}\ }\href {\doibase 10.1063/1.479395} {\bibfield  {journal}
  {\bibinfo  {journal} {J. Chem. Phys.}\ }\textbf {\bibinfo {volume} {111}},\
  \bibinfo {pages} {1366--1372} (\bibinfo {year} {1999})}\BibitemShut {NoStop}%
\bibitem [{\citenamefont {Filatov}\ and\ \citenamefont
  {Cremer}(2003)}]{Filatov03}%
  \BibitemOpen
  \bibfield  {author} {\bibinfo {author} {\bibfnamefont {M.}~\bibnamefont
  {Filatov}}\ and\ \bibinfo {author} {\bibfnamefont {D.}~\bibnamefont
  {Cremer}},\ }\bibfield  {title} {\enquote {\bibinfo {title} {{Representation
  of the exact relativistic electronic Hamiltonian within the regular
  approximation}},}\ }\href {\doibase 10.1063/1.1623473} {\bibfield  {journal}
  {\bibinfo  {journal} {J. Chem. Phys.}\ }\textbf {\bibinfo {volume} {119}},\
  \bibinfo {pages} {11526--11540} (\bibinfo {year} {2003})}\BibitemShut
  {NoStop}%
\bibitem [{\citenamefont {Apr{\`{a}}}\ \emph {et~al.}(2020)\citenamefont
  {Apr{\`{a}}}, \citenamefont {Bylaska}, \citenamefont {de~Jong}, \citenamefont
  {Govind}, \citenamefont {Kowalski}, \citenamefont {Straatsma}, \citenamefont
  {Valiev}, \citenamefont {van Dam}, \citenamefont {Alexeev}, \citenamefont
  {Anchell}, \citenamefont {Anisimov}, \citenamefont {Aquino}, \citenamefont
  {Atta-Fynn}, \citenamefont {Autschbach}, \citenamefont {Bauman},
  \citenamefont {Becca}, \citenamefont {Bernholdt}, \citenamefont
  {Bhaskaran-Nair}, \citenamefont {Bogatko}, \citenamefont {Borowski},
  \citenamefont {Boschen}, \citenamefont {Brabec}, \citenamefont {Bruner},
  \citenamefont {Cau{\"{e}}t}, \citenamefont {Chen}, \citenamefont {Chuev},
  \citenamefont {Cramer}, \citenamefont {Daily}, \citenamefont {Deegan},
  \citenamefont {{Dunning Jr.}}, \citenamefont {Dupuis}, \citenamefont {Dyall},
  \citenamefont {Fann}, \citenamefont {Fischer}, \citenamefont {Fonari},
  \citenamefont {Fr{\"{u}}chtl}, \citenamefont {Gagliardi}, \citenamefont
  {Garza}, \citenamefont {Gawande}, \citenamefont {Ghosh}, \citenamefont
  {Glaesemann}, \citenamefont {G{\"{o}}tz}, \citenamefont {Hammond},
  \citenamefont {Helms}, \citenamefont {Hermes}, \citenamefont {Hirao},
  \citenamefont {Hirata}, \citenamefont {Jacquelin}, \citenamefont {Jensen},
  \citenamefont {Johnson}, \citenamefont {J{\'{o}}nsson}, \citenamefont
  {Kendall}, \citenamefont {Klemm}, \citenamefont {Kobayashi}, \citenamefont
  {Konkov}, \citenamefont {Krishnamoorthy}, \citenamefont {Krishnan},
  \citenamefont {Lin}, \citenamefont {Lins}, \citenamefont {Littlefield},
  \citenamefont {Logsdail}, \citenamefont {Lopata}, \citenamefont {Ma},
  \citenamefont {Marenich}, \citenamefont {{Martin del Campo}}, \citenamefont
  {Mejia-Rodriguez}, \citenamefont {Moore}, \citenamefont {Mullin},
  \citenamefont {Nakajima}, \citenamefont {Nascimento}, \citenamefont
  {Nichols}, \citenamefont {Nichols}, \citenamefont {Nieplocha}, \citenamefont
  {Otero-de-la Roza}, \citenamefont {Palmer}, \citenamefont {Panyala},
  \citenamefont {Pirojsirikul}, \citenamefont {Peng}, \citenamefont {Peverati},
  \citenamefont {Pittner}, \citenamefont {Pollack}, \citenamefont {Richard},
  \citenamefont {Sadayappan}, \citenamefont {Schatz}, \citenamefont {Shelton},
  \citenamefont {Silverstein}, \citenamefont {Smith}, \citenamefont {Soares},
  \citenamefont {Song}, \citenamefont {Swart}, \citenamefont {Taylor},
  \citenamefont {Thomas}, \citenamefont {Tipparaju}, \citenamefont {Truhlar},
  \citenamefont {Tsemekhman}, \citenamefont {{Van Voorhis}}, \citenamefont
  {V{\'{a}}zquez-Mayagoitia}, \citenamefont {Verma}, \citenamefont {Villa},
  \citenamefont {Vishnu}, \citenamefont {Vogiatzis}, \citenamefont {Wang},
  \citenamefont {Weare}, \citenamefont {Williamson}, \citenamefont {Windus},
  \citenamefont {Woli{\'{n}}ski}, \citenamefont {Wong}, \citenamefont {Wu},
  \citenamefont {Yang}, \citenamefont {Yu}, \citenamefont {Zacharias},
  \citenamefont {Zhang}, \citenamefont {Zhao},\ and\ \citenamefont
  {Harrison}}]{Apra20}%
  \BibitemOpen
  \bibfield  {author} {\bibinfo {author} {\bibfnamefont {E.}~\bibnamefont
  {Apr{\`{a}}}}, \bibinfo {author} {\bibfnamefont {E.~J.}\ \bibnamefont
  {Bylaska}}, \bibinfo {author} {\bibfnamefont {W.~A.}\ \bibnamefont
  {de~Jong}}, \bibinfo {author} {\bibfnamefont {N.}~\bibnamefont {Govind}},
  \bibinfo {author} {\bibfnamefont {K.}~\bibnamefont {Kowalski}}, \bibinfo
  {author} {\bibfnamefont {T.~P.}\ \bibnamefont {Straatsma}}, \bibinfo {author}
  {\bibfnamefont {M.}~\bibnamefont {Valiev}}, \bibinfo {author} {\bibfnamefont
  {H.~J.~J.}\ \bibnamefont {van Dam}}, \bibinfo {author} {\bibfnamefont
  {Y.}~\bibnamefont {Alexeev}}, \bibinfo {author} {\bibfnamefont
  {J.}~\bibnamefont {Anchell}}, \bibinfo {author} {\bibfnamefont
  {V.}~\bibnamefont {Anisimov}}, \bibinfo {author} {\bibfnamefont {F.~W.}\
  \bibnamefont {Aquino}}, \bibinfo {author} {\bibfnamefont {R.}~\bibnamefont
  {Atta-Fynn}}, \bibinfo {author} {\bibfnamefont {J.}~\bibnamefont
  {Autschbach}}, \bibinfo {author} {\bibfnamefont {N.~P.}\ \bibnamefont
  {Bauman}}, \bibinfo {author} {\bibfnamefont {J.~C.}\ \bibnamefont {Becca}},
  \bibinfo {author} {\bibfnamefont {D.~E.}\ \bibnamefont {Bernholdt}}, \bibinfo
  {author} {\bibfnamefont {K.}~\bibnamefont {Bhaskaran-Nair}}, \bibinfo
  {author} {\bibfnamefont {S.}~\bibnamefont {Bogatko}}, \bibinfo {author}
  {\bibfnamefont {P.}~\bibnamefont {Borowski}}, \bibinfo {author}
  {\bibfnamefont {J.}~\bibnamefont {Boschen}}, \bibinfo {author} {\bibfnamefont
  {J.}~\bibnamefont {Brabec}}, \bibinfo {author} {\bibfnamefont
  {A.}~\bibnamefont {Bruner}}, \bibinfo {author} {\bibfnamefont
  {E.}~\bibnamefont {Cau{\"{e}}t}}, \bibinfo {author} {\bibfnamefont
  {Y.}~\bibnamefont {Chen}}, \bibinfo {author} {\bibfnamefont {G.~N.}\
  \bibnamefont {Chuev}}, \bibinfo {author} {\bibfnamefont {C.~J.}\ \bibnamefont
  {Cramer}}, \bibinfo {author} {\bibfnamefont {J.}~\bibnamefont {Daily}},
  \bibinfo {author} {\bibfnamefont {M.~J.~O.}\ \bibnamefont {Deegan}}, \bibinfo
  {author} {\bibfnamefont {T.~H.}\ \bibnamefont {{Dunning Jr.}}}, \bibinfo
  {author} {\bibfnamefont {M.}~\bibnamefont {Dupuis}}, \bibinfo {author}
  {\bibfnamefont {K.~G.}\ \bibnamefont {Dyall}}, \bibinfo {author}
  {\bibfnamefont {G.~I.}\ \bibnamefont {Fann}}, \bibinfo {author}
  {\bibfnamefont {S.~A.}\ \bibnamefont {Fischer}}, \bibinfo {author}
  {\bibfnamefont {A.}~\bibnamefont {Fonari}}, \bibinfo {author} {\bibfnamefont
  {H.}~\bibnamefont {Fr{\"{u}}chtl}}, \bibinfo {author} {\bibfnamefont
  {L.}~\bibnamefont {Gagliardi}}, \bibinfo {author} {\bibfnamefont
  {J.}~\bibnamefont {Garza}}, \bibinfo {author} {\bibfnamefont
  {N.}~\bibnamefont {Gawande}}, \bibinfo {author} {\bibfnamefont
  {S.}~\bibnamefont {Ghosh}}, \bibinfo {author} {\bibfnamefont
  {K.}~\bibnamefont {Glaesemann}}, \bibinfo {author} {\bibfnamefont {A.~W.}\
  \bibnamefont {G{\"{o}}tz}}, \bibinfo {author} {\bibfnamefont
  {J.}~\bibnamefont {Hammond}}, \bibinfo {author} {\bibfnamefont
  {V.}~\bibnamefont {Helms}}, \bibinfo {author} {\bibfnamefont {E.~D.}\
  \bibnamefont {Hermes}}, \bibinfo {author} {\bibfnamefont {K.}~\bibnamefont
  {Hirao}}, \bibinfo {author} {\bibfnamefont {S.}~\bibnamefont {Hirata}},
  \bibinfo {author} {\bibfnamefont {M.}~\bibnamefont {Jacquelin}}, \bibinfo
  {author} {\bibfnamefont {L.}~\bibnamefont {Jensen}}, \bibinfo {author}
  {\bibfnamefont {B.~G.}\ \bibnamefont {Johnson}}, \bibinfo {author}
  {\bibfnamefont {H.}~\bibnamefont {J{\'{o}}nsson}}, \bibinfo {author}
  {\bibfnamefont {R.~A.}\ \bibnamefont {Kendall}}, \bibinfo {author}
  {\bibfnamefont {M.}~\bibnamefont {Klemm}}, \bibinfo {author} {\bibfnamefont
  {R.}~\bibnamefont {Kobayashi}}, \bibinfo {author} {\bibfnamefont
  {V.}~\bibnamefont {Konkov}}, \bibinfo {author} {\bibfnamefont
  {S.}~\bibnamefont {Krishnamoorthy}}, \bibinfo {author} {\bibfnamefont
  {M.}~\bibnamefont {Krishnan}}, \bibinfo {author} {\bibfnamefont
  {Z.}~\bibnamefont {Lin}}, \bibinfo {author} {\bibfnamefont {R.~D.}\
  \bibnamefont {Lins}}, \bibinfo {author} {\bibfnamefont {R.~J.}\ \bibnamefont
  {Littlefield}}, \bibinfo {author} {\bibfnamefont {A.~J.}\ \bibnamefont
  {Logsdail}}, \bibinfo {author} {\bibfnamefont {K.}~\bibnamefont {Lopata}},
  \bibinfo {author} {\bibfnamefont {W.}~\bibnamefont {Ma}}, \bibinfo {author}
  {\bibfnamefont {A.~V.}\ \bibnamefont {Marenich}}, \bibinfo {author}
  {\bibfnamefont {J.}~\bibnamefont {{Martin del Campo}}}, \bibinfo {author}
  {\bibfnamefont {D.}~\bibnamefont {Mejia-Rodriguez}}, \bibinfo {author}
  {\bibfnamefont {J.~E.}\ \bibnamefont {Moore}}, \bibinfo {author}
  {\bibfnamefont {J.~M.}\ \bibnamefont {Mullin}}, \bibinfo {author}
  {\bibfnamefont {T.}~\bibnamefont {Nakajima}}, \bibinfo {author}
  {\bibfnamefont {D.~R.}\ \bibnamefont {Nascimento}}, \bibinfo {author}
  {\bibfnamefont {J.~A.}\ \bibnamefont {Nichols}}, \bibinfo {author}
  {\bibfnamefont {P.~J.}\ \bibnamefont {Nichols}}, \bibinfo {author}
  {\bibfnamefont {J.}~\bibnamefont {Nieplocha}}, \bibinfo {author}
  {\bibfnamefont {A.}~\bibnamefont {Otero-de-la Roza}}, \bibinfo {author}
  {\bibfnamefont {B.}~\bibnamefont {Palmer}}, \bibinfo {author} {\bibfnamefont
  {A.}~\bibnamefont {Panyala}}, \bibinfo {author} {\bibfnamefont
  {T.}~\bibnamefont {Pirojsirikul}}, \bibinfo {author} {\bibfnamefont
  {B.}~\bibnamefont {Peng}}, \bibinfo {author} {\bibfnamefont {R.}~\bibnamefont
  {Peverati}}, \bibinfo {author} {\bibfnamefont {J.}~\bibnamefont {Pittner}},
  \bibinfo {author} {\bibfnamefont {L.}~\bibnamefont {Pollack}}, \bibinfo
  {author} {\bibfnamefont {R.~M.}\ \bibnamefont {Richard}}, \bibinfo {author}
  {\bibfnamefont {P.}~\bibnamefont {Sadayappan}}, \bibinfo {author}
  {\bibfnamefont {G.~C.}\ \bibnamefont {Schatz}}, \bibinfo {author}
  {\bibfnamefont {W.~A.}\ \bibnamefont {Shelton}}, \bibinfo {author}
  {\bibfnamefont {D.~W.}\ \bibnamefont {Silverstein}}, \bibinfo {author}
  {\bibfnamefont {D.~M.~A.}\ \bibnamefont {Smith}}, \bibinfo {author}
  {\bibfnamefont {T.~A.}\ \bibnamefont {Soares}}, \bibinfo {author}
  {\bibfnamefont {D.}~\bibnamefont {Song}}, \bibinfo {author} {\bibfnamefont
  {M.}~\bibnamefont {Swart}}, \bibinfo {author} {\bibfnamefont {H.~L.}\
  \bibnamefont {Taylor}}, \bibinfo {author} {\bibfnamefont {G.~S.}\
  \bibnamefont {Thomas}}, \bibinfo {author} {\bibfnamefont {V.}~\bibnamefont
  {Tipparaju}}, \bibinfo {author} {\bibfnamefont {D.~G.}\ \bibnamefont
  {Truhlar}}, \bibinfo {author} {\bibfnamefont {K.}~\bibnamefont {Tsemekhman}},
  \bibinfo {author} {\bibfnamefont {T.}~\bibnamefont {{Van Voorhis}}}, \bibinfo
  {author} {\bibfnamefont {{\'{A}}.}~\bibnamefont {V{\'{a}}zquez-Mayagoitia}},
  \bibinfo {author} {\bibfnamefont {P.}~\bibnamefont {Verma}}, \bibinfo
  {author} {\bibfnamefont {O.}~\bibnamefont {Villa}}, \bibinfo {author}
  {\bibfnamefont {A.}~\bibnamefont {Vishnu}}, \bibinfo {author} {\bibfnamefont
  {K.~D.}\ \bibnamefont {Vogiatzis}}, \bibinfo {author} {\bibfnamefont
  {D.}~\bibnamefont {Wang}}, \bibinfo {author} {\bibfnamefont {J.~H.}\
  \bibnamefont {Weare}}, \bibinfo {author} {\bibfnamefont {M.~J.}\ \bibnamefont
  {Williamson}}, \bibinfo {author} {\bibfnamefont {T.~L.}\ \bibnamefont
  {Windus}}, \bibinfo {author} {\bibfnamefont {K.}~\bibnamefont
  {Woli{\'{n}}ski}}, \bibinfo {author} {\bibfnamefont {A.~T.}\ \bibnamefont
  {Wong}}, \bibinfo {author} {\bibfnamefont {Q.}~\bibnamefont {Wu}}, \bibinfo
  {author} {\bibfnamefont {C.}~\bibnamefont {Yang}}, \bibinfo {author}
  {\bibfnamefont {Q.}~\bibnamefont {Yu}}, \bibinfo {author} {\bibfnamefont
  {M.}~\bibnamefont {Zacharias}}, \bibinfo {author} {\bibfnamefont
  {Z.}~\bibnamefont {Zhang}}, \bibinfo {author} {\bibfnamefont
  {Y.}~\bibnamefont {Zhao}}, \ and\ \bibinfo {author} {\bibfnamefont {R.~J.}\
  \bibnamefont {Harrison}},\ }\bibfield  {title} {\enquote {\bibinfo {title}
  {{NWChem: Past, present, and future}},}\ }\href {\doibase 10.1063/5.0004997}
  {\bibfield  {journal} {\bibinfo  {journal} {J. Chem. Phys.}\ }\textbf
  {\bibinfo {volume} {152}},\ \bibinfo {pages} {184102} (\bibinfo {year}
  {2020})}\BibitemShut {NoStop}%
\bibitem [{\citenamefont {Zhang}\ \emph {et~al.}(2020)\citenamefont {Zhang},
  \citenamefont {Suo}, \citenamefont {Wang}, \citenamefont {Zhang},
  \citenamefont {Li}, \citenamefont {Lei}, \citenamefont {Zou}, \citenamefont
  {Gao}, \citenamefont {Peng}, \citenamefont {Pu}, \citenamefont {Xiao},
  \citenamefont {Sun}, \citenamefont {Wang}, \citenamefont {Ma}, \citenamefont
  {Wang}, \citenamefont {Guo},\ and\ \citenamefont {Liu}}]{Zhang20c}%
  \BibitemOpen
  \bibfield  {author} {\bibinfo {author} {\bibfnamefont {Y.}~\bibnamefont
  {Zhang}}, \bibinfo {author} {\bibfnamefont {B.}~\bibnamefont {Suo}}, \bibinfo
  {author} {\bibfnamefont {Z.}~\bibnamefont {Wang}}, \bibinfo {author}
  {\bibfnamefont {N.}~\bibnamefont {Zhang}}, \bibinfo {author} {\bibfnamefont
  {Z.}~\bibnamefont {Li}}, \bibinfo {author} {\bibfnamefont {Y.}~\bibnamefont
  {Lei}}, \bibinfo {author} {\bibfnamefont {W.}~\bibnamefont {Zou}}, \bibinfo
  {author} {\bibfnamefont {J.}~\bibnamefont {Gao}}, \bibinfo {author}
  {\bibfnamefont {D.}~\bibnamefont {Peng}}, \bibinfo {author} {\bibfnamefont
  {Z.}~\bibnamefont {Pu}}, \bibinfo {author} {\bibfnamefont {Y.}~\bibnamefont
  {Xiao}}, \bibinfo {author} {\bibfnamefont {Q.}~\bibnamefont {Sun}}, \bibinfo
  {author} {\bibfnamefont {F.}~\bibnamefont {Wang}}, \bibinfo {author}
  {\bibfnamefont {Y.}~\bibnamefont {Ma}}, \bibinfo {author} {\bibfnamefont
  {X.}~\bibnamefont {Wang}}, \bibinfo {author} {\bibfnamefont {Y.}~\bibnamefont
  {Guo}}, \ and\ \bibinfo {author} {\bibfnamefont {W.}~\bibnamefont {Liu}},\
  }\bibfield  {title} {\enquote {\bibinfo {title} {{BDF: A relativistic
  electronic structure program package}},}\ }\href {\doibase 10.1063/1.5143173}
  {\bibfield  {journal} {\bibinfo  {journal} {J. Chem. Phys.}\ }\textbf
  {\bibinfo {volume} {152}},\ \bibinfo {pages} {64113} (\bibinfo {year}
  {2020})}\BibitemShut {NoStop}%
\bibitem [{\citenamefont {Saue}\ \emph {et~al.}(2020)\citenamefont {Saue},
  \citenamefont {Bast}, \citenamefont {Gomes}, \citenamefont {Jensen},
  \citenamefont {Visscher}, \citenamefont {Aucar}, \citenamefont {{Di
  Remigio}}, \citenamefont {Dyall}, \citenamefont {Eliav}, \citenamefont
  {Fasshauer}, \citenamefont {Fleig}, \citenamefont {Halbert}, \citenamefont
  {Hedeg{\aa}rd}, \citenamefont {Helmich-Paris}, \citenamefont {Ilia{\v{s}}},
  \citenamefont {Jacob}, \citenamefont {Knecht}, \citenamefont {Laerdahl},
  \citenamefont {Vidal}, \citenamefont {Nayak}, \citenamefont {Olejniczak},
  \citenamefont {Olsen}, \citenamefont {Pernpointner}, \citenamefont {Senjean},
  \citenamefont {Shee}, \citenamefont {Sunaga},\ and\ \citenamefont {van
  Stralen}}]{Saue20}%
  \BibitemOpen
  \bibfield  {author} {\bibinfo {author} {\bibfnamefont {T.}~\bibnamefont
  {Saue}}, \bibinfo {author} {\bibfnamefont {R.}~\bibnamefont {Bast}}, \bibinfo
  {author} {\bibfnamefont {A.~S.~P.}\ \bibnamefont {Gomes}}, \bibinfo {author}
  {\bibfnamefont {H.~J.~{\relax Aa}.}\ \bibnamefont {Jensen}}, \bibinfo
  {author} {\bibfnamefont {L.}~\bibnamefont {Visscher}}, \bibinfo {author}
  {\bibfnamefont {I.~A.}\ \bibnamefont {Aucar}}, \bibinfo {author}
  {\bibfnamefont {R.}~\bibnamefont {{Di Remigio}}}, \bibinfo {author}
  {\bibfnamefont {K.~G.}\ \bibnamefont {Dyall}}, \bibinfo {author}
  {\bibfnamefont {E.}~\bibnamefont {Eliav}}, \bibinfo {author} {\bibfnamefont
  {E.}~\bibnamefont {Fasshauer}}, \bibinfo {author} {\bibfnamefont
  {T.}~\bibnamefont {Fleig}}, \bibinfo {author} {\bibfnamefont
  {L.}~\bibnamefont {Halbert}}, \bibinfo {author} {\bibfnamefont {E.~D.}\
  \bibnamefont {Hedeg{\aa}rd}}, \bibinfo {author} {\bibfnamefont
  {B.}~\bibnamefont {Helmich-Paris}}, \bibinfo {author} {\bibfnamefont
  {M.}~\bibnamefont {Ilia{\v{s}}}}, \bibinfo {author} {\bibfnamefont {C.~R.}\
  \bibnamefont {Jacob}}, \bibinfo {author} {\bibfnamefont {S.}~\bibnamefont
  {Knecht}}, \bibinfo {author} {\bibfnamefont {J.~K.}\ \bibnamefont
  {Laerdahl}}, \bibinfo {author} {\bibfnamefont {M.~L.}\ \bibnamefont {Vidal}},
  \bibinfo {author} {\bibfnamefont {M.~K.}\ \bibnamefont {Nayak}}, \bibinfo
  {author} {\bibfnamefont {M.}~\bibnamefont {Olejniczak}}, \bibinfo {author}
  {\bibfnamefont {J.~M.~H.}\ \bibnamefont {Olsen}}, \bibinfo {author}
  {\bibfnamefont {M.}~\bibnamefont {Pernpointner}}, \bibinfo {author}
  {\bibfnamefont {B.}~\bibnamefont {Senjean}}, \bibinfo {author} {\bibfnamefont
  {A.}~\bibnamefont {Shee}}, \bibinfo {author} {\bibfnamefont {A.}~\bibnamefont
  {Sunaga}}, \ and\ \bibinfo {author} {\bibfnamefont {J.~N.~P.}\ \bibnamefont
  {van Stralen}},\ }\bibfield  {title} {\enquote {\bibinfo {title} {{The DIRAC
  code for relativistic molecular calculations}},}\ }\href {\doibase
  10.1063/5.0004844} {\bibfield  {journal} {\bibinfo  {journal} {J. Chem.
  Phys.}\ }\textbf {\bibinfo {volume} {152}},\ \bibinfo {pages} {204104}
  (\bibinfo {year} {2020})}\BibitemShut {NoStop}%
\bibitem [{\citenamefont {Autschbach}, \citenamefont {Peng},\ and\
  \citenamefont {Reiher}(2012)}]{Autschbach12a}%
  \BibitemOpen
  \bibfield  {author} {\bibinfo {author} {\bibfnamefont {J.}~\bibnamefont
  {Autschbach}}, \bibinfo {author} {\bibfnamefont {D.}~\bibnamefont {Peng}}, \
  and\ \bibinfo {author} {\bibfnamefont {M.}~\bibnamefont {Reiher}},\
  }\bibfield  {title} {\enquote {\bibinfo {title} {Two-component relativistic
  calculations of electric-field gradients using exact decoupling methods:
  spin-orbit and picture-change effects},}\ }\href@noop {} {\bibfield
  {journal} {\bibinfo  {journal} {J. Chem. Theory Comput.}\ }\textbf {\bibinfo
  {volume} {8}},\ \bibinfo {pages} {4239--4248} (\bibinfo {year}
  {2012})}\BibitemShut {NoStop}%
\bibitem [{\citenamefont {Autschbach}(2017)}]{Autschbach17}%
  \BibitemOpen
  \bibfield  {author} {\bibinfo {author} {\bibfnamefont {J.}~\bibnamefont
  {Autschbach}},\ }\bibfield  {title} {\enquote {\bibinfo {title}
  {{Relativistic Effects on Electron–Nucleus Hyperfine Coupling Studied with
  an Exact 2-Component (X2C) Hamiltonian}},}\ }\href {\doibase
  10.1021/acs.jctc.6b01014} {\bibfield  {journal} {\bibinfo  {journal} {J.
  Chem. Theory Comput.}\ }\textbf {\bibinfo {volume} {13}},\ \bibinfo {pages}
  {710--718} (\bibinfo {year} {2017})}\BibitemShut {NoStop}%
\bibitem [{\citenamefont {Matthews}\ \emph {et~al.}(2020)\citenamefont
  {Matthews}, \citenamefont {Cheng}, \citenamefont {Harding}, \citenamefont
  {Lipparini}, \citenamefont {Stopkowicz}, \citenamefont {Jagau}, \citenamefont
  {Szalay}, \citenamefont {Gauss},\ and\ \citenamefont
  {Stanton}}]{Matthews20a}%
  \BibitemOpen
  \bibfield  {author} {\bibinfo {author} {\bibfnamefont {D.~A.}\ \bibnamefont
  {Matthews}}, \bibinfo {author} {\bibfnamefont {L.}~\bibnamefont {Cheng}},
  \bibinfo {author} {\bibfnamefont {M.~E.}\ \bibnamefont {Harding}}, \bibinfo
  {author} {\bibfnamefont {F.}~\bibnamefont {Lipparini}}, \bibinfo {author}
  {\bibfnamefont {S.}~\bibnamefont {Stopkowicz}}, \bibinfo {author}
  {\bibfnamefont {T.-C.}\ \bibnamefont {Jagau}}, \bibinfo {author}
  {\bibfnamefont {P.~G.}\ \bibnamefont {Szalay}}, \bibinfo {author}
  {\bibfnamefont {J.}~\bibnamefont {Gauss}}, \ and\ \bibinfo {author}
  {\bibfnamefont {J.~F.}\ \bibnamefont {Stanton}},\ }\bibfield  {title}
  {\enquote {\bibinfo {title} {{Coupled-cluster techniques for computational
  chemistry: The CFOUR program package}},}\ }\href {\doibase 10.1063/5.0004837}
  {\bibfield  {journal} {\bibinfo  {journal} {J. Chem. Phys.}\ }\textbf
  {\bibinfo {volume} {152}},\ \bibinfo {pages} {214108} (\bibinfo {year}
  {2020})}\BibitemShut {NoStop}%
\bibitem [{\citenamefont {Peng}\ \emph {et~al.}(2013)\citenamefont {Peng},
  \citenamefont {Middendorf}, \citenamefont {Weigend},\ and\ \citenamefont
  {Reiher}}]{Peng13}%
  \BibitemOpen
  \bibfield  {author} {\bibinfo {author} {\bibfnamefont {D.}~\bibnamefont
  {Peng}}, \bibinfo {author} {\bibfnamefont {N.}~\bibnamefont {Middendorf}},
  \bibinfo {author} {\bibfnamefont {F.}~\bibnamefont {Weigend}}, \ and\
  \bibinfo {author} {\bibfnamefont {M.}~\bibnamefont {Reiher}},\ }\bibfield
  {title} {\enquote {\bibinfo {title} {{An efficient implementation of
  two-component relativistic exact-decoupling methods for large molecules}},}\
  }\href {\doibase 10.1063/1.4803693} {\bibfield  {journal} {\bibinfo
  {journal} {J. Chem. Phys.}\ }\textbf {\bibinfo {volume} {138}},\ \bibinfo
  {pages} {184105} (\bibinfo {year} {2013})}\BibitemShut {NoStop}%
\bibitem [{\citenamefont {Feng}, \citenamefont {Duignan},\ and\ \citenamefont
  {Autschbach}(2021)}]{Feng21}%
  \BibitemOpen
  \bibfield  {author} {\bibinfo {author} {\bibfnamefont {R.}~\bibnamefont
  {Feng}}, \bibinfo {author} {\bibfnamefont {T.~J.}\ \bibnamefont {Duignan}}, \
  and\ \bibinfo {author} {\bibfnamefont {J.}~\bibnamefont {Autschbach}},\
  }\bibfield  {title} {\enquote {\bibinfo {title} {{Electron–Nucleus
  Hyperfine Coupling Calculated from Restricted Active Space Wavefunctions and
  an Exact Two-Component Hamiltonian}},}\ }\href {\doibase
  10.1021/acs.jctc.0c01005} {\bibfield  {journal} {\bibinfo  {journal} {J.
  Chem. Theory Comput.}\ }\textbf {\bibinfo {volume} {17}},\ \bibinfo {pages}
  {255--268} (\bibinfo {year} {2021})}\BibitemShut {NoStop}%
\bibitem [{\citenamefont {Aquilante}\ \emph {et~al.}(2020)\citenamefont
  {Aquilante}, \citenamefont {Autschbach}, \citenamefont {Baiardi},
  \citenamefont {Battaglia}, \citenamefont {Borin}, \citenamefont {Chibotaru},
  \citenamefont {Conti}, \citenamefont {{De Vico}}, \citenamefont {Delcey},
  \citenamefont {{Fdez. Galv{\'{a}}n}}, \citenamefont {Ferr{\'{e}}},
  \citenamefont {Freitag}, \citenamefont {Garavelli}, \citenamefont {Gong},
  \citenamefont {Knecht}, \citenamefont {Larsson}, \citenamefont {Lindh},
  \citenamefont {Lundberg}, \citenamefont {Malmqvist}, \citenamefont {Nenov},
  \citenamefont {Norell}, \citenamefont {Odelius}, \citenamefont {Olivucci},
  \citenamefont {Pedersen}, \citenamefont {Pedraza-Gonz{\'{a}}lez},
  \citenamefont {Phung}, \citenamefont {Pierloot}, \citenamefont {Reiher},
  \citenamefont {Schapiro}, \citenamefont {Segarra-Mart{\'{i}}}, \citenamefont
  {Segatta}, \citenamefont {Seijo}, \citenamefont {Sen}, \citenamefont
  {Sergentu}, \citenamefont {Stein}, \citenamefont {Ungur}, \citenamefont
  {Vacher}, \citenamefont {Valentini},\ and\ \citenamefont
  {Veryazov}}]{Aquilante20}%
  \BibitemOpen
  \bibfield  {author} {\bibinfo {author} {\bibfnamefont {F.}~\bibnamefont
  {Aquilante}}, \bibinfo {author} {\bibfnamefont {J.}~\bibnamefont
  {Autschbach}}, \bibinfo {author} {\bibfnamefont {A.}~\bibnamefont {Baiardi}},
  \bibinfo {author} {\bibfnamefont {S.}~\bibnamefont {Battaglia}}, \bibinfo
  {author} {\bibfnamefont {V.~A.}\ \bibnamefont {Borin}}, \bibinfo {author}
  {\bibfnamefont {L.~F.}\ \bibnamefont {Chibotaru}}, \bibinfo {author}
  {\bibfnamefont {I.}~\bibnamefont {Conti}}, \bibinfo {author} {\bibfnamefont
  {L.}~\bibnamefont {{De Vico}}}, \bibinfo {author} {\bibfnamefont
  {M.}~\bibnamefont {Delcey}}, \bibinfo {author} {\bibfnamefont
  {I.}~\bibnamefont {{Fdez. Galv{\'{a}}n}}}, \bibinfo {author} {\bibfnamefont
  {N.}~\bibnamefont {Ferr{\'{e}}}}, \bibinfo {author} {\bibfnamefont
  {L.}~\bibnamefont {Freitag}}, \bibinfo {author} {\bibfnamefont
  {M.}~\bibnamefont {Garavelli}}, \bibinfo {author} {\bibfnamefont
  {X.}~\bibnamefont {Gong}}, \bibinfo {author} {\bibfnamefont {S.}~\bibnamefont
  {Knecht}}, \bibinfo {author} {\bibfnamefont {E.~D.}\ \bibnamefont {Larsson}},
  \bibinfo {author} {\bibfnamefont {R.}~\bibnamefont {Lindh}}, \bibinfo
  {author} {\bibfnamefont {M.}~\bibnamefont {Lundberg}}, \bibinfo {author}
  {\bibfnamefont {P.~{\AA}.}\ \bibnamefont {Malmqvist}}, \bibinfo {author}
  {\bibfnamefont {A.}~\bibnamefont {Nenov}}, \bibinfo {author} {\bibfnamefont
  {J.}~\bibnamefont {Norell}}, \bibinfo {author} {\bibfnamefont
  {M.}~\bibnamefont {Odelius}}, \bibinfo {author} {\bibfnamefont
  {M.}~\bibnamefont {Olivucci}}, \bibinfo {author} {\bibfnamefont {T.~B.}\
  \bibnamefont {Pedersen}}, \bibinfo {author} {\bibfnamefont {L.}~\bibnamefont
  {Pedraza-Gonz{\'{a}}lez}}, \bibinfo {author} {\bibfnamefont {Q.~M.}\
  \bibnamefont {Phung}}, \bibinfo {author} {\bibfnamefont {K.}~\bibnamefont
  {Pierloot}}, \bibinfo {author} {\bibfnamefont {M.}~\bibnamefont {Reiher}},
  \bibinfo {author} {\bibfnamefont {I.}~\bibnamefont {Schapiro}}, \bibinfo
  {author} {\bibfnamefont {J.}~\bibnamefont {Segarra-Mart{\'{i}}}}, \bibinfo
  {author} {\bibfnamefont {F.}~\bibnamefont {Segatta}}, \bibinfo {author}
  {\bibfnamefont {L.}~\bibnamefont {Seijo}}, \bibinfo {author} {\bibfnamefont
  {S.}~\bibnamefont {Sen}}, \bibinfo {author} {\bibfnamefont {D.-C.}\
  \bibnamefont {Sergentu}}, \bibinfo {author} {\bibfnamefont {C.~J.}\
  \bibnamefont {Stein}}, \bibinfo {author} {\bibfnamefont {L.}~\bibnamefont
  {Ungur}}, \bibinfo {author} {\bibfnamefont {M.}~\bibnamefont {Vacher}},
  \bibinfo {author} {\bibfnamefont {A.}~\bibnamefont {Valentini}}, \ and\
  \bibinfo {author} {\bibfnamefont {V.}~\bibnamefont {Veryazov}},\ }\bibfield
  {title} {\enquote {\bibinfo {title} {{Modern quantum chemistry with
  [Open]Molcas}},}\ }\href {\doibase 10.1063/5.0004835} {\bibfield  {journal}
  {\bibinfo  {journal} {J. Chem. Phys.}\ }\textbf {\bibinfo {volume} {152}},\
  \bibinfo {pages} {214117} (\bibinfo {year} {2020})}\BibitemShut {NoStop}%
\bibitem [{\citenamefont {Werner}\ \emph {et~al.}(2020)\citenamefont {Werner},
  \citenamefont {Knowles}, \citenamefont {Manby}, \citenamefont {Black},
  \citenamefont {Doll}, \citenamefont {He{\ss}elmann}, \citenamefont {Kats},
  \citenamefont {K{\"{o}}hn}, \citenamefont {Korona}, \citenamefont {Kreplin},
  \citenamefont {Ma}, \citenamefont {{Miller III}}, \citenamefont
  {Mitrushchenkov}, \citenamefont {Peterson}, \citenamefont {Polyak},
  \citenamefont {Rauhut},\ and\ \citenamefont {Sibaev}}]{Werner20}%
  \BibitemOpen
  \bibfield  {author} {\bibinfo {author} {\bibfnamefont {H.-J.}\ \bibnamefont
  {Werner}}, \bibinfo {author} {\bibfnamefont {P.~J.}\ \bibnamefont {Knowles}},
  \bibinfo {author} {\bibfnamefont {F.~R.}\ \bibnamefont {Manby}}, \bibinfo
  {author} {\bibfnamefont {J.~A.}\ \bibnamefont {Black}}, \bibinfo {author}
  {\bibfnamefont {K.}~\bibnamefont {Doll}}, \bibinfo {author} {\bibfnamefont
  {A.}~\bibnamefont {He{\ss}elmann}}, \bibinfo {author} {\bibfnamefont
  {D.}~\bibnamefont {Kats}}, \bibinfo {author} {\bibfnamefont {A.}~\bibnamefont
  {K{\"{o}}hn}}, \bibinfo {author} {\bibfnamefont {T.}~\bibnamefont {Korona}},
  \bibinfo {author} {\bibfnamefont {D.~A.}\ \bibnamefont {Kreplin}}, \bibinfo
  {author} {\bibfnamefont {Q.}~\bibnamefont {Ma}}, \bibinfo {author}
  {\bibfnamefont {T.~F.}\ \bibnamefont {{Miller III}}}, \bibinfo {author}
  {\bibfnamefont {A.}~\bibnamefont {Mitrushchenkov}}, \bibinfo {author}
  {\bibfnamefont {K.~A.}\ \bibnamefont {Peterson}}, \bibinfo {author}
  {\bibfnamefont {I.}~\bibnamefont {Polyak}}, \bibinfo {author} {\bibfnamefont
  {G.}~\bibnamefont {Rauhut}}, \ and\ \bibinfo {author} {\bibfnamefont
  {M.}~\bibnamefont {Sibaev}},\ }\bibfield  {title} {\enquote {\bibinfo {title}
  {{The Molpro quantum chemistry package}},}\ }\href {\doibase
  10.1063/5.0005081} {\bibfield  {journal} {\bibinfo  {journal} {J. Chem.
  Phys.}\ }\textbf {\bibinfo {volume} {152}},\ \bibinfo {pages} {144107}
  (\bibinfo {year} {2020})}\BibitemShut {NoStop}%
\bibitem [{\citenamefont {Franzke}\ and\ \citenamefont
  {Holzer}(2023)}]{Franzke23}%
  \BibitemOpen
  \bibfield  {author} {\bibinfo {author} {\bibfnamefont {Y.~J.}\ \bibnamefont
  {Franzke}}\ and\ \bibinfo {author} {\bibfnamefont {C.}~\bibnamefont
  {Holzer}},\ }\bibfield  {title} {\enquote {\bibinfo {title} {{Exact
  two-component theory becoming an efficient tool for NMR shieldings and shifts
  with spin–orbit coupling}},}\ }\href {\doibase 10.1063/5.0171509}
  {\bibfield  {journal} {\bibinfo  {journal} {J. Chem. Phys.}\ }\textbf
  {\bibinfo {volume} {159}},\ \bibinfo {pages} {184102} (\bibinfo {year}
  {2023})}\BibitemShut {NoStop}%
\bibitem [{\citenamefont {Balasubramani}\ \emph {et~al.}(2020)\citenamefont
  {Balasubramani}, \citenamefont {Chen}, \citenamefont {Coriani}, \citenamefont
  {Diedenhofen}, \citenamefont {Frank}, \citenamefont {Franzke}, \citenamefont
  {Furche}, \citenamefont {Grotjahn}, \citenamefont {Harding}, \citenamefont
  {H{\"{a}}ttig}, \citenamefont {Hellweg}, \citenamefont {Helmich-Paris},
  \citenamefont {Holzer}, \citenamefont {Huniar}, \citenamefont {Kaupp},
  \citenamefont {{Marefat Khah}}, \citenamefont {{Karbalaei Khani}},
  \citenamefont {M{\"{u}}ller}, \citenamefont {Mack}, \citenamefont {Nguyen},
  \citenamefont {Parker}, \citenamefont {Perlt}, \citenamefont {Rappoport},
  \citenamefont {Reiter}, \citenamefont {Roy}, \citenamefont {R{\"{u}}ckert},
  \citenamefont {Schmitz}, \citenamefont {Sierka}, \citenamefont {Tapavicza},
  \citenamefont {Tew}, \citenamefont {van W{\"{u}}llen}, \citenamefont {Voora},
  \citenamefont {Weigend}, \citenamefont {Wody{\'{n}}ski},\ and\ \citenamefont
  {Yu}}]{Balasubramani20}%
  \BibitemOpen
  \bibfield  {author} {\bibinfo {author} {\bibfnamefont {S.~G.}\ \bibnamefont
  {Balasubramani}}, \bibinfo {author} {\bibfnamefont {G.~P.}\ \bibnamefont
  {Chen}}, \bibinfo {author} {\bibfnamefont {S.}~\bibnamefont {Coriani}},
  \bibinfo {author} {\bibfnamefont {M.}~\bibnamefont {Diedenhofen}}, \bibinfo
  {author} {\bibfnamefont {M.~S.}\ \bibnamefont {Frank}}, \bibinfo {author}
  {\bibfnamefont {Y.~J.}\ \bibnamefont {Franzke}}, \bibinfo {author}
  {\bibfnamefont {F.}~\bibnamefont {Furche}}, \bibinfo {author} {\bibfnamefont
  {R.}~\bibnamefont {Grotjahn}}, \bibinfo {author} {\bibfnamefont {M.~E.}\
  \bibnamefont {Harding}}, \bibinfo {author} {\bibfnamefont {C.}~\bibnamefont
  {H{\"{a}}ttig}}, \bibinfo {author} {\bibfnamefont {A.}~\bibnamefont
  {Hellweg}}, \bibinfo {author} {\bibfnamefont {B.}~\bibnamefont
  {Helmich-Paris}}, \bibinfo {author} {\bibfnamefont {C.}~\bibnamefont
  {Holzer}}, \bibinfo {author} {\bibfnamefont {U.}~\bibnamefont {Huniar}},
  \bibinfo {author} {\bibfnamefont {M.}~\bibnamefont {Kaupp}}, \bibinfo
  {author} {\bibfnamefont {A.}~\bibnamefont {{Marefat Khah}}}, \bibinfo
  {author} {\bibfnamefont {S.}~\bibnamefont {{Karbalaei Khani}}}, \bibinfo
  {author} {\bibfnamefont {T.}~\bibnamefont {M{\"{u}}ller}}, \bibinfo {author}
  {\bibfnamefont {F.}~\bibnamefont {Mack}}, \bibinfo {author} {\bibfnamefont
  {B.~D.}\ \bibnamefont {Nguyen}}, \bibinfo {author} {\bibfnamefont {S.~M.}\
  \bibnamefont {Parker}}, \bibinfo {author} {\bibfnamefont {E.}~\bibnamefont
  {Perlt}}, \bibinfo {author} {\bibfnamefont {D.}~\bibnamefont {Rappoport}},
  \bibinfo {author} {\bibfnamefont {K.}~\bibnamefont {Reiter}}, \bibinfo
  {author} {\bibfnamefont {S.}~\bibnamefont {Roy}}, \bibinfo {author}
  {\bibfnamefont {M.}~\bibnamefont {R{\"{u}}ckert}}, \bibinfo {author}
  {\bibfnamefont {G.}~\bibnamefont {Schmitz}}, \bibinfo {author} {\bibfnamefont
  {M.}~\bibnamefont {Sierka}}, \bibinfo {author} {\bibfnamefont
  {E.}~\bibnamefont {Tapavicza}}, \bibinfo {author} {\bibfnamefont {D.~P.}\
  \bibnamefont {Tew}}, \bibinfo {author} {\bibfnamefont {C.}~\bibnamefont {van
  W{\"{u}}llen}}, \bibinfo {author} {\bibfnamefont {V.~K.}\ \bibnamefont
  {Voora}}, \bibinfo {author} {\bibfnamefont {F.}~\bibnamefont {Weigend}},
  \bibinfo {author} {\bibfnamefont {A.}~\bibnamefont {Wody{\'{n}}ski}}, \ and\
  \bibinfo {author} {\bibfnamefont {J.~M.}\ \bibnamefont {Yu}},\ }\bibfield
  {title} {\enquote {\bibinfo {title} {{TURBOMOLE: Modular program suite for ab
  initio quantum-chemical and condensed-matter simulations}},}\ }\href
  {\doibase 10.1063/5.0004635} {\bibfield  {journal} {\bibinfo  {journal} {J.
  Chem. Phys.}\ }\textbf {\bibinfo {volume} {152}},\ \bibinfo {pages} {184107}
  (\bibinfo {year} {2020})}\BibitemShut {NoStop}%
\bibitem [{\citenamefont {Verma}, \citenamefont {Derricotte},\ and\
  \citenamefont {Evangelista}(2016)}]{Verma16}%
  \BibitemOpen
  \bibfield  {author} {\bibinfo {author} {\bibfnamefont {P.}~\bibnamefont
  {Verma}}, \bibinfo {author} {\bibfnamefont {W.~D.}\ \bibnamefont
  {Derricotte}}, \ and\ \bibinfo {author} {\bibfnamefont {F.~A.}\ \bibnamefont
  {Evangelista}},\ }\bibfield  {title} {\enquote {\bibinfo {title} {{Predicting
  Near Edge X-ray Absorption Spectra with the Spin-Free Exact-Two-Component
  Hamiltonian and Orthogonality Constrained Density Functional Theory}},}\
  }\href {\doibase 10.1021/acs.jctc.5b00817} {\bibfield  {journal} {\bibinfo
  {journal} {J. Chem. Theor. Comp.}\ }\textbf {\bibinfo {volume} {12}},\
  \bibinfo {pages} {144--156} (\bibinfo {year} {2016})}\BibitemShut {NoStop}%
\bibitem [{\citenamefont {Smith}\ \emph {et~al.}(2020)\citenamefont {Smith},
  \citenamefont {Burns}, \citenamefont {Simmonett}, \citenamefont {Parrish},
  \citenamefont {Schieber}, \citenamefont {Galvelis}, \citenamefont {Kraus},
  \citenamefont {Kruse}, \citenamefont {{Di Remigio}}, \citenamefont
  {Alenaizan}, \citenamefont {James}, \citenamefont {Lehtola}, \citenamefont
  {Misiewicz}, \citenamefont {Scheurer}, \citenamefont {Shaw}, \citenamefont
  {Schriber}, \citenamefont {Xie}, \citenamefont {Glick}, \citenamefont
  {Sirianni}, \citenamefont {O'Brien}, \citenamefont {Waldrop}, \citenamefont
  {Kumar}, \citenamefont {Hohenstein}, \citenamefont {Pritchard}, \citenamefont
  {Brooks}, \citenamefont {{Schaefer III}}, \citenamefont {Sokolov},
  \citenamefont {Patkowski}, \citenamefont {{DePrince III}}, \citenamefont
  {Bozkaya}, \citenamefont {King}, \citenamefont {Evangelista}, \citenamefont
  {Turney}, \citenamefont {Crawford},\ and\ \citenamefont
  {Sherrill}}]{Smith20}%
  \BibitemOpen
  \bibfield  {author} {\bibinfo {author} {\bibfnamefont {D.~G.~A.}\
  \bibnamefont {Smith}}, \bibinfo {author} {\bibfnamefont {L.~A.}\ \bibnamefont
  {Burns}}, \bibinfo {author} {\bibfnamefont {A.~C.}\ \bibnamefont
  {Simmonett}}, \bibinfo {author} {\bibfnamefont {R.~M.}\ \bibnamefont
  {Parrish}}, \bibinfo {author} {\bibfnamefont {M.~C.}\ \bibnamefont
  {Schieber}}, \bibinfo {author} {\bibfnamefont {R.}~\bibnamefont {Galvelis}},
  \bibinfo {author} {\bibfnamefont {P.}~\bibnamefont {Kraus}}, \bibinfo
  {author} {\bibfnamefont {H.}~\bibnamefont {Kruse}}, \bibinfo {author}
  {\bibfnamefont {R.}~\bibnamefont {{Di Remigio}}}, \bibinfo {author}
  {\bibfnamefont {A.}~\bibnamefont {Alenaizan}}, \bibinfo {author}
  {\bibfnamefont {A.~M.}\ \bibnamefont {James}}, \bibinfo {author}
  {\bibfnamefont {S.}~\bibnamefont {Lehtola}}, \bibinfo {author} {\bibfnamefont
  {J.~P.}\ \bibnamefont {Misiewicz}}, \bibinfo {author} {\bibfnamefont
  {M.}~\bibnamefont {Scheurer}}, \bibinfo {author} {\bibfnamefont {R.~A.}\
  \bibnamefont {Shaw}}, \bibinfo {author} {\bibfnamefont {J.~B.}\ \bibnamefont
  {Schriber}}, \bibinfo {author} {\bibfnamefont {Y.}~\bibnamefont {Xie}},
  \bibinfo {author} {\bibfnamefont {Z.~L.}\ \bibnamefont {Glick}}, \bibinfo
  {author} {\bibfnamefont {D.~A.}\ \bibnamefont {Sirianni}}, \bibinfo {author}
  {\bibfnamefont {J.~S.}\ \bibnamefont {O'Brien}}, \bibinfo {author}
  {\bibfnamefont {J.~M.}\ \bibnamefont {Waldrop}}, \bibinfo {author}
  {\bibfnamefont {A.}~\bibnamefont {Kumar}}, \bibinfo {author} {\bibfnamefont
  {E.~G.}\ \bibnamefont {Hohenstein}}, \bibinfo {author} {\bibfnamefont
  {B.~P.}\ \bibnamefont {Pritchard}}, \bibinfo {author} {\bibfnamefont {B.~R.}\
  \bibnamefont {Brooks}}, \bibinfo {author} {\bibfnamefont {H.~F.}\
  \bibnamefont {{Schaefer III}}}, \bibinfo {author} {\bibfnamefont {A.~Y.}\
  \bibnamefont {Sokolov}}, \bibinfo {author} {\bibfnamefont {K.}~\bibnamefont
  {Patkowski}}, \bibinfo {author} {\bibfnamefont {A.~E.}\ \bibnamefont
  {{DePrince III}}}, \bibinfo {author} {\bibfnamefont {U.}~\bibnamefont
  {Bozkaya}}, \bibinfo {author} {\bibfnamefont {R.~A.}\ \bibnamefont {King}},
  \bibinfo {author} {\bibfnamefont {F.~A.}\ \bibnamefont {Evangelista}},
  \bibinfo {author} {\bibfnamefont {J.~M.}\ \bibnamefont {Turney}}, \bibinfo
  {author} {\bibfnamefont {T.~D.}\ \bibnamefont {Crawford}}, \ and\ \bibinfo
  {author} {\bibfnamefont {C.~D.}\ \bibnamefont {Sherrill}},\ }\bibfield
  {title} {\enquote {\bibinfo {title} {{PSI4 1.4: Open-source software for
  high-throughput quantum chemistry}},}\ }\href {\doibase 10.1063/5.0006002}
  {\bibfield  {journal} {\bibinfo  {journal} {J. Chem. Phys.}\ }\textbf
  {\bibinfo {volume} {152}},\ \bibinfo {pages} {184108} (\bibinfo {year}
  {2020})}\BibitemShut {NoStop}%
\bibitem [{\citenamefont {Goings}\ \emph {et~al.}(2016)\citenamefont {Goings},
  \citenamefont {Kasper}, \citenamefont {Egidi}, \citenamefont {Sun},\ and\
  \citenamefont {Li}}]{Goings16}%
  \BibitemOpen
  \bibfield  {author} {\bibinfo {author} {\bibfnamefont {J.~J.}\ \bibnamefont
  {Goings}}, \bibinfo {author} {\bibfnamefont {J.~M.}\ \bibnamefont {Kasper}},
  \bibinfo {author} {\bibfnamefont {F.}~\bibnamefont {Egidi}}, \bibinfo
  {author} {\bibfnamefont {S.}~\bibnamefont {Sun}}, \ and\ \bibinfo {author}
  {\bibfnamefont {X.}~\bibnamefont {Li}},\ }\bibfield  {title} {\enquote
  {\bibinfo {title} {{Real time propagation of the exact two component
  time-dependent density functional theory}},}\ }\href {\doibase
  10.1063/1.4962422} {\bibfield  {journal} {\bibinfo  {journal} {J. Chem.
  Phys.}\ }\textbf {\bibinfo {volume} {145}},\ \bibinfo {pages} {104107}
  (\bibinfo {year} {2016})}\BibitemShut {NoStop}%
\bibitem [{\citenamefont {Egidi}\ \emph {et~al.}(2017)\citenamefont {Egidi},
  \citenamefont {Sun}, \citenamefont {Goings}, \citenamefont {Scalmani},
  \citenamefont {Frisch},\ and\ \citenamefont {Li}}]{Egidi17}%
  \BibitemOpen
  \bibfield  {author} {\bibinfo {author} {\bibfnamefont {F.}~\bibnamefont
  {Egidi}}, \bibinfo {author} {\bibfnamefont {S.}~\bibnamefont {Sun}}, \bibinfo
  {author} {\bibfnamefont {J.~J.}\ \bibnamefont {Goings}}, \bibinfo {author}
  {\bibfnamefont {G.}~\bibnamefont {Scalmani}}, \bibinfo {author}
  {\bibfnamefont {M.~J.}\ \bibnamefont {Frisch}}, \ and\ \bibinfo {author}
  {\bibfnamefont {X.}~\bibnamefont {Li}},\ }\bibfield  {title} {\enquote
  {\bibinfo {title} {Two-component noncollinear time-dependent spin density
  functional theory for excited state calculations},}\ }\href@noop {}
  {\bibfield  {journal} {\bibinfo  {journal} {J. Chem. Theory Comput.}\
  }\textbf {\bibinfo {volume} {13}},\ \bibinfo {pages} {2591--2603} (\bibinfo
  {year} {2017})}\BibitemShut {NoStop}%
\bibitem [{\citenamefont {Koulias}\ \emph {et~al.}(2019)\citenamefont
  {Koulias}, \citenamefont {Williams-Young}, \citenamefont {Nascimento},
  \citenamefont {DePrince},\ and\ \citenamefont {Li}}]{Koulias19}%
  \BibitemOpen
  \bibfield  {author} {\bibinfo {author} {\bibfnamefont {L.~N.}\ \bibnamefont
  {Koulias}}, \bibinfo {author} {\bibfnamefont {D.~B.}\ \bibnamefont
  {Williams-Young}}, \bibinfo {author} {\bibfnamefont {D.~R.}\ \bibnamefont
  {Nascimento}}, \bibinfo {author} {\bibfnamefont {A.~E.}\ \bibnamefont
  {DePrince}}, \ and\ \bibinfo {author} {\bibfnamefont {X.}~\bibnamefont
  {Li}},\ }\bibfield  {title} {\enquote {\bibinfo {title} {{Relativistic
  Real-Time Time-Dependent Equation-of-Motion Coupled-Cluster}},}\ }\href
  {\doibase 10.1021/acs.jctc.9b00729} {\bibfield  {journal} {\bibinfo
  {journal} {J. Chem. Theory Comput.}\ }\textbf {\bibinfo {volume} {15}},\
  \bibinfo {pages} {6617--6624} (\bibinfo {year} {2019})}\BibitemShut {NoStop}%
\bibitem [{\citenamefont {Sharma}\ \emph {et~al.}(2022)\citenamefont {Sharma},
  \citenamefont {Jenkins}, \citenamefont {Scalmani}, \citenamefont {Frisch},
  \citenamefont {Truhlar}, \citenamefont {Gagliardi},\ and\ \citenamefont
  {Li}}]{SharmaP22}%
  \BibitemOpen
  \bibfield  {author} {\bibinfo {author} {\bibfnamefont {P.}~\bibnamefont
  {Sharma}}, \bibinfo {author} {\bibfnamefont {A.~J.}\ \bibnamefont {Jenkins}},
  \bibinfo {author} {\bibfnamefont {G.}~\bibnamefont {Scalmani}}, \bibinfo
  {author} {\bibfnamefont {M.~J.}\ \bibnamefont {Frisch}}, \bibinfo {author}
  {\bibfnamefont {D.~G.}\ \bibnamefont {Truhlar}}, \bibinfo {author}
  {\bibfnamefont {L.}~\bibnamefont {Gagliardi}}, \ and\ \bibinfo {author}
  {\bibfnamefont {X.}~\bibnamefont {Li}},\ }\bibfield  {title} {\enquote
  {\bibinfo {title} {{Exact-Two-Component Multiconfiguration Pair-Density
  Functional Theory}},}\ }\href {\doibase 10.1021/acs.jctc.2c00062} {\bibfield
  {journal} {\bibinfo  {journal} {J. Chem. Theory Comput.}\ }\textbf {\bibinfo
  {volume} {18}},\ \bibinfo {pages} {2947--2954} (\bibinfo {year}
  {2022})}\BibitemShut {NoStop}%
\bibitem [{\citenamefont {Zhang}\ \emph {et~al.}(2024)\citenamefont {Zhang},
  \citenamefont {Banerjee}, \citenamefont {Koulias}, \citenamefont {Valeev},
  \citenamefont {DePrince},\ and\ \citenamefont {Li}}]{ZhangT24}%
  \BibitemOpen
  \bibfield  {author} {\bibinfo {author} {\bibfnamefont {T.}~\bibnamefont
  {Zhang}}, \bibinfo {author} {\bibfnamefont {S.}~\bibnamefont {Banerjee}},
  \bibinfo {author} {\bibfnamefont {L.~N.}\ \bibnamefont {Koulias}}, \bibinfo
  {author} {\bibfnamefont {E.~F.}\ \bibnamefont {Valeev}}, \bibinfo {author}
  {\bibfnamefont {A.~E. I. I.~I.}\ \bibnamefont {DePrince}}, \ and\ \bibinfo
  {author} {\bibfnamefont {X.}~\bibnamefont {Li}},\ }\bibfield  {title}
  {\enquote {\bibinfo {title} {{Dirac–Coulomb–Breit Molecular Mean-Field
  Exact-Two-Component Relativistic Equation-of-Motion Coupled-Cluster
  Theory}},}\ }\href {\doibase 10.1021/acs.jpca.3c08167} {\bibfield  {journal}
  {\bibinfo  {journal} {J. Phys. Chem. A}\ }\textbf {\bibinfo {volume} {128}},\
  \bibinfo {pages} {3408--3418} (\bibinfo {year} {2024})}\BibitemShut {NoStop}%
\bibitem [{\citenamefont {Kovtun}\ \emph {et~al.}(2024)\citenamefont {Kovtun},
  \citenamefont {Lambros}, \citenamefont {Liu}, \citenamefont {Tang},
  \citenamefont {Williams-Young},\ and\ \citenamefont {Li}}]{Kovtun24}%
  \BibitemOpen
  \bibfield  {author} {\bibinfo {author} {\bibfnamefont {M.}~\bibnamefont
  {Kovtun}}, \bibinfo {author} {\bibfnamefont {E.}~\bibnamefont {Lambros}},
  \bibinfo {author} {\bibfnamefont {A.}~\bibnamefont {Liu}}, \bibinfo {author}
  {\bibfnamefont {D.}~\bibnamefont {Tang}}, \bibinfo {author} {\bibfnamefont
  {D.~B.}\ \bibnamefont {Williams-Young}}, \ and\ \bibinfo {author}
  {\bibfnamefont {X.}~\bibnamefont {Li}},\ }\bibfield  {title} {\enquote
  {\bibinfo {title} {{Accelerating Relativistic Exact-Two-Component Density
  Functional Theory Calculations with Graphical Processing Units}},}\ }\href
  {\doibase 10.1021/acs.jctc.4c00843} {\bibfield  {journal} {\bibinfo
  {journal} {J. Chem. Theory Comput.}\ }\textbf {\bibinfo {volume} {20}},\
  \bibinfo {pages} {7694--7699} (\bibinfo {year} {2024})}\BibitemShut {NoStop}%
\bibitem [{\citenamefont {Williams-Young}\ \emph {et~al.}(2020)\citenamefont
  {Williams-Young}, \citenamefont {Petrone}, \citenamefont {Sun}, \citenamefont
  {Stetina}, \citenamefont {Lestrange}, \citenamefont {Hoyer}, \citenamefont
  {Nascimento}, \citenamefont {Koulias}, \citenamefont {Wildman}, \citenamefont
  {Kasper}, \citenamefont {Goings}, \citenamefont {Ding}, \citenamefont
  {{DePrince III}}, \citenamefont {Valeev},\ and\ \citenamefont
  {Li}}]{WilliamsYoung20}%
  \BibitemOpen
  \bibfield  {author} {\bibinfo {author} {\bibfnamefont {D.~B.}\ \bibnamefont
  {Williams-Young}}, \bibinfo {author} {\bibfnamefont {A.}~\bibnamefont
  {Petrone}}, \bibinfo {author} {\bibfnamefont {S.}~\bibnamefont {Sun}},
  \bibinfo {author} {\bibfnamefont {T.~F.}\ \bibnamefont {Stetina}}, \bibinfo
  {author} {\bibfnamefont {P.}~\bibnamefont {Lestrange}}, \bibinfo {author}
  {\bibfnamefont {C.~E.}\ \bibnamefont {Hoyer}}, \bibinfo {author}
  {\bibfnamefont {D.~R.}\ \bibnamefont {Nascimento}}, \bibinfo {author}
  {\bibfnamefont {L.}~\bibnamefont {Koulias}}, \bibinfo {author} {\bibfnamefont
  {A.}~\bibnamefont {Wildman}}, \bibinfo {author} {\bibfnamefont
  {J.}~\bibnamefont {Kasper}}, \bibinfo {author} {\bibfnamefont {J.~J.}\
  \bibnamefont {Goings}}, \bibinfo {author} {\bibfnamefont {F.}~\bibnamefont
  {Ding}}, \bibinfo {author} {\bibfnamefont {A.~E.}\ \bibnamefont {{DePrince
  III}}}, \bibinfo {author} {\bibfnamefont {E.~F.}\ \bibnamefont {Valeev}}, \
  and\ \bibinfo {author} {\bibfnamefont {X.}~\bibnamefont {Li}},\ }\bibfield
  {title} {\enquote {\bibinfo {title} {{The Chronus Quantum software
  package}},}\ }\href {\doibase https://doi.org/10.1002/wcms.1436} {\bibfield
  {journal} {\bibinfo  {journal} {WIREs Comput. Mol. Sci.}\ }\textbf {\bibinfo
  {volume} {10}},\ \bibinfo {pages} {e1436} (\bibinfo {year}
  {2020})}\BibitemShut {NoStop}%
\bibitem [{\citenamefont {Guo}\ \emph {et~al.}(2016)\citenamefont {Guo},
  \citenamefont {Watson}, \citenamefont {Hu}, \citenamefont {Sun},\ and\
  \citenamefont {Chan}}]{Guo16}%
  \BibitemOpen
  \bibfield  {author} {\bibinfo {author} {\bibfnamefont {S.}~\bibnamefont
  {Guo}}, \bibinfo {author} {\bibfnamefont {M.~A.}\ \bibnamefont {Watson}},
  \bibinfo {author} {\bibfnamefont {W.}~\bibnamefont {Hu}}, \bibinfo {author}
  {\bibfnamefont {Q.}~\bibnamefont {Sun}}, \ and\ \bibinfo {author}
  {\bibfnamefont {G.~K.-L.}\ \bibnamefont {Chan}},\ }\bibfield  {title}
  {\enquote {\bibinfo {title} {N-{{Electron Valence State Perturbation Theory
  Based}} on a {{Density Matrix Renormalization Group Reference Function}},
  with {{Applications}} to the {{Chromium Dimer}} and a {{Trimer Model}} of
  {{Poly}}(p-{{Phenylenevinylene}}).}}\ }\href {\doibase
  10.1021/acs.jctc.5b01225} {\bibfield  {journal} {\bibinfo  {journal} {J.
  Chem. Theory Comput.}\ }\textbf {\bibinfo {volume} {12}},\ \bibinfo {pages}
  {1583--1591} (\bibinfo {year} {2016})}\BibitemShut {NoStop}%
\bibitem [{\citenamefont {Mussard}\ and\ \citenamefont
  {Sharma}(2018)}]{Mussard18}%
  \BibitemOpen
  \bibfield  {author} {\bibinfo {author} {\bibfnamefont {B.}~\bibnamefont
  {Mussard}}\ and\ \bibinfo {author} {\bibfnamefont {S.}~\bibnamefont
  {Sharma}},\ }\bibfield  {title} {\enquote {\bibinfo {title} {One-step
  treatment of spin-orbit coupling and electron correlation in large active
  spaces},}\ }\href@noop {} {\bibfield  {journal} {\bibinfo  {journal} {J.
  Chem. Theory Comput.}\ }\textbf {\bibinfo {volume} {14}},\ \bibinfo {pages}
  {154--165} (\bibinfo {year} {2018})}\BibitemShut {NoStop}%
\bibitem [{\citenamefont {Yeh}\ \emph {et~al.}(2022)\citenamefont {Yeh},
  \citenamefont {Shee}, \citenamefont {Sun}, \citenamefont {Gull},\ and\
  \citenamefont {Zgid}}]{Yeh22}%
  \BibitemOpen
  \bibfield  {author} {\bibinfo {author} {\bibfnamefont {C.-N.}\ \bibnamefont
  {Yeh}}, \bibinfo {author} {\bibfnamefont {A.}~\bibnamefont {Shee}}, \bibinfo
  {author} {\bibfnamefont {Q.}~\bibnamefont {Sun}}, \bibinfo {author}
  {\bibfnamefont {E.}~\bibnamefont {Gull}}, \ and\ \bibinfo {author}
  {\bibfnamefont {D.}~\bibnamefont {Zgid}},\ }\bibfield  {title} {\enquote
  {\bibinfo {title} {{Relativistic self-consistent $GW$: Exact two-component
  formalism with one-electron approximation for solids}},}\ }\href {\doibase
  10.1103/PhysRevB.106.085121} {\bibfield  {journal} {\bibinfo  {journal}
  {Phys. Rev. B}\ }\textbf {\bibinfo {volume} {106}},\ \bibinfo {pages} {85121}
  (\bibinfo {year} {2022})}\BibitemShut {NoStop}%
\bibitem [{\citenamefont {Majumder}\ and\ \citenamefont
  {Sokolov}(2024)}]{Majumder24}%
  \BibitemOpen
  \bibfield  {author} {\bibinfo {author} {\bibfnamefont {R.}~\bibnamefont
  {Majumder}}\ and\ \bibinfo {author} {\bibfnamefont {A.~Y.}\ \bibnamefont
  {Sokolov}},\ }\bibfield  {title} {\enquote {\bibinfo {title} {{Consistent
  Second-Order Treatment of Spin–Orbit Coupling and Dynamic Correlation in
  Quasidegenerate N-Electron Valence Perturbation Theory}},}\ }\href {\doibase
  10.1021/acs.jctc.4c00458} {\bibfield  {journal} {\bibinfo  {journal} {J.
  Chem. Theory Comput.}\ }\textbf {\bibinfo {volume} {20}},\ \bibinfo {pages}
  {4676--4688} (\bibinfo {year} {2024})}\BibitemShut {NoStop}%
\bibitem [{\citenamefont {Sun}\ \emph {et~al.}(2020)\citenamefont {Sun},
  \citenamefont {Zhang}, \citenamefont {Banerjee}, \citenamefont {Bao},
  \citenamefont {Barbry}, \citenamefont {Blunt}, \citenamefont {Bogdanov},
  \citenamefont {Booth}, \citenamefont {Chen}, \citenamefont {Cui},
  \citenamefont {Eriksen}, \citenamefont {Gao}, \citenamefont {Guo},
  \citenamefont {Hermann}, \citenamefont {Hermes}, \citenamefont {Koh},
  \citenamefont {Koval}, \citenamefont {Lehtola}, \citenamefont {Li},
  \citenamefont {Liu}, \citenamefont {Mardirossian}, \citenamefont {McClain},
  \citenamefont {Motta}, \citenamefont {Mussard}, \citenamefont {Pham},
  \citenamefont {Pulkin}, \citenamefont {Purwanto}, \citenamefont {Robinson},
  \citenamefont {Ronca}, \citenamefont {Sayfutyarova}, \citenamefont
  {Scheurer}, \citenamefont {Schurkus}, \citenamefont {Smith}, \citenamefont
  {Sun}, \citenamefont {Sun}, \citenamefont {Upadhyay}, \citenamefont {Wagner},
  \citenamefont {Wang}, \citenamefont {White}, \citenamefont {Whitfield},
  \citenamefont {Williamson}, \citenamefont {Wouters}, \citenamefont {Yang},
  \citenamefont {Yu}, \citenamefont {Zhu}, \citenamefont {Berkelbach},
  \citenamefont {Sharma}, \citenamefont {Sokolov},\ and\ \citenamefont
  {Chan}}]{Sun20}%
  \BibitemOpen
  \bibfield  {author} {\bibinfo {author} {\bibfnamefont {Q.}~\bibnamefont
  {Sun}}, \bibinfo {author} {\bibfnamefont {X.}~\bibnamefont {Zhang}}, \bibinfo
  {author} {\bibfnamefont {S.}~\bibnamefont {Banerjee}}, \bibinfo {author}
  {\bibfnamefont {P.}~\bibnamefont {Bao}}, \bibinfo {author} {\bibfnamefont
  {M.}~\bibnamefont {Barbry}}, \bibinfo {author} {\bibfnamefont {N.~S.}\
  \bibnamefont {Blunt}}, \bibinfo {author} {\bibfnamefont {N.~A.}\ \bibnamefont
  {Bogdanov}}, \bibinfo {author} {\bibfnamefont {G.~H.}\ \bibnamefont {Booth}},
  \bibinfo {author} {\bibfnamefont {J.}~\bibnamefont {Chen}}, \bibinfo {author}
  {\bibfnamefont {Z.-H.}\ \bibnamefont {Cui}}, \bibinfo {author} {\bibfnamefont
  {J.~J.}\ \bibnamefont {Eriksen}}, \bibinfo {author} {\bibfnamefont
  {Y.}~\bibnamefont {Gao}}, \bibinfo {author} {\bibfnamefont {S.}~\bibnamefont
  {Guo}}, \bibinfo {author} {\bibfnamefont {J.}~\bibnamefont {Hermann}},
  \bibinfo {author} {\bibfnamefont {M.~R.}\ \bibnamefont {Hermes}}, \bibinfo
  {author} {\bibfnamefont {K.}~\bibnamefont {Koh}}, \bibinfo {author}
  {\bibfnamefont {P.}~\bibnamefont {Koval}}, \bibinfo {author} {\bibfnamefont
  {S.}~\bibnamefont {Lehtola}}, \bibinfo {author} {\bibfnamefont
  {Z.}~\bibnamefont {Li}}, \bibinfo {author} {\bibfnamefont {J.}~\bibnamefont
  {Liu}}, \bibinfo {author} {\bibfnamefont {N.}~\bibnamefont {Mardirossian}},
  \bibinfo {author} {\bibfnamefont {J.~D.}\ \bibnamefont {McClain}}, \bibinfo
  {author} {\bibfnamefont {M.}~\bibnamefont {Motta}}, \bibinfo {author}
  {\bibfnamefont {B.}~\bibnamefont {Mussard}}, \bibinfo {author} {\bibfnamefont
  {H.~Q.}\ \bibnamefont {Pham}}, \bibinfo {author} {\bibfnamefont
  {A.}~\bibnamefont {Pulkin}}, \bibinfo {author} {\bibfnamefont
  {W.}~\bibnamefont {Purwanto}}, \bibinfo {author} {\bibfnamefont {P.~J.}\
  \bibnamefont {Robinson}}, \bibinfo {author} {\bibfnamefont {E.}~\bibnamefont
  {Ronca}}, \bibinfo {author} {\bibfnamefont {E.~R.}\ \bibnamefont
  {Sayfutyarova}}, \bibinfo {author} {\bibfnamefont {M.}~\bibnamefont
  {Scheurer}}, \bibinfo {author} {\bibfnamefont {H.~F.}\ \bibnamefont
  {Schurkus}}, \bibinfo {author} {\bibfnamefont {J.~E.~T.}\ \bibnamefont
  {Smith}}, \bibinfo {author} {\bibfnamefont {C.}~\bibnamefont {Sun}}, \bibinfo
  {author} {\bibfnamefont {S.-N.}\ \bibnamefont {Sun}}, \bibinfo {author}
  {\bibfnamefont {S.}~\bibnamefont {Upadhyay}}, \bibinfo {author}
  {\bibfnamefont {L.~K.}\ \bibnamefont {Wagner}}, \bibinfo {author}
  {\bibfnamefont {X.}~\bibnamefont {Wang}}, \bibinfo {author} {\bibfnamefont
  {A.}~\bibnamefont {White}}, \bibinfo {author} {\bibfnamefont {J.~D.}\
  \bibnamefont {Whitfield}}, \bibinfo {author} {\bibfnamefont {M.~J.}\
  \bibnamefont {Williamson}}, \bibinfo {author} {\bibfnamefont
  {S.}~\bibnamefont {Wouters}}, \bibinfo {author} {\bibfnamefont
  {J.}~\bibnamefont {Yang}}, \bibinfo {author} {\bibfnamefont {J.~M.}\
  \bibnamefont {Yu}}, \bibinfo {author} {\bibfnamefont {T.}~\bibnamefont
  {Zhu}}, \bibinfo {author} {\bibfnamefont {T.~C.}\ \bibnamefont {Berkelbach}},
  \bibinfo {author} {\bibfnamefont {S.}~\bibnamefont {Sharma}}, \bibinfo
  {author} {\bibfnamefont {A.~Y.}\ \bibnamefont {Sokolov}}, \ and\ \bibinfo
  {author} {\bibfnamefont {G.~K.-L.}\ \bibnamefont {Chan}},\ }\bibfield
  {title} {\enquote {\bibinfo {title} {Recent developments in the {{PySCF}}
  program package},}\ }\href {\doibase 10.1063/5.0006074} {\bibfield  {journal}
  {\bibinfo  {journal} {J. Chem. Phys.}\ }\textbf {\bibinfo {volume} {153}},\
  \bibinfo {pages} {024109} (\bibinfo {year} {2020})}\BibitemShut {NoStop}%
\bibitem [{\citenamefont {Repisky}\ \emph {et~al.}(2020)\citenamefont
  {Repisky}, \citenamefont {Komorovsky}, \citenamefont {Kadek}, \citenamefont
  {Konecny}, \citenamefont {Ekstr{\"{o}}m}, \citenamefont {Malkin},
  \citenamefont {Kaupp}, \citenamefont {Ruud}, \citenamefont {Malkina},\ and\
  \citenamefont {Malkin}}]{Repisky20}%
  \BibitemOpen
  \bibfield  {author} {\bibinfo {author} {\bibfnamefont {M.}~\bibnamefont
  {Repisky}}, \bibinfo {author} {\bibfnamefont {S.}~\bibnamefont {Komorovsky}},
  \bibinfo {author} {\bibfnamefont {M.}~\bibnamefont {Kadek}}, \bibinfo
  {author} {\bibfnamefont {L.}~\bibnamefont {Konecny}}, \bibinfo {author}
  {\bibfnamefont {U.}~\bibnamefont {Ekstr{\"{o}}m}}, \bibinfo {author}
  {\bibfnamefont {E.}~\bibnamefont {Malkin}}, \bibinfo {author} {\bibfnamefont
  {M.}~\bibnamefont {Kaupp}}, \bibinfo {author} {\bibfnamefont
  {K.}~\bibnamefont {Ruud}}, \bibinfo {author} {\bibfnamefont {O.~L.}\
  \bibnamefont {Malkina}}, \ and\ \bibinfo {author} {\bibfnamefont {V.~G.}\
  \bibnamefont {Malkin}},\ }\bibfield  {title} {\enquote {\bibinfo {title}
  {{ReSpect: Relativistic spectroscopy DFT program package}},}\ }\href
  {\doibase 10.1063/5.0005094} {\bibfield  {journal} {\bibinfo  {journal} {J.
  Chem. Phys.}\ }\textbf {\bibinfo {volume} {152}},\ \bibinfo {pages} {184101}
  (\bibinfo {year} {2020})}\BibitemShut {NoStop}%
\bibitem [{\citenamefont {Birnoschi}\ and\ \citenamefont
  {Chilton}(2022)}]{Birnoschi22}%
  \BibitemOpen
  \bibfield  {author} {\bibinfo {author} {\bibfnamefont {L.}~\bibnamefont
  {Birnoschi}}\ and\ \bibinfo {author} {\bibfnamefont {N.~F.}\ \bibnamefont
  {Chilton}},\ }\bibfield  {title} {\enquote {\bibinfo {title} {{Hyperion: A
  New Computational Tool for Relativistic Ab Initio Hyperfine Coupling}},}\
  }\href {\doibase 10.1021/acs.jctc.2c00257} {\bibfield  {journal} {\bibinfo
  {journal} {J. Chem. Theory Comput.}\ }\textbf {\bibinfo {volume} {18}},\
  \bibinfo {pages} {4719--4732} (\bibinfo {year} {2022})}\BibitemShut {NoStop}%
\bibitem [{\citenamefont {Cunha}\ \emph {et~al.}(2022)\citenamefont {Cunha},
  \citenamefont {Hait}, \citenamefont {Kang}, \citenamefont {Mao},\ and\
  \citenamefont {Head-Gordon}}]{Cunha22}%
  \BibitemOpen
  \bibfield  {author} {\bibinfo {author} {\bibfnamefont {L.~A.}\ \bibnamefont
  {Cunha}}, \bibinfo {author} {\bibfnamefont {D.}~\bibnamefont {Hait}},
  \bibinfo {author} {\bibfnamefont {R.}~\bibnamefont {Kang}}, \bibinfo {author}
  {\bibfnamefont {Y.}~\bibnamefont {Mao}}, \ and\ \bibinfo {author}
  {\bibfnamefont {M.}~\bibnamefont {Head-Gordon}},\ }\bibfield  {title}
  {\enquote {\bibinfo {title} {{Relativistic Orbital-Optimized Density
  Functional Theory for Accurate Core-Level Spectroscopy}},}\ }\href {\doibase
  10.1021/acs.jpclett.2c00578} {\bibfield  {journal} {\bibinfo  {journal} {J.
  Phys. Chem. Lett.}\ }\textbf {\bibinfo {volume} {13}},\ \bibinfo {pages}
  {3438--3449} (\bibinfo {year} {2022})}\BibitemShut {NoStop}%
\bibitem [{\citenamefont {Epifanovsky}\ \emph {et~al.}(2021)\citenamefont
  {Epifanovsky}, \citenamefont {Gilbert}, \citenamefont {Feng}, \citenamefont
  {Lee}, \citenamefont {Mao}, \citenamefont {Mardirossian}, \citenamefont
  {Pokhilko}, \citenamefont {White}, \citenamefont {Coons}, \citenamefont
  {Dempwolff}, \citenamefont {Gan}, \citenamefont {Hait}, \citenamefont {Horn},
  \citenamefont {Jacobson}, \citenamefont {Kaliman}, \citenamefont {Kussmann},
  \citenamefont {Lange}, \citenamefont {Lao}, \citenamefont {Levine},
  \citenamefont {Liu}, \citenamefont {McKenzie}, \citenamefont {Morrison},
  \citenamefont {Nanda}, \citenamefont {Plasser}, \citenamefont {Rehn},
  \citenamefont {Vidal}, \citenamefont {You}, \citenamefont {Zhu},
  \citenamefont {Alam}, \citenamefont {Albrecht}, \citenamefont {Aldossary},
  \citenamefont {Alguire}, \citenamefont {Andersen}, \citenamefont {Athavale},
  \citenamefont {Barton}, \citenamefont {Begam}, \citenamefont {Behn},
  \citenamefont {Bellonzi}, \citenamefont {Bernard}, \citenamefont {Berquist},
  \citenamefont {Burton}, \citenamefont {Carreras}, \citenamefont
  {Carter-Fenk}, \citenamefont {Chakraborty}, \citenamefont {Chien},
  \citenamefont {Closser}, \citenamefont {Cofer-Shabica}, \citenamefont
  {Dasgupta}, \citenamefont {de~Wergifosse}, \citenamefont {Deng},
  \citenamefont {Diedenhofen}, \citenamefont {Do}, \citenamefont {Ehlert},
  \citenamefont {Fang}, \citenamefont {Fatehi}, \citenamefont {Feng},
  \citenamefont {Friedhoff}, \citenamefont {Gayvert}, \citenamefont {Ge},
  \citenamefont {Gidofalvi}, \citenamefont {Goldey}, \citenamefont {Gomes},
  \citenamefont {Gonz{\'{a}}lez-Espinoza}, \citenamefont {Gulania},
  \citenamefont {Gunina}, \citenamefont {Hanson-Heine}, \citenamefont
  {Harbach}, \citenamefont {Hauser}, \citenamefont {Herbst}, \citenamefont
  {{Hern{\'{a}}ndez Vera}}, \citenamefont {Hodecker}, \citenamefont {Holden},
  \citenamefont {Houck}, \citenamefont {Huang}, \citenamefont {Hui},
  \citenamefont {Huynh}, \citenamefont {Ivanov}, \citenamefont {J{\'{a}}sz},
  \citenamefont {Ji}, \citenamefont {Jiang}, \citenamefont {Kaduk},
  \citenamefont {K{\"{a}}hler}, \citenamefont {Khistyaev}, \citenamefont {Kim},
  \citenamefont {Kis}, \citenamefont {Klunzinger}, \citenamefont
  {Koczor-Benda}, \citenamefont {Koh}, \citenamefont {Kosenkov}, \citenamefont
  {Koulias}, \citenamefont {Kowalczyk}, \citenamefont {Krauter}, \citenamefont
  {Kue}, \citenamefont {Kunitsa}, \citenamefont {Kus}, \citenamefont
  {Ladj{\'{a}}nszki}, \citenamefont {Landau}, \citenamefont {Lawler},
  \citenamefont {Lefrancois}, \citenamefont {Lehtola}, \citenamefont {Li},
  \citenamefont {Li}, \citenamefont {Liang}, \citenamefont {Liebenthal},
  \citenamefont {Lin}, \citenamefont {Lin}, \citenamefont {Liu}, \citenamefont
  {Liu}, \citenamefont {Loipersberger}, \citenamefont {Luenser}, \citenamefont
  {Manjanath}, \citenamefont {Manohar}, \citenamefont {Mansoor}, \citenamefont
  {Manzer}, \citenamefont {Mao}, \citenamefont {Marenich}, \citenamefont
  {Markovich}, \citenamefont {Mason}, \citenamefont {Maurer}, \citenamefont
  {McLaughlin}, \citenamefont {Menger}, \citenamefont {Mewes}, \citenamefont
  {Mewes}, \citenamefont {Morgante}, \citenamefont {Mullinax}, \citenamefont
  {Oosterbaan}, \citenamefont {Paran}, \citenamefont {Paul}, \citenamefont
  {Paul}, \citenamefont {Pavo{\v{s}}evi{\'{c}}}, \citenamefont {Pei},
  \citenamefont {Prager}, \citenamefont {Proynov}, \citenamefont {R{\'{a}}k},
  \citenamefont {Ramos-Cordoba}, \citenamefont {Rana}, \citenamefont {Rask},
  \citenamefont {Rettig}, \citenamefont {Richard}, \citenamefont {Rob},
  \citenamefont {Rossomme}, \citenamefont {Scheele}, \citenamefont {Scheurer},
  \citenamefont {Schneider}, \citenamefont {Sergueev}, \citenamefont {Sharada},
  \citenamefont {Skomorowski}, \citenamefont {Small}, \citenamefont {Stein},
  \citenamefont {Su}, \citenamefont {Sundstrom}, \citenamefont {Tao},
  \citenamefont {Thirman}, \citenamefont {Tornai}, \citenamefont {Tsuchimochi},
  \citenamefont {Tubman}, \citenamefont {Veccham}, \citenamefont {Vydrov},
  \citenamefont {Wenzel}, \citenamefont {Witte}, \citenamefont {Yamada},
  \citenamefont {Yao}, \citenamefont {Yeganeh}, \citenamefont {Yost},
  \citenamefont {Zech}, \citenamefont {Zhang}, \citenamefont {Zhang},
  \citenamefont {Zhang}, \citenamefont {Zuev}, \citenamefont {Aspuru-Guzik},
  \citenamefont {Bell}, \citenamefont {Besley}, \citenamefont {Bravaya},
  \citenamefont {Brooks}, \citenamefont {Casanova}, \citenamefont {Chai},
  \citenamefont {Coriani}, \citenamefont {Cramer}, \citenamefont {Cserey},
  \citenamefont {{DePrince III}}, \citenamefont {{DiStasio Jr.}}, \citenamefont
  {Dreuw}, \citenamefont {Dunietz}, \citenamefont {Furlani}, \citenamefont
  {{Goddard III}}, \citenamefont {Hammes-Schiffer}, \citenamefont
  {Head-Gordon}, \citenamefont {Hehre}, \citenamefont {Hsu}, \citenamefont
  {Jagau}, \citenamefont {Jung}, \citenamefont {Klamt}, \citenamefont {Kong},
  \citenamefont {Lambrecht}, \citenamefont {Liang}, \citenamefont {Mayhall},
  \citenamefont {McCurdy}, \citenamefont {Neaton}, \citenamefont {Ochsenfeld},
  \citenamefont {Parkhill}, \citenamefont {Peverati}, \citenamefont {Rassolov},
  \citenamefont {Shao}, \citenamefont {Slipchenko}, \citenamefont {Stauch},
  \citenamefont {Steele}, \citenamefont {Subotnik}, \citenamefont {Thom},
  \citenamefont {Tkatchenko}, \citenamefont {Truhlar}, \citenamefont {{Van
  Voorhis}}, \citenamefont {Wesolowski}, \citenamefont {Whaley}, \citenamefont
  {{Woodcock III}}, \citenamefont {Zimmerman}, \citenamefont {Faraji},
  \citenamefont {Gill}, \citenamefont {Head-Gordon}, \citenamefont {Herbert},\
  and\ \citenamefont {Krylov}}]{Epifanovsky21}%
  \BibitemOpen
  \bibfield  {author} {\bibinfo {author} {\bibfnamefont {E.}~\bibnamefont
  {Epifanovsky}}, \bibinfo {author} {\bibfnamefont {A.~T.~B.}\ \bibnamefont
  {Gilbert}}, \bibinfo {author} {\bibfnamefont {X.}~\bibnamefont {Feng}},
  \bibinfo {author} {\bibfnamefont {J.}~\bibnamefont {Lee}}, \bibinfo {author}
  {\bibfnamefont {Y.}~\bibnamefont {Mao}}, \bibinfo {author} {\bibfnamefont
  {N.}~\bibnamefont {Mardirossian}}, \bibinfo {author} {\bibfnamefont
  {P.}~\bibnamefont {Pokhilko}}, \bibinfo {author} {\bibfnamefont {A.~F.}\
  \bibnamefont {White}}, \bibinfo {author} {\bibfnamefont {M.~P.}\ \bibnamefont
  {Coons}}, \bibinfo {author} {\bibfnamefont {A.~L.}\ \bibnamefont
  {Dempwolff}}, \bibinfo {author} {\bibfnamefont {Z.}~\bibnamefont {Gan}},
  \bibinfo {author} {\bibfnamefont {D.}~\bibnamefont {Hait}}, \bibinfo {author}
  {\bibfnamefont {P.~R.}\ \bibnamefont {Horn}}, \bibinfo {author}
  {\bibfnamefont {L.~D.}\ \bibnamefont {Jacobson}}, \bibinfo {author}
  {\bibfnamefont {I.}~\bibnamefont {Kaliman}}, \bibinfo {author} {\bibfnamefont
  {J.}~\bibnamefont {Kussmann}}, \bibinfo {author} {\bibfnamefont {A.~W.}\
  \bibnamefont {Lange}}, \bibinfo {author} {\bibfnamefont {K.~U.}\ \bibnamefont
  {Lao}}, \bibinfo {author} {\bibfnamefont {D.~S.}\ \bibnamefont {Levine}},
  \bibinfo {author} {\bibfnamefont {J.}~\bibnamefont {Liu}}, \bibinfo {author}
  {\bibfnamefont {S.~C.}\ \bibnamefont {McKenzie}}, \bibinfo {author}
  {\bibfnamefont {A.~F.}\ \bibnamefont {Morrison}}, \bibinfo {author}
  {\bibfnamefont {K.~D.}\ \bibnamefont {Nanda}}, \bibinfo {author}
  {\bibfnamefont {F.}~\bibnamefont {Plasser}}, \bibinfo {author} {\bibfnamefont
  {D.~R.}\ \bibnamefont {Rehn}}, \bibinfo {author} {\bibfnamefont {M.~L.}\
  \bibnamefont {Vidal}}, \bibinfo {author} {\bibfnamefont {Z.-Q.}\ \bibnamefont
  {You}}, \bibinfo {author} {\bibfnamefont {Y.}~\bibnamefont {Zhu}}, \bibinfo
  {author} {\bibfnamefont {B.}~\bibnamefont {Alam}}, \bibinfo {author}
  {\bibfnamefont {B.~J.}\ \bibnamefont {Albrecht}}, \bibinfo {author}
  {\bibfnamefont {A.}~\bibnamefont {Aldossary}}, \bibinfo {author}
  {\bibfnamefont {E.}~\bibnamefont {Alguire}}, \bibinfo {author} {\bibfnamefont
  {J.~H.}\ \bibnamefont {Andersen}}, \bibinfo {author} {\bibfnamefont
  {V.}~\bibnamefont {Athavale}}, \bibinfo {author} {\bibfnamefont
  {D.}~\bibnamefont {Barton}}, \bibinfo {author} {\bibfnamefont
  {K.}~\bibnamefont {Begam}}, \bibinfo {author} {\bibfnamefont
  {A.}~\bibnamefont {Behn}}, \bibinfo {author} {\bibfnamefont {N.}~\bibnamefont
  {Bellonzi}}, \bibinfo {author} {\bibfnamefont {Y.~A.}\ \bibnamefont
  {Bernard}}, \bibinfo {author} {\bibfnamefont {E.~J.}\ \bibnamefont
  {Berquist}}, \bibinfo {author} {\bibfnamefont {H.~G.~A.}\ \bibnamefont
  {Burton}}, \bibinfo {author} {\bibfnamefont {A.}~\bibnamefont {Carreras}},
  \bibinfo {author} {\bibfnamefont {K.}~\bibnamefont {Carter-Fenk}}, \bibinfo
  {author} {\bibfnamefont {R.}~\bibnamefont {Chakraborty}}, \bibinfo {author}
  {\bibfnamefont {A.~D.}\ \bibnamefont {Chien}}, \bibinfo {author}
  {\bibfnamefont {K.~D.}\ \bibnamefont {Closser}}, \bibinfo {author}
  {\bibfnamefont {V.}~\bibnamefont {Cofer-Shabica}}, \bibinfo {author}
  {\bibfnamefont {S.}~\bibnamefont {Dasgupta}}, \bibinfo {author}
  {\bibfnamefont {M.}~\bibnamefont {de~Wergifosse}}, \bibinfo {author}
  {\bibfnamefont {J.}~\bibnamefont {Deng}}, \bibinfo {author} {\bibfnamefont
  {M.}~\bibnamefont {Diedenhofen}}, \bibinfo {author} {\bibfnamefont
  {H.}~\bibnamefont {Do}}, \bibinfo {author} {\bibfnamefont {S.}~\bibnamefont
  {Ehlert}}, \bibinfo {author} {\bibfnamefont {P.-T.}\ \bibnamefont {Fang}},
  \bibinfo {author} {\bibfnamefont {S.}~\bibnamefont {Fatehi}}, \bibinfo
  {author} {\bibfnamefont {Q.}~\bibnamefont {Feng}}, \bibinfo {author}
  {\bibfnamefont {T.}~\bibnamefont {Friedhoff}}, \bibinfo {author}
  {\bibfnamefont {J.}~\bibnamefont {Gayvert}}, \bibinfo {author} {\bibfnamefont
  {Q.}~\bibnamefont {Ge}}, \bibinfo {author} {\bibfnamefont {G.}~\bibnamefont
  {Gidofalvi}}, \bibinfo {author} {\bibfnamefont {M.}~\bibnamefont {Goldey}},
  \bibinfo {author} {\bibfnamefont {J.}~\bibnamefont {Gomes}}, \bibinfo
  {author} {\bibfnamefont {C.~E.}\ \bibnamefont {Gonz{\'{a}}lez-Espinoza}},
  \bibinfo {author} {\bibfnamefont {S.}~\bibnamefont {Gulania}}, \bibinfo
  {author} {\bibfnamefont {A.~O.}\ \bibnamefont {Gunina}}, \bibinfo {author}
  {\bibfnamefont {M.~W.~D.}\ \bibnamefont {Hanson-Heine}}, \bibinfo {author}
  {\bibfnamefont {P.~H.~P.}\ \bibnamefont {Harbach}}, \bibinfo {author}
  {\bibfnamefont {A.}~\bibnamefont {Hauser}}, \bibinfo {author} {\bibfnamefont
  {M.~F.}\ \bibnamefont {Herbst}}, \bibinfo {author} {\bibfnamefont
  {M.}~\bibnamefont {{Hern{\'{a}}ndez Vera}}}, \bibinfo {author} {\bibfnamefont
  {M.}~\bibnamefont {Hodecker}}, \bibinfo {author} {\bibfnamefont {Z.~C.}\
  \bibnamefont {Holden}}, \bibinfo {author} {\bibfnamefont {S.}~\bibnamefont
  {Houck}}, \bibinfo {author} {\bibfnamefont {X.}~\bibnamefont {Huang}},
  \bibinfo {author} {\bibfnamefont {K.}~\bibnamefont {Hui}}, \bibinfo {author}
  {\bibfnamefont {B.~C.}\ \bibnamefont {Huynh}}, \bibinfo {author}
  {\bibfnamefont {M.}~\bibnamefont {Ivanov}}, \bibinfo {author} {\bibfnamefont
  {{\'{A}}.}~\bibnamefont {J{\'{a}}sz}}, \bibinfo {author} {\bibfnamefont
  {H.}~\bibnamefont {Ji}}, \bibinfo {author} {\bibfnamefont {H.}~\bibnamefont
  {Jiang}}, \bibinfo {author} {\bibfnamefont {B.}~\bibnamefont {Kaduk}},
  \bibinfo {author} {\bibfnamefont {S.}~\bibnamefont {K{\"{a}}hler}}, \bibinfo
  {author} {\bibfnamefont {K.}~\bibnamefont {Khistyaev}}, \bibinfo {author}
  {\bibfnamefont {J.}~\bibnamefont {Kim}}, \bibinfo {author} {\bibfnamefont
  {G.}~\bibnamefont {Kis}}, \bibinfo {author} {\bibfnamefont {P.}~\bibnamefont
  {Klunzinger}}, \bibinfo {author} {\bibfnamefont {Z.}~\bibnamefont
  {Koczor-Benda}}, \bibinfo {author} {\bibfnamefont {J.~H.}\ \bibnamefont
  {Koh}}, \bibinfo {author} {\bibfnamefont {D.}~\bibnamefont {Kosenkov}},
  \bibinfo {author} {\bibfnamefont {L.}~\bibnamefont {Koulias}}, \bibinfo
  {author} {\bibfnamefont {T.}~\bibnamefont {Kowalczyk}}, \bibinfo {author}
  {\bibfnamefont {C.~M.}\ \bibnamefont {Krauter}}, \bibinfo {author}
  {\bibfnamefont {K.}~\bibnamefont {Kue}}, \bibinfo {author} {\bibfnamefont
  {A.}~\bibnamefont {Kunitsa}}, \bibinfo {author} {\bibfnamefont
  {T.}~\bibnamefont {Kus}}, \bibinfo {author} {\bibfnamefont {I.}~\bibnamefont
  {Ladj{\'{a}}nszki}}, \bibinfo {author} {\bibfnamefont {A.}~\bibnamefont
  {Landau}}, \bibinfo {author} {\bibfnamefont {K.~V.}\ \bibnamefont {Lawler}},
  \bibinfo {author} {\bibfnamefont {D.}~\bibnamefont {Lefrancois}}, \bibinfo
  {author} {\bibfnamefont {S.}~\bibnamefont {Lehtola}}, \bibinfo {author}
  {\bibfnamefont {R.~R.}\ \bibnamefont {Li}}, \bibinfo {author} {\bibfnamefont
  {Y.-P.}\ \bibnamefont {Li}}, \bibinfo {author} {\bibfnamefont
  {J.}~\bibnamefont {Liang}}, \bibinfo {author} {\bibfnamefont
  {M.}~\bibnamefont {Liebenthal}}, \bibinfo {author} {\bibfnamefont {H.-H.}\
  \bibnamefont {Lin}}, \bibinfo {author} {\bibfnamefont {Y.-S.}\ \bibnamefont
  {Lin}}, \bibinfo {author} {\bibfnamefont {F.}~\bibnamefont {Liu}}, \bibinfo
  {author} {\bibfnamefont {K.-Y.}\ \bibnamefont {Liu}}, \bibinfo {author}
  {\bibfnamefont {M.}~\bibnamefont {Loipersberger}}, \bibinfo {author}
  {\bibfnamefont {A.}~\bibnamefont {Luenser}}, \bibinfo {author} {\bibfnamefont
  {A.}~\bibnamefont {Manjanath}}, \bibinfo {author} {\bibfnamefont
  {P.}~\bibnamefont {Manohar}}, \bibinfo {author} {\bibfnamefont
  {E.}~\bibnamefont {Mansoor}}, \bibinfo {author} {\bibfnamefont {S.~F.}\
  \bibnamefont {Manzer}}, \bibinfo {author} {\bibfnamefont {S.-P.}\
  \bibnamefont {Mao}}, \bibinfo {author} {\bibfnamefont {A.~V.}\ \bibnamefont
  {Marenich}}, \bibinfo {author} {\bibfnamefont {T.}~\bibnamefont {Markovich}},
  \bibinfo {author} {\bibfnamefont {S.}~\bibnamefont {Mason}}, \bibinfo
  {author} {\bibfnamefont {S.~A.}\ \bibnamefont {Maurer}}, \bibinfo {author}
  {\bibfnamefont {P.~F.}\ \bibnamefont {McLaughlin}}, \bibinfo {author}
  {\bibfnamefont {M.~F. S.~J.}\ \bibnamefont {Menger}}, \bibinfo {author}
  {\bibfnamefont {J.-M.}\ \bibnamefont {Mewes}}, \bibinfo {author}
  {\bibfnamefont {S.~A.}\ \bibnamefont {Mewes}}, \bibinfo {author}
  {\bibfnamefont {P.}~\bibnamefont {Morgante}}, \bibinfo {author}
  {\bibfnamefont {J.~W.}\ \bibnamefont {Mullinax}}, \bibinfo {author}
  {\bibfnamefont {K.~J.}\ \bibnamefont {Oosterbaan}}, \bibinfo {author}
  {\bibfnamefont {G.}~\bibnamefont {Paran}}, \bibinfo {author} {\bibfnamefont
  {A.~C.}\ \bibnamefont {Paul}}, \bibinfo {author} {\bibfnamefont {S.~K.}\
  \bibnamefont {Paul}}, \bibinfo {author} {\bibfnamefont {F.}~\bibnamefont
  {Pavo{\v{s}}evi{\'{c}}}}, \bibinfo {author} {\bibfnamefont {Z.}~\bibnamefont
  {Pei}}, \bibinfo {author} {\bibfnamefont {S.}~\bibnamefont {Prager}},
  \bibinfo {author} {\bibfnamefont {E.~I.}\ \bibnamefont {Proynov}}, \bibinfo
  {author} {\bibfnamefont {{\'{A}}.}~\bibnamefont {R{\'{a}}k}}, \bibinfo
  {author} {\bibfnamefont {E.}~\bibnamefont {Ramos-Cordoba}}, \bibinfo {author}
  {\bibfnamefont {B.}~\bibnamefont {Rana}}, \bibinfo {author} {\bibfnamefont
  {A.~E.}\ \bibnamefont {Rask}}, \bibinfo {author} {\bibfnamefont
  {A.}~\bibnamefont {Rettig}}, \bibinfo {author} {\bibfnamefont {R.~M.}\
  \bibnamefont {Richard}}, \bibinfo {author} {\bibfnamefont {F.}~\bibnamefont
  {Rob}}, \bibinfo {author} {\bibfnamefont {E.}~\bibnamefont {Rossomme}},
  \bibinfo {author} {\bibfnamefont {T.}~\bibnamefont {Scheele}}, \bibinfo
  {author} {\bibfnamefont {M.}~\bibnamefont {Scheurer}}, \bibinfo {author}
  {\bibfnamefont {M.}~\bibnamefont {Schneider}}, \bibinfo {author}
  {\bibfnamefont {N.}~\bibnamefont {Sergueev}}, \bibinfo {author}
  {\bibfnamefont {S.~M.}\ \bibnamefont {Sharada}}, \bibinfo {author}
  {\bibfnamefont {W.}~\bibnamefont {Skomorowski}}, \bibinfo {author}
  {\bibfnamefont {D.~W.}\ \bibnamefont {Small}}, \bibinfo {author}
  {\bibfnamefont {C.~J.}\ \bibnamefont {Stein}}, \bibinfo {author}
  {\bibfnamefont {Y.-C.}\ \bibnamefont {Su}}, \bibinfo {author} {\bibfnamefont
  {E.~J.}\ \bibnamefont {Sundstrom}}, \bibinfo {author} {\bibfnamefont
  {Z.}~\bibnamefont {Tao}}, \bibinfo {author} {\bibfnamefont {J.}~\bibnamefont
  {Thirman}}, \bibinfo {author} {\bibfnamefont {G.~J.}\ \bibnamefont {Tornai}},
  \bibinfo {author} {\bibfnamefont {T.}~\bibnamefont {Tsuchimochi}}, \bibinfo
  {author} {\bibfnamefont {N.~M.}\ \bibnamefont {Tubman}}, \bibinfo {author}
  {\bibfnamefont {S.~P.}\ \bibnamefont {Veccham}}, \bibinfo {author}
  {\bibfnamefont {O.}~\bibnamefont {Vydrov}}, \bibinfo {author} {\bibfnamefont
  {J.}~\bibnamefont {Wenzel}}, \bibinfo {author} {\bibfnamefont
  {J.}~\bibnamefont {Witte}}, \bibinfo {author} {\bibfnamefont
  {A.}~\bibnamefont {Yamada}}, \bibinfo {author} {\bibfnamefont
  {K.}~\bibnamefont {Yao}}, \bibinfo {author} {\bibfnamefont {S.}~\bibnamefont
  {Yeganeh}}, \bibinfo {author} {\bibfnamefont {S.~R.}\ \bibnamefont {Yost}},
  \bibinfo {author} {\bibfnamefont {A.}~\bibnamefont {Zech}}, \bibinfo {author}
  {\bibfnamefont {I.~Y.}\ \bibnamefont {Zhang}}, \bibinfo {author}
  {\bibfnamefont {X.}~\bibnamefont {Zhang}}, \bibinfo {author} {\bibfnamefont
  {Y.}~\bibnamefont {Zhang}}, \bibinfo {author} {\bibfnamefont
  {D.}~\bibnamefont {Zuev}}, \bibinfo {author} {\bibfnamefont {A.}~\bibnamefont
  {Aspuru-Guzik}}, \bibinfo {author} {\bibfnamefont {A.~T.}\ \bibnamefont
  {Bell}}, \bibinfo {author} {\bibfnamefont {N.~A.}\ \bibnamefont {Besley}},
  \bibinfo {author} {\bibfnamefont {K.~B.}\ \bibnamefont {Bravaya}}, \bibinfo
  {author} {\bibfnamefont {B.~R.}\ \bibnamefont {Brooks}}, \bibinfo {author}
  {\bibfnamefont {D.}~\bibnamefont {Casanova}}, \bibinfo {author}
  {\bibfnamefont {J.-D.}\ \bibnamefont {Chai}}, \bibinfo {author}
  {\bibfnamefont {S.}~\bibnamefont {Coriani}}, \bibinfo {author} {\bibfnamefont
  {C.~J.}\ \bibnamefont {Cramer}}, \bibinfo {author} {\bibfnamefont
  {G.}~\bibnamefont {Cserey}}, \bibinfo {author} {\bibfnamefont {A.~E.}\
  \bibnamefont {{DePrince III}}}, \bibinfo {author} {\bibfnamefont {R.~A.}\
  \bibnamefont {{DiStasio Jr.}}}, \bibinfo {author} {\bibfnamefont
  {A.}~\bibnamefont {Dreuw}}, \bibinfo {author} {\bibfnamefont {B.~D.}\
  \bibnamefont {Dunietz}}, \bibinfo {author} {\bibfnamefont {T.~R.}\
  \bibnamefont {Furlani}}, \bibinfo {author} {\bibfnamefont {W.~A.}\
  \bibnamefont {{Goddard III}}}, \bibinfo {author} {\bibfnamefont
  {S.}~\bibnamefont {Hammes-Schiffer}}, \bibinfo {author} {\bibfnamefont
  {T.}~\bibnamefont {Head-Gordon}}, \bibinfo {author} {\bibfnamefont {W.~J.}\
  \bibnamefont {Hehre}}, \bibinfo {author} {\bibfnamefont {C.-P.}\ \bibnamefont
  {Hsu}}, \bibinfo {author} {\bibfnamefont {T.-C.}\ \bibnamefont {Jagau}},
  \bibinfo {author} {\bibfnamefont {Y.}~\bibnamefont {Jung}}, \bibinfo {author}
  {\bibfnamefont {A.}~\bibnamefont {Klamt}}, \bibinfo {author} {\bibfnamefont
  {J.}~\bibnamefont {Kong}}, \bibinfo {author} {\bibfnamefont {D.~S.}\
  \bibnamefont {Lambrecht}}, \bibinfo {author} {\bibfnamefont {W.}~\bibnamefont
  {Liang}}, \bibinfo {author} {\bibfnamefont {N.~J.}\ \bibnamefont {Mayhall}},
  \bibinfo {author} {\bibfnamefont {C.~W.}\ \bibnamefont {McCurdy}}, \bibinfo
  {author} {\bibfnamefont {J.~B.}\ \bibnamefont {Neaton}}, \bibinfo {author}
  {\bibfnamefont {C.}~\bibnamefont {Ochsenfeld}}, \bibinfo {author}
  {\bibfnamefont {J.~A.}\ \bibnamefont {Parkhill}}, \bibinfo {author}
  {\bibfnamefont {R.}~\bibnamefont {Peverati}}, \bibinfo {author}
  {\bibfnamefont {V.~A.}\ \bibnamefont {Rassolov}}, \bibinfo {author}
  {\bibfnamefont {Y.}~\bibnamefont {Shao}}, \bibinfo {author} {\bibfnamefont
  {L.~V.}\ \bibnamefont {Slipchenko}}, \bibinfo {author} {\bibfnamefont
  {T.}~\bibnamefont {Stauch}}, \bibinfo {author} {\bibfnamefont {R.~P.}\
  \bibnamefont {Steele}}, \bibinfo {author} {\bibfnamefont {J.~E.}\
  \bibnamefont {Subotnik}}, \bibinfo {author} {\bibfnamefont {A.~J.~W.}\
  \bibnamefont {Thom}}, \bibinfo {author} {\bibfnamefont {A.}~\bibnamefont
  {Tkatchenko}}, \bibinfo {author} {\bibfnamefont {D.~G.}\ \bibnamefont
  {Truhlar}}, \bibinfo {author} {\bibfnamefont {T.}~\bibnamefont {{Van
  Voorhis}}}, \bibinfo {author} {\bibfnamefont {T.~A.}\ \bibnamefont
  {Wesolowski}}, \bibinfo {author} {\bibfnamefont {K.~B.}\ \bibnamefont
  {Whaley}}, \bibinfo {author} {\bibfnamefont {H.~L.}\ \bibnamefont {{Woodcock
  III}}}, \bibinfo {author} {\bibfnamefont {P.~M.}\ \bibnamefont {Zimmerman}},
  \bibinfo {author} {\bibfnamefont {S.}~\bibnamefont {Faraji}}, \bibinfo
  {author} {\bibfnamefont {P.~M.~W.}\ \bibnamefont {Gill}}, \bibinfo {author}
  {\bibfnamefont {M.}~\bibnamefont {Head-Gordon}}, \bibinfo {author}
  {\bibfnamefont {J.~M.}\ \bibnamefont {Herbert}}, \ and\ \bibinfo {author}
  {\bibfnamefont {A.~I.}\ \bibnamefont {Krylov}},\ }\bibfield  {title}
  {\enquote {\bibinfo {title} {{Software for the frontiers of quantum
  chemistry: An overview of developments in the Q-Chem 5 package}},}\ }\href
  {\doibase 10.1063/5.0055522} {\bibfield  {journal} {\bibinfo  {journal} {J.
  Chem. Phys.}\ }\textbf {\bibinfo {volume} {155}},\ \bibinfo {pages} {84801}
  (\bibinfo {year} {2021})}\BibitemShut {NoStop}%
\bibitem [{\citenamefont {Neese}(2022)}]{Neese22}%
  \BibitemOpen
  \bibfield  {author} {\bibinfo {author} {\bibfnamefont {F.}~\bibnamefont
  {Neese}},\ }\bibfield  {title} {\enquote {\bibinfo {title} {{Software update:
  The ORCA program system—Version 5.0}},}\ }\href {\doibase
  https://doi.org/10.1002/wcms.1606} {\bibfield  {journal} {\bibinfo  {journal}
  {WIREs Comput. Mol. Sci.}\ }\textbf {\bibinfo {volume} {12}},\ \bibinfo
  {pages} {e1606} (\bibinfo {year} {2022})}\BibitemShut {NoStop}%
\bibitem [{\citenamefont {Seino}\ and\ \citenamefont {Nakai}(2012)}]{Seino12}%
  \BibitemOpen
  \bibfield  {author} {\bibinfo {author} {\bibfnamefont {J.}~\bibnamefont
  {Seino}}\ and\ \bibinfo {author} {\bibfnamefont {H.}~\bibnamefont {Nakai}},\
  }\bibfield  {title} {\enquote {\bibinfo {title} {{Local unitary
  transformation method for large-scale two-component relativistic
  calculations. II. Extension to two-electron Coulomb interaction}},}\ }\href
  {\doibase 10.1063/1.4757263} {\bibfield  {journal} {\bibinfo  {journal} {J.
  Chem. Phys.}\ }\textbf {\bibinfo {volume} {137}},\ \bibinfo {pages} {144101}
  (\bibinfo {year} {2012})}\BibitemShut {NoStop}%
\bibitem [{\citenamefont {Zhang}, \citenamefont {Kasper},\ and\ \citenamefont
  {Li}(2020)}]{ZhangT20}%
  \BibitemOpen
  \bibfield  {author} {\bibinfo {author} {\bibfnamefont {T.}~\bibnamefont
  {Zhang}}, \bibinfo {author} {\bibfnamefont {J.~M.}\ \bibnamefont {Kasper}}, \
  and\ \bibinfo {author} {\bibfnamefont {X.}~\bibnamefont {Li}},\ }\bibfield
  {title} {\enquote {\bibinfo {title} {{Chapter Two - Localized relativistic
  two-component methods for ground and excited state calculations}},}\ \
  }(\bibinfo  {publisher} {Elsevier},\ \bibinfo {year} {2020})\ pp.\ \bibinfo
  {pages} {17--37}\BibitemShut {NoStop}%
\bibitem [{\citenamefont {Zhao}\ \emph {et~al.}(2016)\citenamefont {Zhao},
  \citenamefont {Zhang}, \citenamefont {Xiao},\ and\ \citenamefont
  {Liu}}]{Zhao16}%
  \BibitemOpen
  \bibfield  {author} {\bibinfo {author} {\bibfnamefont {R.}~\bibnamefont
  {Zhao}}, \bibinfo {author} {\bibfnamefont {Y.}~\bibnamefont {Zhang}},
  \bibinfo {author} {\bibfnamefont {Y.}~\bibnamefont {Xiao}}, \ and\ \bibinfo
  {author} {\bibfnamefont {W.}~\bibnamefont {Liu}},\ }\bibfield  {title}
  {\enquote {\bibinfo {title} {{Exact two-component relativistic energy band
  theory and application}},}\ }\href {\doibase 10.1063/1.4940140} {\bibfield
  {journal} {\bibinfo  {journal} {J. Chem. Phys.}\ }\textbf {\bibinfo {volume}
  {144}},\ \bibinfo {pages} {44105} (\bibinfo {year} {2016})}\BibitemShut
  {NoStop}%
\bibitem [{\citenamefont {Stanton}\ and\ \citenamefont
  {Havriliak}(1984)}]{Stanton84}%
  \BibitemOpen
  \bibfield  {author} {\bibinfo {author} {\bibfnamefont {R.~E.}\ \bibnamefont
  {Stanton}}\ and\ \bibinfo {author} {\bibfnamefont {S.}~\bibnamefont
  {Havriliak}},\ }\bibfield  {title} {\enquote {\bibinfo {title} {Kinetic
  balance: A partial solution to the problem of variational safety in dirac
  calculations},}\ }\href@noop {} {\bibfield  {journal} {\bibinfo  {journal}
  {J. Chem. Phys.}\ }\textbf {\bibinfo {volume} {81}},\ \bibinfo {pages}
  {1910--1918} (\bibinfo {year} {1984})}\BibitemShut {NoStop}%
\bibitem [{\citenamefont {van Lenthe}, \citenamefont {Snijders},\ and\
  \citenamefont {Baerends}(1996)}]{VanLenthe96a}%
  \BibitemOpen
  \bibfield  {author} {\bibinfo {author} {\bibfnamefont {E.}~\bibnamefont {van
  Lenthe}}, \bibinfo {author} {\bibfnamefont {J.~G.}\ \bibnamefont {Snijders}},
  \ and\ \bibinfo {author} {\bibfnamefont {E.~J.}\ \bibnamefont {Baerends}},\
  }\bibfield  {title} {\enquote {\bibinfo {title} {{The zero‐order regular
  approximation for relativistic effects: The effect of spin–orbit coupling
  in closed shell molecules}},}\ }\href {\doibase 10.1063/1.472460} {\bibfield
  {journal} {\bibinfo  {journal} {J. Chem. Phys.}\ }\textbf {\bibinfo {volume}
  {105}},\ \bibinfo {pages} {6505--6516} (\bibinfo {year} {1996})}\BibitemShut
  {NoStop}%
\bibitem [{\citenamefont {Hess}\ and\ \citenamefont {Kaldor}(2000)}]{Hess00}%
  \BibitemOpen
  \bibfield  {author} {\bibinfo {author} {\bibfnamefont {B.~A.}\ \bibnamefont
  {Hess}}\ and\ \bibinfo {author} {\bibfnamefont {U.}~\bibnamefont {Kaldor}},\
  }\bibfield  {title} {\enquote {\bibinfo {title} {{Relativistic all-electron
  coupled-cluster calculations on Au2 in the framework of the Douglas–Kroll
  transformation}},}\ }\href {\doibase 10.1063/1.480744} {\bibfield  {journal}
  {\bibinfo  {journal} {J. Chem. Phys.}\ }\textbf {\bibinfo {volume} {112}},\
  \bibinfo {pages} {1809--1813} (\bibinfo {year} {2000})}\BibitemShut {NoStop}%
\bibitem [{\citenamefont {Neese}(2005)}]{Neese05}%
  \BibitemOpen
  \bibfield  {author} {\bibinfo {author} {\bibfnamefont {F.}~\bibnamefont
  {Neese}},\ }\bibfield  {title} {\enquote {\bibinfo {title} {Efficient and
  accurate approximations to the molecular spin-orbit coupling operator and
  their use in molecular $g$-tensor calculations},}\ }\href@noop {} {\bibfield
  {journal} {\bibinfo  {journal} {J. Chem. Phys.}\ }\textbf {\bibinfo {volume}
  {122}},\ \bibinfo {pages} {034107} (\bibinfo {year} {2005})}\BibitemShut
  {NoStop}%
\bibitem [{\citenamefont {Epifanovsky}\ \emph {et~al.}(2015)\citenamefont
  {Epifanovsky}, \citenamefont {Klein}, \citenamefont {Stopkowicz},
  \citenamefont {Gauss},\ and\ \citenamefont {Krylov}}]{Epifanovsky15}%
  \BibitemOpen
  \bibfield  {author} {\bibinfo {author} {\bibfnamefont {E.}~\bibnamefont
  {Epifanovsky}}, \bibinfo {author} {\bibfnamefont {K.}~\bibnamefont {Klein}},
  \bibinfo {author} {\bibfnamefont {S.}~\bibnamefont {Stopkowicz}}, \bibinfo
  {author} {\bibfnamefont {J.}~\bibnamefont {Gauss}}, \ and\ \bibinfo {author}
  {\bibfnamefont {A.~I.}\ \bibnamefont {Krylov}},\ }\bibfield  {title}
  {\enquote {\bibinfo {title} {{Spin-orbit couplings within the
  equation-of-motion coupled-cluster framework: Theory, implementation, and
  benchmark calculations}},}\ }\href {\doibase
  http://dx.doi.org/10.1063/1.4927785} {\bibfield  {journal} {\bibinfo
  {journal} {J. Chem. Phys.}\ }\textbf {\bibinfo {volume} {143}},\ \bibinfo
  {pages} {064102} (\bibinfo {year} {2015})}\BibitemShut {NoStop}%
\bibitem [{\citenamefont {Cao}\ \emph {et~al.}(2017)\citenamefont {Cao},
  \citenamefont {Li}, \citenamefont {Wang},\ and\ \citenamefont {Liu}}]{Cao17}%
  \BibitemOpen
  \bibfield  {author} {\bibinfo {author} {\bibfnamefont {Z.}~\bibnamefont
  {Cao}}, \bibinfo {author} {\bibfnamefont {Z.}~\bibnamefont {Li}}, \bibinfo
  {author} {\bibfnamefont {F.}~\bibnamefont {Wang}}, \ and\ \bibinfo {author}
  {\bibfnamefont {W.}~\bibnamefont {Liu}},\ }\bibfield  {title} {\enquote
  {\bibinfo {title} {{Combining the spin-separated exact two-component
  relativistic Hamiltonian with the equation-of-motion coupled-cluster method
  for the treatment of spin–orbit splittings of light and heavy elements}},}\
  }\href {\doibase 10.1039/C6CP07588F} {\bibfield  {journal} {\bibinfo
  {journal} {Phys. Chem. Chem. Phys.}\ }\textbf {\bibinfo {volume} {19}},\
  \bibinfo {pages} {3713--3721} (\bibinfo {year} {2017})}\BibitemShut {NoStop}%
\bibitem [{\citenamefont {Perera}\ \emph {et~al.}(2017)\citenamefont {Perera},
  \citenamefont {Gauss}, \citenamefont {Verma},\ and\ \citenamefont
  {Morales}}]{Perera17}%
  \BibitemOpen
  \bibfield  {author} {\bibinfo {author} {\bibfnamefont {A.}~\bibnamefont
  {Perera}}, \bibinfo {author} {\bibfnamefont {J.}~\bibnamefont {Gauss}},
  \bibinfo {author} {\bibfnamefont {P.}~\bibnamefont {Verma}}, \ and\ \bibinfo
  {author} {\bibfnamefont {J.~A.}\ \bibnamefont {Morales}},\ }\bibfield
  {title} {\enquote {\bibinfo {title} {Benchmark coupled-cluster g-tensor
  calculations with full inclusion of the two-particle spin-orbit
  contributions},}\ }\href@noop {} {\bibfield  {journal} {\bibinfo  {journal}
  {J. Chem. Phys.}\ }\textbf {\bibinfo {volume} {146}},\ \bibinfo {pages}
  {164104} (\bibinfo {year} {2017})}\BibitemShut {NoStop}%
\bibitem [{\citenamefont {Cheng}\ \emph {et~al.}(2018)\citenamefont {Cheng},
  \citenamefont {Wang}, \citenamefont {Stanton},\ and\ \citenamefont
  {Gauss}}]{Cheng18a}%
  \BibitemOpen
  \bibfield  {author} {\bibinfo {author} {\bibfnamefont {L.}~\bibnamefont
  {Cheng}}, \bibinfo {author} {\bibfnamefont {F.}~\bibnamefont {Wang}},
  \bibinfo {author} {\bibfnamefont {J.}~\bibnamefont {Stanton}}, \ and\
  \bibinfo {author} {\bibfnamefont {J.}~\bibnamefont {Gauss}},\ }\bibfield
  {title} {\enquote {\bibinfo {title} {{Perturbative treatment of
  spin-orbit-coupling within spin-free exact two-component theory using
  equation-of-motion coupled-cluster methods}},}\ }\href@noop {} {\bibfield
  {journal} {\bibinfo  {journal} {J. Chem. Phys.}\ }\textbf {\bibinfo {volume}
  {148}},\ \bibinfo {pages} {044108} (\bibinfo {year} {2018})}\BibitemShut
  {NoStop}%
\bibitem [{\citenamefont {Berning}\ \emph {et~al.}(2000)\citenamefont
  {Berning}, \citenamefont {Schweizer}, \citenamefont {Werner}, \citenamefont
  {Knowles},\ and\ \citenamefont {Palmieri}}]{Berning00}%
  \BibitemOpen
  \bibfield  {author} {\bibinfo {author} {\bibfnamefont {A.}~\bibnamefont
  {Berning}}, \bibinfo {author} {\bibfnamefont {M.}~\bibnamefont {Schweizer}},
  \bibinfo {author} {\bibfnamefont {H.-J.}\ \bibnamefont {Werner}}, \bibinfo
  {author} {\bibfnamefont {P.~J.}\ \bibnamefont {Knowles}}, \ and\ \bibinfo
  {author} {\bibfnamefont {P.}~\bibnamefont {Palmieri}},\ }\bibfield  {title}
  {\enquote {\bibinfo {title} {{Spin-orbit matrix elements for internally
  contracted multireference configuration interaction wavefunctions}},}\ }\href
  {\doibase 10.1080/00268970009483386} {\bibfield  {journal} {\bibinfo
  {journal} {Mol. Phys.}\ }\textbf {\bibinfo {volume} {98}},\ \bibinfo {pages}
  {1823--1833} (\bibinfo {year} {2000})}\BibitemShut {NoStop}%
\bibitem [{\citenamefont {Netz}, \citenamefont {Mitrushchenkov},\ and\
  \citenamefont {K\"ohn}(2021)}]{Netz21}%
  \BibitemOpen
  \bibfield  {author} {\bibinfo {author} {\bibfnamefont {J.}~\bibnamefont
  {Netz}}, \bibinfo {author} {\bibfnamefont {A.~O.}\ \bibnamefont
  {Mitrushchenkov}}, \ and\ \bibinfo {author} {\bibfnamefont {A.}~\bibnamefont
  {K\"ohn}},\ }\bibfield  {title} {\enquote {\bibinfo {title} {On the accuracy
  of mean-field spin-orbit operators for 3d transition-metal systems},}\
  }\href@noop {} {\bibfield  {journal} {\bibinfo  {journal} {J. Chem. Theory
  Comput.}\ }\textbf {\bibinfo {volume} {17}},\ \bibinfo {pages} {5530--5537}
  (\bibinfo {year} {2021})}\BibitemShut {NoStop}%
\bibitem [{\citenamefont {Liu}\ \emph {et~al.}(2018)\citenamefont {Liu},
  \citenamefont {Shen}, \citenamefont {Asthana},\ and\ \citenamefont
  {Cheng}}]{Liu18b}%
  \BibitemOpen
  \bibfield  {author} {\bibinfo {author} {\bibfnamefont {J.}~\bibnamefont
  {Liu}}, \bibinfo {author} {\bibfnamefont {Y.}~\bibnamefont {Shen}}, \bibinfo
  {author} {\bibfnamefont {A.}~\bibnamefont {Asthana}}, \ and\ \bibinfo
  {author} {\bibfnamefont {L.}~\bibnamefont {Cheng}},\ }\bibfield  {title}
  {\enquote {\bibinfo {title} {Two-component relativistic coupled-cluster
  methods using mean-field spin-orbit integrals},}\ }\href@noop {} {\bibfield
  {journal} {\bibinfo  {journal} {J. Chem. Phys.}\ }\textbf {\bibinfo {volume}
  {148}},\ \bibinfo {pages} {034106} (\bibinfo {year} {2018})}\BibitemShut
  {NoStop}%
\bibitem [{\citenamefont {Huzinaga}\ and\ \citenamefont
  {Cantu}(1971)}]{Huzinaga71}%
  \BibitemOpen
  \bibfield  {author} {\bibinfo {author} {\bibfnamefont {S.}~\bibnamefont
  {Huzinaga}}\ and\ \bibinfo {author} {\bibfnamefont {A.~A.}\ \bibnamefont
  {Cantu}},\ }\bibfield  {title} {\enquote {\bibinfo {title} {{Theory of
  Separability of Many‐Electron Systems}},}\ }\href {\doibase
  10.1063/1.1675720} {\bibfield  {journal} {\bibinfo  {journal} {J. Chem.
  Phys.}\ }\textbf {\bibinfo {volume} {55}},\ \bibinfo {pages} {5543--5549}
  (\bibinfo {year} {1971})}\BibitemShut {NoStop}%
\bibitem [{\citenamefont {Stanton}\ \emph {et~al.}()\citenamefont {Stanton},
  \citenamefont {Gauss}, \citenamefont {Cheng}, \citenamefont {Harding},
  \citenamefont {Matthews},\ and\ \citenamefont {Szalay}}]{CFOURfull}%
  \BibitemOpen
  \bibfield  {author} {\bibinfo {author} {\bibfnamefont {J.~F.}\ \bibnamefont
  {Stanton}}, \bibinfo {author} {\bibfnamefont {J.}~\bibnamefont {Gauss}},
  \bibinfo {author} {\bibfnamefont {L.}~\bibnamefont {Cheng}}, \bibinfo
  {author} {\bibfnamefont {M.~E.}\ \bibnamefont {Harding}}, \bibinfo {author}
  {\bibfnamefont {D.~A.}\ \bibnamefont {Matthews}}, \ and\ \bibinfo {author}
  {\bibfnamefont {P.~G.}\ \bibnamefont {Szalay}},\ }\href@noop {} {\enquote
  {\bibinfo {title} {{CFOUR, Coupled-Cluster techniques for Computational
  Chemistry, a quantum-chemical program package}},}\ }\bibinfo {note} {{W}ith
  contributions from {A}.{A}. {A}uer, {A}. {A}sthana, {R}.{J}. {B}artlett, {U}.
  {B}enedikt, {C}. {B}erger, {D}.{E}. {B}ernholdt, {S.} {B}laschke, {Y}. {J}.
  {B}omble, {S.} {B}urger, {O}. {C}hristiansen, {D.} Datta, {F}. Engel, {R}.
  Faber, {J.} {G}reiner, {M}. {H}eckert, {O}. {H}eun, {M}. Hilgenberg, {C}.
  {H}uber, {T}.-{C}. {J}agau, {D}. {J}onsson, {J}. {J}us{\'e}lius, {T}. Kirsch,
  {K}. {K}lein, {G}.{M.} Kopper{W}.{J}. {L}auderdale, {F}. {L}ipparini, {J}.
  {L}iu, {T}. {M}etzroth, {L}.{A}. {M}{\"u}ck, {D}.{P}. {O}'{N}eill, {T.}
  {N}ottoli, {D}.{R}. {P}rice, {E}. {P}rochnow, {C}. {P}uzzarini, {K}. {R}uud,
  {F}. {S}chiffmann, {W}. {S}chwalbach, {C}. {S}immons, {S}. {S}topkowicz, {A}.
  {T}ajti, {J}. {V}{\'a}zquez, {F}. {W}ang, {J}.{D}. {W}atts, {C}. {Z}hang,
  {X}. {Z}heng, and the integral packages {MOLECULE} ({J}. {A}lml{\"o}f and
  {P}.{R}. {T}aylor), {PROPS} ({P}.{R}. {T}aylor), {ABACUS} ({T}. {H}elgaker,
  {H}.{J}. {A}a. {J}ensen, {P}. {J}{\o}rgensen, and {J}. {O}lsen), and {ECP}
  routines by {A}. {V}. {M}itin and {C}. van {W}{\"u}llen. {F}or the current
  version, see http://www.cfour.de.}\BibitemShut {Stop}%
\bibitem [{\citenamefont {Sun}(2024)}]{Sun24}%
  \BibitemOpen
  \bibfield  {author} {\bibinfo {author} {\bibfnamefont {Q.}~\bibnamefont
  {Sun}},\ }\bibfield  {title} {\enquote {\bibinfo {title} {The updates in
  {{Libcint}} 6: {{More}} integrals, {{API}} refinements, and {{SIMD}}
  optimization techniques},}\ }\href {\doibase 10.1063/5.0200293} {\bibfield
  {journal} {\bibinfo  {journal} {J. Chem. Phys.}\ }\textbf {\bibinfo {volume}
  {160}},\ \bibinfo {pages} {174116} (\bibinfo {year} {2024})}\BibitemShut
  {NoStop}%
\bibitem [{\citenamefont {Zhang}\ and\ \citenamefont
  {Cheng}(2022{\natexlab{b}})}]{Zhang22b}%
  \BibitemOpen
  \bibfield  {author} {\bibinfo {author} {\bibfnamefont {C.}~\bibnamefont
  {Zhang}}\ and\ \bibinfo {author} {\bibfnamefont {L.}~\bibnamefont {Cheng}},\
  }\bibfield  {title} {\enquote {\bibinfo {title} {{Route to Chemical Accuracy
  for Computational Uranium Thermochemistry}},}\ }\href {\doibase
  10.1021/acs.jctc.2c00812} {\bibfield  {journal} {\bibinfo  {journal} {J.
  Chem. Theory Comput.}\ }\textbf {\bibinfo {volume} {18}},\ \bibinfo {pages}
  {6732--6741} (\bibinfo {year} {2022}{\natexlab{b}})}\BibitemShut {NoStop}%
\bibitem [{\citenamefont {F{\ae}gri}(2001)}]{Faegri01}%
  \BibitemOpen
  \bibfield  {author} {\bibinfo {author} {\bibfnamefont {K.}~\bibnamefont
  {F{\ae}gri}},\ }\bibfield  {title} {\enquote {\bibinfo {title} {{Relativistic
  Gaussian basis sets for the elements K - Uuo}},}\ }\href@noop {} {\bibfield
  {journal} {\bibinfo  {journal} {Theor. Chem. Acc.}\ }\textbf {\bibinfo
  {volume} {105}},\ \bibinfo {pages} {252--258} (\bibinfo {year}
  {2001})}\BibitemShut {NoStop}%
\bibitem [{\citenamefont {Roos}\ \emph {et~al.}(2005)\citenamefont {Roos},
  \citenamefont {Lindh}, \citenamefont {Malmqvist}, \citenamefont {Veryazov},\
  and\ \citenamefont {Widmark}}]{Roos05a}%
  \BibitemOpen
  \bibfield  {author} {\bibinfo {author} {\bibfnamefont {B.~O.}\ \bibnamefont
  {Roos}}, \bibinfo {author} {\bibfnamefont {R.}~\bibnamefont {Lindh}},
  \bibinfo {author} {\bibfnamefont {P.-{\AA}.}\ \bibnamefont {Malmqvist}},
  \bibinfo {author} {\bibfnamefont {V.}~\bibnamefont {Veryazov}}, \ and\
  \bibinfo {author} {\bibfnamefont {P.-O.}\ \bibnamefont {Widmark}},\
  }\bibfield  {title} {\enquote {\bibinfo {title} {{New relativistic ANO basis
  sets for actinide atoms}},}\ }\href {\doibase
  https://doi.org/10.1016/j.cplett.2005.05.011} {\bibfield  {journal} {\bibinfo
   {journal} {Chem. Phys. Lett.}\ }\textbf {\bibinfo {volume} {409}},\ \bibinfo
  {pages} {295--299} (\bibinfo {year} {2005})}\BibitemShut {NoStop}%
\bibitem [{\citenamefont {Peterson}(2015)}]{Peterson15}%
  \BibitemOpen
  \bibfield  {author} {\bibinfo {author} {\bibfnamefont {K.~A.}\ \bibnamefont
  {Peterson}},\ }\bibfield  {title} {\enquote {\bibinfo {title} {{Correlation
  consistent basis sets for actinides. I. The Th and U atoms}},}\ }\href
  {\doibase 10.1063/1.4907596} {\bibfield  {journal} {\bibinfo  {journal} {J.
  Chem. Phys.}\ }\textbf {\bibinfo {volume} {142}},\ \bibinfo {pages} {074105}
  (\bibinfo {year} {2015})}\BibitemShut {NoStop}%
\bibitem [{\citenamefont {Dyall}(2002)}]{Dyall02_dyalltz_4p5p6p}%
  \BibitemOpen
  \bibfield  {author} {\bibinfo {author} {\bibfnamefont {K.~G.}\ \bibnamefont
  {Dyall}},\ }\bibfield  {title} {\enquote {\bibinfo {title} {Relativistic and
  nonrelativistic finite nucleus optimized triple-zeta basis sets for the 4p,
  5p and 6p elements},}\ }\href {\doibase 10.1007/s00214-002-0388-0} {\bibfield
   {journal} {\bibinfo  {journal} {Theor. Chem. Acc.}\ }\textbf {\bibinfo
  {volume} {108}},\ \bibinfo {pages} {335--340} (\bibinfo {year}
  {2002})}\BibitemShut {NoStop}%
\bibitem [{\citenamefont {Dyall}(2006)}]{Dyall06_dyalltz_revise_4p5p6p}%
  \BibitemOpen
  \bibfield  {author} {\bibinfo {author} {\bibfnamefont {K.~G.}\ \bibnamefont
  {Dyall}},\ }\bibfield  {title} {\enquote {\bibinfo {title} {Relativistic
  {{Quadruple-Zeta}} and {{Revised Triple-Zeta}} and {{Double-Zeta Basis Sets}}
  for the 4p, 5p, and 6p {{Elements}}},}\ }\href {\doibase
  10.1007/s00214-006-0126-0} {\bibfield  {journal} {\bibinfo  {journal} {Theor.
  Chem. Acc.}\ }\textbf {\bibinfo {volume} {115}},\ \bibinfo {pages} {441--447}
  (\bibinfo {year} {2006})}\BibitemShut {NoStop}%
\bibitem [{\citenamefont {Dyall}(2004)}]{Dyall04_dz_tz_qz_5d}%
  \BibitemOpen
  \bibfield  {author} {\bibinfo {author} {\bibfnamefont {K.~G.}\ \bibnamefont
  {Dyall}},\ }\bibfield  {title} {\enquote {\bibinfo {title} {Relativistic
  double-zeta, triple-zeta, and quadruple-zeta basis sets for the 5d elements
  {{Hf}}--{{Hg}}},}\ }\href {\doibase 10.1007/s00214-004-0607-y} {\bibfield
  {journal} {\bibinfo  {journal} {Theor. Chem. Acc.}\ }\textbf {\bibinfo
  {volume} {112}},\ \bibinfo {pages} {403--409} (\bibinfo {year}
  {2004})}\BibitemShut {NoStop}%
\bibitem [{\citenamefont {Dyall}(2007)}]{Dyall07_4d_dtqz}%
  \BibitemOpen
  \bibfield  {author} {\bibinfo {author} {\bibfnamefont {K.~G.}\ \bibnamefont
  {Dyall}},\ }\bibfield  {title} {\enquote {\bibinfo {title} {Relativistic
  double-zeta, triple-zeta, and quadruple-zeta basis sets for the 4d elements
  {{Y}}--{{Cd}}},}\ }\href {\doibase 10.1007/s00214-006-0174-5} {\bibfield
  {journal} {\bibinfo  {journal} {Theor. Chem. Acc.}\ }\textbf {\bibinfo
  {volume} {117}},\ \bibinfo {pages} {483--489} (\bibinfo {year}
  {2007})}\BibitemShut {NoStop}%
\bibitem [{\citenamefont {Dyall}\ and\ \citenamefont
  {Gomes}(2009)}]{Dyall09_revised_5d}%
  \BibitemOpen
  \bibfield  {author} {\bibinfo {author} {\bibfnamefont {K.~G.}\ \bibnamefont
  {Dyall}}\ and\ \bibinfo {author} {\bibfnamefont {A.~S.~P.}\ \bibnamefont
  {Gomes}},\ }\bibfield  {title} {\enquote {\bibinfo {title} {Revised
  relativistic basis sets for the 5d elements {{Hf}}--{{Hg}}},}\ }\href
  {\doibase 10.1007/s00214-009-0717-7} {\bibfield  {journal} {\bibinfo
  {journal} {Theor. Chem. Acc.}\ }\textbf {\bibinfo {volume} {125}},\ \bibinfo
  {pages} {97} (\bibinfo {year} {2009})}\BibitemShut {NoStop}%
\bibitem [{\citenamefont {Dyall}, \citenamefont {Tecmer},\ and\ \citenamefont
  {Sunaga}(2023)}]{Dyall23_dyatz_5d}%
  \BibitemOpen
  \bibfield  {author} {\bibinfo {author} {\bibfnamefont {K.~G.}\ \bibnamefont
  {Dyall}}, \bibinfo {author} {\bibfnamefont {P.}~\bibnamefont {Tecmer}}, \
  and\ \bibinfo {author} {\bibfnamefont {A.}~\bibnamefont {Sunaga}},\
  }\bibfield  {title} {\enquote {\bibinfo {title} {Diffuse {{Basis Functions}}
  for {{Relativistic}} s and d {{Block Gaussian Basis Sets}}},}\ }\href
  {\doibase 10.1021/acs.jctc.2c01050} {\bibfield  {journal} {\bibinfo
  {journal} {J. Chem. Theory Comput.}\ }\textbf {\bibinfo {volume} {19}},\
  \bibinfo {pages} {198--210} (\bibinfo {year} {2023})}\BibitemShut {NoStop}%
\bibitem [{\citenamefont {{Dunning, Jr.}}(1989)}]{Dunning89}%
  \BibitemOpen
  \bibfield  {author} {\bibinfo {author} {\bibfnamefont {T.~H.}\ \bibnamefont
  {{Dunning, Jr.}}},\ }\bibfield  {title} {\enquote {\bibinfo {title} {Gaussian
  basis sets for use in correlated molecular calculations. {I. The} atoms boron
  through neon and hydrogen},}\ }\href@noop {} {\bibfield  {journal} {\bibinfo
  {journal} {J. Chem. Phys.}\ }\textbf {\bibinfo {volume} {90}},\ \bibinfo
  {pages} {1007--1023} (\bibinfo {year} {1989})}\BibitemShut {NoStop}%
\bibitem [{\citenamefont {Kendall}, \citenamefont {{Dunning Jr.}},\ and\
  \citenamefont {Harrison}(1992)}]{Kendall92}%
  \BibitemOpen
  \bibfield  {author} {\bibinfo {author} {\bibfnamefont {R.~A.}\ \bibnamefont
  {Kendall}}, \bibinfo {author} {\bibfnamefont {T.~H.}\ \bibnamefont {{Dunning
  Jr.}}}, \ and\ \bibinfo {author} {\bibfnamefont {R.~J.}\ \bibnamefont
  {Harrison}},\ }\bibfield  {title} {\enquote {\bibinfo {title} {{Electron
  affinities of the first‐row atoms revisited. Systematic basis sets and wave
  functions}},}\ }\href {\doibase 10.1063/1.462569} {\bibfield  {journal}
  {\bibinfo  {journal} {J. Chem. Phys.}\ }\textbf {\bibinfo {volume} {96}},\
  \bibinfo {pages} {6796--6806} (\bibinfo {year} {1992})}\BibitemShut {NoStop}%
\bibitem [{\citenamefont {Woon}\ and\ \citenamefont {{Dunning
  Jr.}}(1993)}]{Woon93}%
  \BibitemOpen
  \bibfield  {author} {\bibinfo {author} {\bibfnamefont {D.~E.}\ \bibnamefont
  {Woon}}\ and\ \bibinfo {author} {\bibfnamefont {T.~H.}\ \bibnamefont
  {{Dunning Jr.}}},\ }\bibfield  {title} {\enquote {\bibinfo {title}
  {{Gaussian-Basis Sets for Use in Correlated Molecular Calculations. III. the
  Atoms Aluminum Through Argon}},}\ }\href {\doibase 10.1063/1.464303}
  {\bibfield  {journal} {\bibinfo  {journal} {J. Chem. Phys.}\ }\textbf
  {\bibinfo {volume} {98}},\ \bibinfo {pages} {1358--1371} (\bibinfo {year}
  {1993})}\BibitemShut {NoStop}%
\end{thebibliography}%






\end{document}